\documentclass[twocolumn, twocolappendix]{aastex7}
\usepackage{amsmath}
\usepackage{amssymb}
\usepackage{textcomp}
\usepackage{tabularx}
\usepackage{longtable}
\usepackage{multirow}
\usepackage{array}
\usepackage{color}
\usepackage{subfigure}
\usepackage{times}
\usepackage{epsfig}
\usepackage[T1]{fontenc}
\usepackage{ae,aecompl}

\received{January 1, 2020}
\revised{January 1, 2020}
\accepted{\today}

\submitjournal{AJ}

\usepackage{multirow,hhline,graphicx,array}
\newcolumntype{M}[1]{>{\centering\arraybackslash}m{#1}}
\usepackage{rotating} 

\graphicspath{{./}{figures/}}

\newcommand{\Teff}{\mbox{\,\em T$_{\rm eff}$}}         
\newcommand{\kelvin}{\,\mbox{K}}                       

\begin{document}

\title[Metal-poor Stars in the Solar Neighborhood]{Stellar Ancestry Unlocked: Chemical and Orbital Clues Link Metal-Poor Stars to Globular Clusters and the Galaxy’s Thick Disk}
\author[0000-0001-9279-5995]{Gizay Yolalan}
\affiliation{Department of Space Sciences and Technologies, Faculty of Science, Akdeniz University, 07058, Antalya, T\"{u}rkiye}
\email{gizayyolalan@gmail.com}

\author[0000-0002-0296-233X]{Timur \c Sahin}
\affiliation{Department of Space Sciences and Technologies, Faculty of Science, Akdeniz University, 07058, Antalya, T\"{u}rkiye}
\email[show]{timursahin@akdeniz.edu.tr}

\author[0000-0003-3510-1509]{Sel\c{c}uk Bilir}
\affiliation{Department of Astronomy and Space Sciences, Faculty of Science, Istanbul University, 34119, Istanbul, T\"{u}rkiye}
\email{sbilir@istanbul.edu.tr}

\author[0000-0002-0435-4493]{Olcay Plevne}
\affiliation{Department of Astronomy and Space Sciences, Faculty of Science, Istanbul University, 34119, Istanbul, T\"{u}rkiye} 
\email{olcayplevne@istanbul.edu.tr}

\begin{abstract}
This study presents a detailed chemical, kinematic, and orbital dynamic analysis of five metal-poor stars in the solar neighborhood: HD\,2665, HD\,5916, HD\,122956, HD\,189349, and HD\,218857. Using high-resolution spectroscopic data from the ELODIE and ESPaDOnS instruments, we derived elemental abundances for 29 species (25 elements: C, O, Na, Mg, Al, Si, S, Ca, Sc, Ti, V, Cr, Mn, Fe, Co, Ni, Cu, Zn, Sr, Y, Zr, Ba, Ce, Nd, and Sm) via LTE-based analysis with ATLAS9 model atmospheres. Notably, we report first-time detections of Ce and Nd in HD\,2665; Al, V, Sm, and Mn in HD\,5916; Al in HD\,122956; and C, O, S, Sc, Mn, Co, Cu, Zn, Sr, Zr, Nd, and Sm in HD\,189349. Dynamical and chemical diagnostics reveal distinct origins: HD\,2665 shows strong orbital and chemical similarity to GC NGC\,5139 ($\omega$\,Cen), while HD\,218857 exhibits chemodynamic signatures consistent with NGC\,5634. HD\,122956 aligns with NGC\,6864 (M75), though intriguingly shares age, metallicity ([Fe/H]), and [Mg/Fe] ratios with NGC\,6517—a cluster whose reported abundances are derived solely from one star with APOGEE $H$-band spectroscopic measurements, as no optical spectroscopic data exist for its members. In contrast, HD\,5916 and HD\,189349 exhibit kinematic and chemical properties consistent with the field-star population and are classified as thick disk members.  
\end{abstract}


\section{Introduction}

The Milky Way (MW) comprises distinct structural components, including the thin and thick disks, the bulge, and the halo, each characterized by unique chemodynamic signatures that reflect variations in the stellar chemical composition and dynamic behavior. A comprehensive understanding of the MW is critical not only for elucidating its role as the astrophysical environment in which the Solar System originated \citep{Gonzalez2001, Stojković2019} but also for establishing it as a foundational archetype for the study of galactic morphology and evolutionary mechanisms. 

Numerous spectroscopic studies conducted on thin- and thick-disk stars since the early 2000s \citep{Prochaska2000, Feltzing2003, Reddy2003, Bensby2003, Bensby2004, Bensby2005, Mishenina2004} have significantly advanced our understanding of the chemical compositions of the Galactic disk components, the thin and thick disks \citep{Yoshii1982, Gilmore1983}. A hallmark of thick- and thin-disk stars in the MW is the abundance of distinct $\alpha$ elements (such as O, Mg, Si, Ca, and Ti). Thick disk stars consistently exhibit higher [$\alpha$/Fe] ratios than thin disk stars, often accompanied by subsolar metallicities of approximately -0.5 dex. This suggests that the thick disk formed on a shorter time scale from gas enriched with $\alpha$-elements produced by massive stars \citep{Fuhrmann1998, Bensby2003, Bensby2005, Bensby2014, Reddy2003, Reddy2006, Nissen2011, Adibekyan2012, Adibekyan2013, Recio-Blanco2014, Karaali2019, Yaz2017, Plevne2020, Doner2023, Akbaba2024}. An intermediate $\alpha$ population has also been identified at higher metallicities \citep{Bensby2007, Hayden2015, Mikolaitis2017}. Despite extensive research, the formation mechanisms of thin and thick disks remain the subject of ongoing debate \citep{Rix2013}. While several theories have been proposed, none can fully explain all the observed properties of the thick disk. 

Recent cosmological simulations, such as GALACTICA and NEWHORIZON, have not yet provided definitive evidence for the formation mechanism of thick disks \citep{Park2021}. These simulations explored various factors influencing disk formation, such as gas dynamics, star formation rates, and the role of external accretion events. However, the complex interplay between these factors complicates the identification of a single underlying mechanism.

In this context, early models of Galactic formation, such as the monolithic collapse model proposed by \citet{eggen1962}, provided foundational insights into how the Galaxy may have formed. According to this framework, the Galaxy's disk is formed from a single, rapid collapse of primordial gas. This model was later refined by \citet{searle1978}, who suggested that the Galactic halo comprises two distinct components: an inner halo formed via dissipational collapse with prograde motion, and an outer halo formed by the accretion of dwarf galaxies. The latter component, which is now recognized as being more consistent with cosmological models, plays a crucial role in the formation of the Galactic halo. \citet{Nissen2010} empirically demonstrated this dichotomy using stellar abundance and kinematic analyses. Their study argued that stars with [Fe/H] $< -0.8$ dex and low-$\alpha$ abundances originated from disrupted dwarf galaxies, while high-$\alpha$ stars formed in situ. Notably, a 0.2 dex difference in $\alpha$ abundance was found between the two populations.

The Galactic halo, primarily composed of stars accreted from disrupted dwarf galaxies and globular clusters (GCs), reflects the complex assembly history of the Milky Way. These accreted stars maintain their kinematic and spatial coherence for some time before evolving, governed by their orbital and internal dynamics. Although orbital information has proven highly valuable in tracing such structures, it is important to recognize its limitations. Recent high-resolution simulations by \citet{Pagnini2023} demonstrated that orbital dynamics alone cannot reliably reconstruct the full accretion history of the Galaxy. In particular, GCs accreted during major mergers, such as those analogous to the Gaia-Sausage-Enceladus (GSE) event, tend to lose their dynamical coherence over time and do not remain clustered in integrals-of-motion spaces (e.g., energy–angular momentum). Instead, they become broadly distributed, overlapping significantly with both in-situ clusters and debris from other progenitors, particularly in the deeply bound regions of the phase space. Therefore, identifying the origin of stellar systems based solely on kinematics can be misleading. Complementary diagnostics, particularly chemical abundances and age-metallicity relations, are essential for mitigating these ambiguities. This point is further emphasized by \citet{Thomas2025}, who highlight the interpretive challenges inherent in using integrals of motion alone to reconstruct ancient accretion events.

The GSE merger is considered a major accretion event that significantly influenced the early assembly and structure of the proto-Galaxy, particularly the Galactic halo and disk \citep{helmi2018, helmi2020, dodd2023, ceccarelli2024, fukushima2025}. This event involved the accretion of a massive dwarf galaxy, and its remnants continue to influence stellar populations in the solar neighborhood. Although the halo is predominantly formed from accreted substructures, the stellar populations in the solar vicinity reflect a mixture of in situ and accreted components. Most stars in this region belong to the dynamically ``cold'' thin disk; however, a significant fraction is associated with the thick disk. Thick disk stars have orbits with higher eccentricities. They exhibit distinct kinematic properties, such as higher velocity dispersions, a larger rotational lag, and chemically evolved signatures, including elevated $\alpha$-element ratios, compared with thin disk stars \citep[see e.g.][]{Chiba2000, Soubiran2003, Trevisan2011, Bensby2011, Bland-Hawthorn2016}. A small fraction of stars in the solar vicinity, identified by their old age, low metallicity, and retrograde kinematics, have been classified as halo interlopers. These stars likely represent debris from ancient accretion events that spatially overlap with the disk \citep{Bilir2006, Bilir2008, Bilir2012}. However, a metal-rich $\alpha$-enhanced stellar population, potentially originating from the inner disk, has also been observed in the solar neighborhood \citep[for example][]{Adibekyan2011}.  

A clear definition of these populations is essential to better understand the formation and evolution of Galactic stellar populations. This is crucial not only for testing theories of Galactic chemical and dynamical evolution but also for refining models of early nucleosynthesis.

Modern large-scale stellar surveys, including GALAH \citep{Heijmans2012, desilva2015}, GES \citep{Gilmore2012}, LAMOST \citep{Zhao2012}, APOGEE \citep{Majewski2017}, SEGUE \citep{Yanny2009}, and RAVE \citep{Steinmetz2006} have revolutionized our understanding of the chemical evolution of the Galaxy by delivering high-precision stellar parameters for late-type, metal-poor stars (e.g., F-, G-, and K-type dwarfs). These stars serve as critical tracers of early galactic enrichment due to their unblended spectral lines, which enable robust medium-resolution chemical analyses. Building on previous research on F dwarfs \citep{Sahin2020} and G stars \citep{Sahin2023, Marismak2024, Cinar2025}, this study presents new data on metallicities and $\alpha$-element abundances for a sample of G-type, metal-poor, and high-proper motion (HPM) stars from the ELODIE library \citep{Prugniel2001}. In addition, we determined the ages, kinematic properties, and galactic orbits of the stars, focusing on precise metallicity measurements, which are crucial for accurate age determination of the stars.

This study primarily aims to investigate the astrophysical and kinematic properties of metal-poor, HPM G-type stars in the solar neighborhood and highlight their significance in understanding the structure and evolution of the Milky Way. Typically associated with the thick disk or halo components of the Galaxy, these stars provide valuable insights into the early evolutionary history of the Galaxy. Owing to their solar-like masses and temperatures, G-type stars are ideal targets for detailed spectroscopic analysis, which allows for the accurate determination of their atmospheric parameters and chemical compositions. Studying metal-poor G-type stars is also important for understanding the Galaxy's chemical enrichment processes. Furthermore, stars with HPM are generally linked to older Galactic populations and play a pivotal role in mapping the Milky Way's dynamical structure \citep[c.f.,][]{eggen1962, Sandage1969, Carney1994}. \citet{Bensby2003, Bensby2005, Bensby2014}, \citet{Nissen2010}, and \citet{Karatas2005} have shown that these stars differ significantly in both their chemical abundance patterns and their orbital parameters. With the advent of precise astrometric data from the {\it Gaia} mission \citep{Gaia_Mission, Gaia2018, Gaia2022}, it has become possible to trace the Galactic orbits of these stars and thus infer their dynamical origins with greater accuracy. Another objective of the present study is to determine the Galactic origins of these stars through detailed orbital analysis. Combining chemical abundance information with kinematic parameters allows us to infer the environments in which these stars formed and gain insight into past merger events in the Milky Way's history. Specifically, the joint analysis of the [$\alpha$/Fe] ratios and orbital eccentricities \citep[see][]{Hayes2018, Mackereth2019} allows for a more reliable distinction between different Galactic structures. The present study expands on our group's earlier research on metal-poor G-type HPM stars in the solar neighborhood, as outlined by \citet{Marismak2024} and \citet{Cinar2025}. It also builds upon the earlier analysis of F-type HPM stars presented by \citet{Sahin2020}. In this context, the present study aims to characterize a small yet representative sample of metal-poor HPM G-type stars. This study will assess their likely Galactic origins using a joint chemo-dynamical approach. These efforts will contribute to a broader understanding of the assembly history of the Milky Way and the nature of its oldest stellar components.

The remainder of this paper is organized as follows: Section 2 provides information on data selection and observations. Section 3 details the methodologies employed: Section 3.1 presents the spectroscopic analysis, including the determination of model parameters, chemical abundance analyses, and corrections for non-local thermodynamic equilibrium (non-LTE) effects, i.e., departures from local thermodynamic equilibrium (LTE) for the studied elements; Section 3.2 describes the stellar age estimation process; and Section 3.3 discusses the kinematic properties, orbital dynamics, and $E$–$L_{\rm z}$ plane positions of the HPM stars with halo substructures. Section 4 synthesizes the results and broader implications: Section 4.1 summarizes the spectral analysis outcomes for individual stars (HD\,2665, HD\,189349, HD\,5916, HD\,122956, and HD\,218857), with subsections dedicated to each object (4.1.1–4.1.5); Section 4.2 interprets the [Mg\,{\sc i}/Mn\,{\sc i}] vs. [Al\,{\sc i}/Fe] diagram in a chemodynamic context; Section 4.3 explores the Galactic origins of the stars, including encounter probabilities with candidate GCs; and Section 4.4 outlines prospects for future work.

\section{Data}

To identify suitable target stars, we selected G-type stars from the ELODIE library \citep{Soubiran2003}. Of the 545 G spectral-type stars listed in ELODIE, 173 were included in the {\it Gaia} DR3 catalog \citep{Gaia2023}. Next, we focused on HPM stars within the G spectral-type group. A thorough search was performed to ensure the reliability of our analysis. Our final sample of HPM stars comprised 90 G-type stars, with effective temperatures ranging from $4\,285 < T_{\rm eff}~(\kelvin)< 6\,174$, surface gravities between $0.20<\log g~({\rm cgs})<4.71$, metallicities between -$2.49< {\rm [Fe/H]~(dex)}<-0.44$, and distances between $4<d~(\rm pc)<1\,680$.

To further refine our sample, we imposed an RUWE cutoff of 1.4 for the remaining samples. RUWE is a dimensionless value that assesses the quality of astrometric solutions and is calculated by comparing the observed position of a star with its expected position based on its astrometric solution. A RUWE value less than 1.4 is indicative of a high-quality astrometric solution, suggesting that the source is well-behaved and consistent with the single-star model, as recommended by \cite{vanLeeuwen2021}.

\begin{figure*}
 \centering 
 \includegraphics[width=1.0\linewidth]{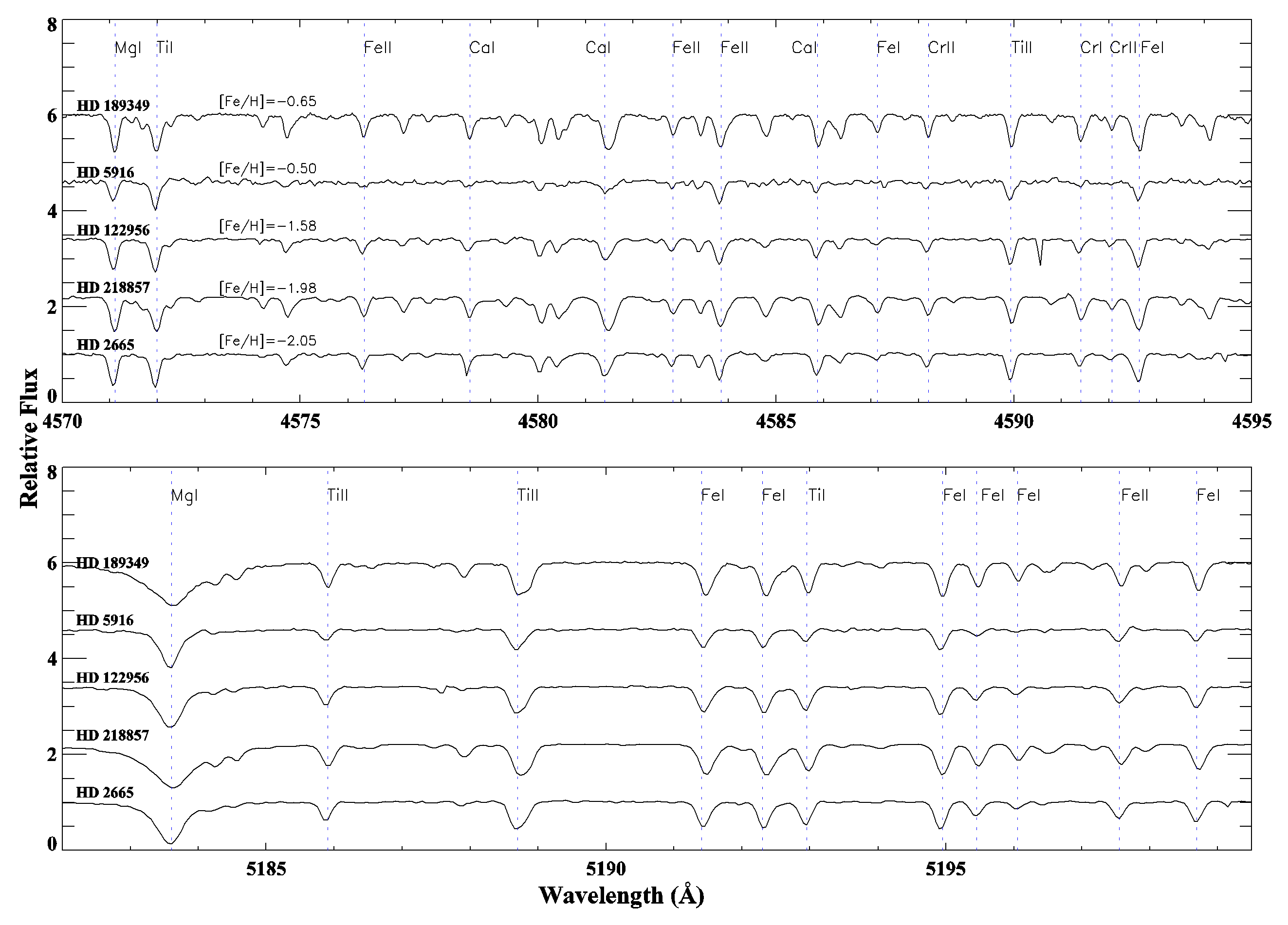}
 \caption{A small region of the spectrum for five HPM sample stars. Identified lines are also indicated.}
 \label{fig:1}
 \end{figure*}
 
\begin{table}
\small
\centering
\renewcommand{\arraystretch}{1}
\setlength{\tabcolsep}{1.5pt}
\caption{Log of observations for the HPM sample stars. The {\it S/N} values in the raw spectra are reported near 5550 \AA.}
\label{tab:1}
\begin{tabular}{lcccccc}
\hline
\hline
Star & Sp.  & Ref.$^{\rm a}$ & Exp. & S/N  &  $V_{\rm hel}$    &    MJD   \\
     & Type &                & (s)  &      & (km s$^{\rm -1}$) & (2400000+)\\
\hline
HD\,002665 & G5III & 1 & 3000 & 135 & -382.34 & 51531.78673 \\
HD\,005916 & G8III-IV & 3 & 1800 & 131 &  -68.17 & 50363.05958 \\
HD\,122956 & G6IV  & 2 & 1500 &  83 &  165.35 & 50565.98404 \\
HD\,189349 & G4III-IV & 3 &   60 & 190 & -104.2$^{\rm b}$  & 57650.25224 \\
HD\,218857 & G5    & 2 & 3601 &  68 & -170.17 & 50687.03151 \\
\hline
\hline
\end{tabular}
\\
($^{\rm a}$) References for spectral type: (1) \citet{Bond1970}, (2) \citet{Houk1988}, and (3) \citet{Molenda2013}. ($^{\rm b}$) Radial velocity from the least-squares deconvolution (LSD) profile.
\end{table}

By applying these rigorous selection criteria, we assembled a sample of 74 G-type HPM stars within the metallicity range of -2.5 to -0.5 dex. We then prioritized stars at various evolutionary stages to ensure a diverse sample. Additionally, the availability and quality of observational data, such as high-resolution spectra, were crucial in our selection. Finally, we prioritized stars with accurate parallax and proper motion measurements from {\it Gaia} DR3 \citep{Gaia2023}, which is particularly important for deriving reliable fundamental parameters. Spectroscopic methods are commonly used to derive stellar parameters such as $T_{\rm eff}$, $\log g$, and [Fe/H]. However, degeneracies between these parameters, particularly those involving $\log g$, can limit the reliability of the results because $\log g$ is not directly observable and is strongly correlated with other parameters of the star. High-precision trigonometric parallax measurements provided by the {\it Gaia} mission \citep{Gaia_Mission} enable the accurate determination of stellar distances and, consequently, stellar luminosities ($L$). Combining these astrometric data with spectroscopically or photometrically derived parameters, such as $T_{\rm eff}$ and stellar radius ($R$), significantly improves the determination of $\log g$. Therefore, astrometric measurements play a critical role in overcoming parameter degeneracies in spectroscopic analyses and enhancing the accuracy and reliability of fundamental stellar parameter determinations. By carefully considering these factors, we selected a representative sample of five metal-poor stars in the solar neighborhood. Table \ref{tab:1} lists the complete sample used in this study.

High-resolution ($R=42\,000$) and high signal-to-noise ratio (S/N; up to 135) spectra for HD\,2665, HD\,5916, HD\,122956, and HD\,218857 obtained from the 1.93 m telescope at the Haute Provence Observatory and the {\sc ELODIE} fiber-fed cross-dispersed echelle spectrograph were processed using the data reduction pipeline at the telescope, which included continuum normalization, wavelength calibration, and radial velocity corrections. However, issues in the continuum placement were identified in some of the pipeline-normalized spectra. To ensure consistency and accuracy in the abundance analysis, all {\sc ELODIE} spectra were renormalized using an interactive tool, {\sc LIME} \citep{Sahin2017}, developed in {\sc IDL}. The final processed spectra cover the wavelength range 3895--6815\,\AA.

For HD\,189349, we use the ESPaDOnS spectrum (star only, $R\sim 80\,000$) from the PolarBASE\footnote{\url{http://polarbase.irap.omp.eu/}} archive \citep{Petit2014}. As overlapping regions with low signal-to-noise ratios in normalized one-dimensional (1D) ESPaDOnS echelle spectra adversely affect spectral analyses, high-resolution ESPaDOnS spectra were obtained from the Canadian Astronomy Data Centre \citep[CADC\footnote{\url{https://www.cadc-ccda.hia-iha.nrc-cnrc.gc.ca/en/search/?collection=CFHT&noexec=true}};][]{Crabtree1994} data centered on the raw two-dimensional (2D) data format. Although 2D spectra have wavelength calibration, fundamental operations such as radial velocity correction, normalization, and order merging have not yet been applied. Thirty-seven orders were exported to the external environment in ASCII format using a Python interface. Subsequently, continuum normalization and radial-velocity correction were performed in a Python environment. The order-merging process was performed using the STARLINK-DIPSO package \citep{Howarth1998}. The characteristics of the target star spectra are shown in Figure \ref{fig:1}. Several lines in the spectra that were suitable for abundance analysis were unblended.

In addition to normalizing the spectra, LIME was used for the line identification. It provides the most likely identifications for the lines of interest and lists recent atomic data (e.g., Rowland Multiplet Number-RMT and $\log gf$ lower-level excitation potential (LEP)) compiled from literature sources (from the {\sc NIST} database). The equivalent widths (EWs) were obtained using the SPECTRE \citep{Sneden1973} and LIME codes. The results for a representative sample of weak and strong areas were in good agreement, with a difference of $\pm 4$ m\AA.

\section{Results}

\subsection{Spectroscopic Analysis}

For the abundance analysis of the five selected G-type stars in this study—hereafter referred to as the program stars— we used the {\sc ATLAS9} model atmospheres \citep{Castelli2004} assuming LTE with ODFNEW opacity distributions. The {\sc LTE} line analysis code MOOG \citep{Sneden1973}\footnote{The source code of the MOOG can be downloaded from \\ \url{http://www.as.utexas.edu/~chris/moog.html}} was used to calculate the elemental abundances of the program stars. The procedures adopted for the abundance analysis, along with the sources of atomic data, are consistent with those employed in earlier studies by \citet{Sahin2009}, \citet{Sahin2011, Sahin2016}, \citet{Sahin2020}, \citet{Sahin2023, Sahin2024}, \citet{Marismak2024}, \citet{Senturk2024}, and \citet{Cinar2025}. In the following sections, we discuss the line list, atomic data, and derivation of stellar parameters.

We conducted a thorough search for unblended lines by calculating synthetic spectra for the observed wavelength regions using MOOG. In only a small percentage of cases, we used spectrum synthesis instead of a direct EW estimate. When selecting iron lines, we were careful to consider that the 3D hydrodynamical model atmospheres of cool stars suggest a lower-level energy dependence of 3D abundance corrections for neutral iron lines in metal-poor stars \citep[e.g.][]{Dobrovolskas2013}, and that low-excitation (i.e. $E_{\rm exc}<$1.2 eV) Fe\,{\sc i} lines may provide higher abundances compared to lines with higher excitation energies \citep[e.g.][]{Lai2008}. Our analysis of the mean abundances from the low-excitation lines of Fe\,{\sc i} showed a difference of approximately 0.05 dex in the mean neutral iron abundances among the program stars. We found that the 3D effect on \Teff\, was negligible in this case. The $gf$-values for the selected lines of Fe\,{\sc i} and Fe\,{\sc ii} were obtained from the compilation of \citet{Fuhr2006}. A list of identified lines and the most up-to-date atomic data were provided by \citet{Sahin2023, Sahin2024}.

To ensure the accuracy of $gf$values and reduce systematic errors, we derived solar abundances using stellar lines. We measured the lines from the solar flux atlas of \citet{Kurucz1984} and analyzed them using the solar model atmosphere from the \citet{Castelli2004} grid, which had $\Teff\,=$ 5790 K, $\log g\,=$ 4.40 cgs, [Fe/H]= 0 dex, and a microturbulence of $\xi$=0.66 km s$^{-1}$. The abundances are listed in Table \ref{tab:solarabund}. We found that the solar abundances were recreated fairly well and compared them to those reported by \citet{Asplund2009} in their review. We used our derived solar abundances as the reference when computing stellar abundances; thus, stellar abundances in this study are presented differentially with respect to our own solar values, rather than those reported in the literature. This approach helps reduce errors caused by uncertainties in oscillator strengths, the influence of spectrograph characteristics, and deviations from LTE. Errors in the effective temperatures obtained from the spectroscopic excitation technique can also stem from systematic errors in the oscillator strength as a function of the excitation potential. The same is true for the errors in the equivalent widths.

\begin{table}
\scriptsize
\centering
\setlength{\tabcolsep}{4pt}
\renewcommand{\arraystretch}{1.2}
\caption{Solar abundances obtained by employing the solar model atmosphere from \citet{Castelli2004} compared to the photospheric abundances from \citet{Asplund2009}. Elements measured via spectral synthesis are marked with a, while the remaining abundances were calculated using the line EWs. $\Delta {\rm log} \epsilon_{\rm \odot}{\rm (X) = log} \epsilon_{\rm \odot} {\rm (X)}_{\rm This\,study} - {\rm log} \epsilon_{\rm \odot} {\rm (X)}_{\rm Asplund}$  }
\centering
\begin{tabular}{l|c|c|c|c}
\hline
     	&   This study &  &   \citet{Asplund2009} \\
\cline{2-2}
\cline{4-4}
Species   &  $\log\epsilon_{\rm \odot}$(X) & $N$ &   $\log\epsilon_{\rm \odot}$(X) & $\Delta\log\epsilon_{\rm \odot}$(X) \\
\cline{2-2}
\cline{4-4}
		&  (dex) &&  (dex)&  (dex)\\
 \hline
\textbf{C\,{\sc i}}\textsuperscript{a}   & 8.48 $\pm$ 0.11 & 2   & 8.43 $\pm$ 0.05 & 0.05  \\
\textbf{O\,{\sc i}}\textsuperscript{a}   & 8.81 $\pm$ 0.03 & 3   & 8.69 $\pm$ 0.05 & 0.12  \\
Na\,{\sc i}           & 6.17 $\pm$ 0.09 & 3	  & 6.24 $\pm$ 0.04 & -0.07 \\
\textbf{Mg\,{\sc i}}\textsuperscript{a}  & 7.62 $\pm$ 0.03 & 5	  & 7.60 $\pm$ 0.04 & 0.02  \\
\textbf{Al\,\,{\sc i}}\textsuperscript{a}& 6.45 $\pm$ 0.02 & 8   & 6.45 $\pm$ 0.03 & 0.00  \\
Si\,{\sc i}  	      & 7.50 $\pm$ 0.07 & 12  & 7.51 $\pm$ 0.03 & -0.01 \\
Ca\,{\sc i}  	      & 6.34 $\pm$ 0.08 & 18  & 6.34 $\pm$ 0.04 & 0.00  \\
Sc\,{\sc i} 	      & 3.12 $\pm$ 0.00 & 1	  & 3.15 $\pm$ 0.04 & -0.03 \\
Sc\,{\sc ii} 	      & 3.23 $\pm$ 0.08 & 7	  & 3.15 $\pm$ 0.04 & 0.08  \\
Ti\,{\sc i}  	      & 4.96 $\pm$ 0.09 & 43  & 4.95 $\pm$ 0.05 & 0.01  \\
Ti\,{\sc ii} 	      & 4.99 $\pm$ 0.08 & 12  & 4.95 $\pm$ 0.05 & 0.04  \\
\textbf{V\,{\sc i}}\textsuperscript{a}   & 3.90 $\pm$ 0.03 & 5	  & 3.93 $\pm$ 0.08 & -0.03 \\
Cr\,{\sc i}  	      & 5.71 $\pm$ 0.07 & 19  & 5.64 $\pm$ 0.04 & 0.07  \\
Cr\,{\sc ii} 	      & 5.64 $\pm$ 0.14 & 3	  & 5.64 $\pm$ 0.04 & 0.00  \\
\textbf{Mn\,{\sc i}}\textsuperscript{a}  & 5.62 $\pm$ 0.13 & 13  & 5.43 $\pm$ 0.05 & 0.19  \\
Fe\,{\sc i}  	      & 7.54 $\pm$ 0.09 & 132 & 7.50 $\pm$ 0.04 & 0.04  \\
Fe\,{\sc ii}          & 7.51 $\pm$ 0.04 & 17  & 7.50 $\pm$ 0.04 & 0.01  \\
\textbf{Co\,{\sc i}}\textsuperscript{a}  & 4.90 $\pm$ 0.14 & 6	  & 4.99 $\pm$ 0.07 & -0.05 \\
Ni\,{\sc i}  	      & 6.28 $\pm$ 0.09 & 54  & 6.22 $\pm$ 0.04 & 0.06  \\
\textbf{Zn\,{\sc i}}\textsuperscript{a}  & 4.60 $\pm$ 0.00 & 2	  & 4.56 $\pm$ 0.05 & 0.04  \\
\textbf{Sr\,{\sc i}}\textsuperscript{a}  & 2.86 $\pm$ 0.00 & 1   & 2.87 $\pm$ 0.07 & -0.01 \\
\textbf{Y\,{\sc ii}}\textsuperscript{a}  & 2.20 $\pm$ 0.04 & 2	  & 2.21 $\pm$ 0.05 & -0.01 \\
\textbf{Zr\,{\sc ii}}\textsuperscript{a} & 2.70 $\pm$ 0.00 & 1	  & 2.58 $\pm$ 0.04 & 0.12  \\
\textbf{Ba\,{\sc ii}}\textsuperscript{a} & 2.52 $\pm$ 0.19 & 4	  & 2.18 $\pm$ 0.09 & 0.32  \\
\textbf{Ce\,{\sc ii}}\textsuperscript{a} & 1.58 $\pm$ 0.02 & 2	  & 1.58 $\pm$ 0.04 & 0.00  \\
\textbf{Nd\,{\sc ii}}\textsuperscript{a} & 1.33 $\pm$ 0.09 & 3	  & 1.42 $\pm$ 0.04 & -0.09 \\
\textbf{Sm\,{\sc ii}}\textsuperscript{a} & 0.95 $\pm$ 0.00 & 1	  & 0.96 $\pm$ 0.04 & -0.01 \\
\hline
\multicolumn{5}{l}{\textsuperscript{a}Derived using synthesis.} 
\end{tabular}
\label{tab:solarabund}
\end{table}

To ensure the accuracy of the $gf$values used in this study, we also compared them with those included in the {\it Gaia}-ESO line list v.6 provided by the GES collaboration \citep{Heiter2021}. This line list includes the recommended lines and atomic data (hyperfine structure-corrected $gf$ values) for the analysis of FGK stars. It should be noted that some lines in the spectra of FGK stars are either unidentified \citep{Heiter2015} or not suggested for abundance analysis \citep{Heiter2021}. The GES line list (v.6) contains 141\,233 atomic transitions for 206 species (78 elements) over the wavelength range 4\,200 – 10\,000 \AA. Within this same wavelength interval, 52\,786 lines were available for 24 elements (28 species) that overlapped with the elements analyzed in this study. This range corresponds to the coverage of the {\sc ELODIE} spectra used for the diagnostics. From our primary line list, 552 lines were initially identified within the broader 4\,000-10\,000 \AA\, interval as potential matches to the GES line list. Among these, 381 lines—spanning 4020 - 9120 \AA\, were found to be common. Of these, 352 lines lie within the 4200–6800 \AA\, range, aligning precisely with the GES line list coverage and satisfying a one-to-one match in terms of wavelength, atomic species, $\log gf$ and lower excitation potential (LEP). These matched lines were used to compute the elemental abundances reported in Table \ref{tab:solarabund}. The difference between the $\log gf$ values\footnote{$\log gf_{\rm This\, study}$ - $\log gf_{\rm GES}$} for the 128 common Fe\,{\sc i} lines (out of 132) in the GES line list was -0.02$\pm$0.12 dex. For the 16 Fe\,{\sc ii} lines (out of 17), the mean difference is 0.02$\pm$0.08 dex. Differences in $\log gf$-values for other elements are summarized as follows (number of lines in parentheses): –0.02 dex for C {\sc i} (2); 0.02 dex for Na {\sc i} (3); 0.32 dex for Mg\,{\sc i} (5); 0.29 dex for Al {\sc i} (8); -0.03 dex for Si\,{\sc i} (12) and Zn\,{\sc i} (2); –0.01 dex for Ca {\sc i} (18); 0.05 dex for Sc\,{\sc ii} (7); 0.03 dex for Ti\,{\sc ii} (12); 0.10 dex for Cr {\sc i} (19); 0.69 dex for Mn {\sc i} (11); 1.06 dex for Co {\sc i} (6); –0.05 dex for Ni {\sc i} (54); 0.71 dex for Cu {\sc i} (1) and V {\sc i} (5);  0.14 dex for Cr\,{\sc ii} (19); -0.05 dex for Ni\,{\sc i} (54); –0.07 dex for Y {\sc ii} (2); 0.27 dex for Ba\,{\sc ii} (4); no difference for O {\sc i} (3), Ti {\sc i} (41), Sr {\sc i} (1), Ce {\sc ii} (1), Nd {\sc ii} (2), and Sm {\sc ii} (1).

\begin{table}
\small
\centering
\setlength{\tabcolsep}{3.5pt}
\renewcommand{\arraystretch}{1.2}
\caption{Model atmospheric parameters of the five program stars and Sun.}
\centering
\begin{tabular}{lcccc}
\hline
Star	  &	$T_{\rm eff}$	      &	$\log g$ 	           &	[Fe/H]	             &  $\xi$               \\
\cline{2-5}
          & (K)                   & (cgs)                  &     (dex)               & (km s$^{-1}$)        \\
\hline
HD\,002665 & 5010$\pm$70 & 2.39$\pm$0.10 & -2.05$\pm$0.09 & 1.96$\pm$0.50 \\
HD\,005916 & 5200$\pm$45 & 3.06$\pm$0.06 & -0.50$\pm$0.06 & 1.63$\pm$0.50 \\
HD\,122956 & 4700$\pm$89 & 1.70$\pm$0.27 & -1.58$\pm$0.10 & 1.83$\pm$0.50 \\
HD\,189349 & 5000$\pm$125 & 2.60$\pm$0.21 & -0.65$\pm$0.13 & 1.41$\pm$0.50 \\
HD\,218857 & 5080$\pm$95 & 2.40$\pm$0.24 & -1.98$\pm$0.10 & 2.01$\pm$0.50 \\
Sun & 5790$\pm$45 & 4.40$\pm$0.09 & 0.00$\pm$0.04 & 0.66$\pm$0.50\\

\hline
\end{tabular}
  \label{tab:atmosphere_parameters}
\end{table}

Because the stellar parameters reported in the literature for some program stars presented large variations (see Table \ref{tab:lit1}), we decided to obtain new spectroscopic measurements of the model atmospheric parameters in this study. To determine stellar parameters such as the effective temperature ($T_{\rm eff}$), surface gravity ($\log g$), metallicity ([Fe/H]), and microturbulence ($\xi$), we used the neutral and ionized Fe lines reported in \citet{Sahin2023, Sahin2024}. The abundances of Fe\,{\sc i} and Fe\,{\sc ii} were used to determine the atmospheric parameters of the model. 

\begin{figure*}
\centering
\includegraphics[width=1\linewidth]{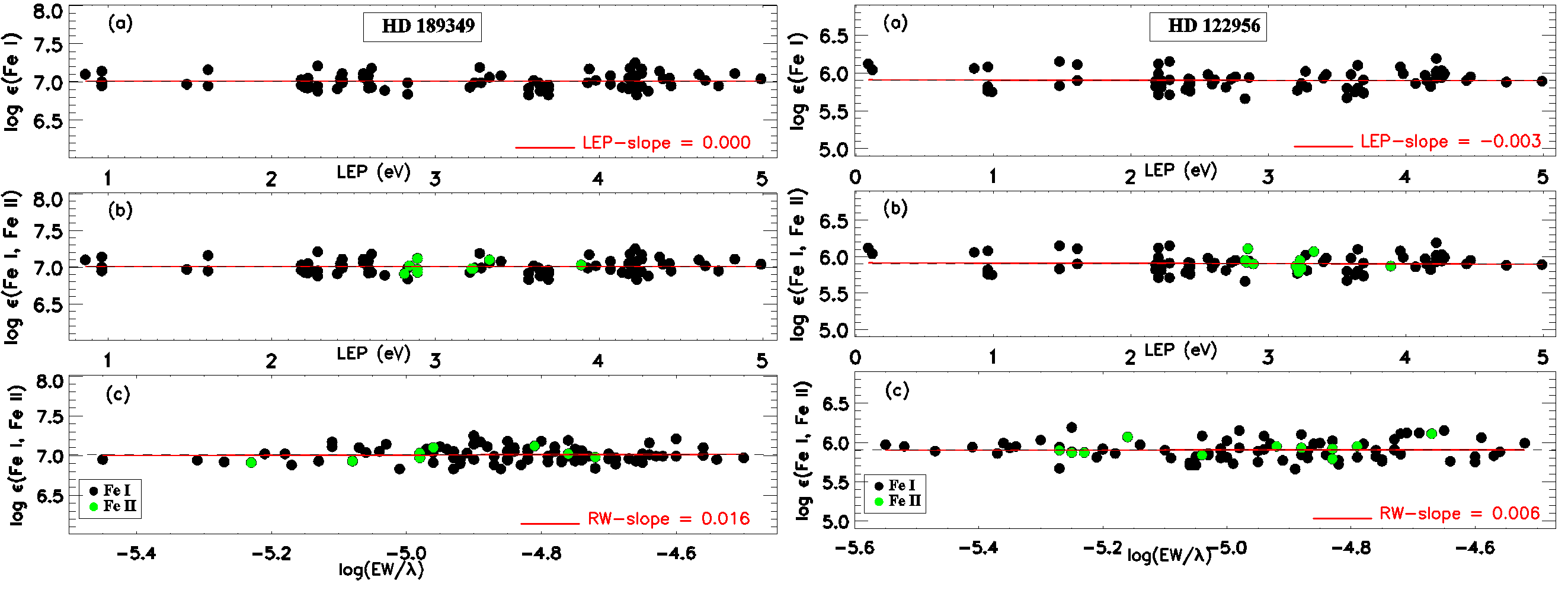}
\caption{An example for the determination of atmospheric parameters $T_{\rm eff}$ and $\xi$ using abundance ($\log \epsilon$) as a function of both LEP (panels a and b) and reduced EW (REW; log (EW/$\lambda$), panels c) for HD\,189349 and HD\,122956. The solid line in all panels is the least-squares fit to the data.}
\label{fig:moog_results_1}
\end{figure*}
 
The effective temperature was initially estimated by imposing the condition that the derived abundance was independent of the LEP. Microturbulence ($\xi$) was determined by assuming that the derived abundance was independent of the reduced equivalent width (EW) and that all lines had the same LEP and a similar wavelength. The precision in determining the microturbulent velocity is $\pm$0.5 km s$^{-1}$ (see Figure~\ref{fig:moog_results_1} and Figure \ref{fig:moog_results_2}). Surface gravity ($\log g$) was determined by requiring ionization equilibrium. For example, the Fe\,{\sc i} and Fe\,{\sc ii} lines provide the same Fe abundances. Iron lines are abundant, even in very metal-poor stellar samples. Owing to the interdependence of the atmospheric model parameters, an iterative procedure was employed, with small adjustments made to the model parameters at each step. We also verified that no significant trend in iron abundance with wavelength was observed. The resulting stellar model parameters, along with the solar model parameters, are listed in Table~\ref{tab:atmosphere_parameters}.

In {\it Gaia} DR2 \citep{Gaia2018}, only effective temperatures were reported for program stars, and {\it Gaia} DR3 \citep{Gaia2023} additionally includes surface gravity and metallicity values. According to the {\it Gaia} DR2 data, the effective temperatures determined for the program stars HD\,2665, HD\,5916, HD\,122956, HD\,189349, and HD\,218857 are 4986, 5029, 4712, 5114, and 5194 K, respectively. However, there was no consistency between the temperatures reported for these stars in the two data releases. In {\it Gaia} DR3, the derived effective temperatures and $\log g$ values (excluding HD\,122956) are 6073 K / 2.90 dex, 5488 K / 2.91 dex, 5557 K / 2.95 dex, and 5973 K / 3.14 dex, respectively. The {\it Gaia} DR2 method relies on RV template stars, similar to the target stars, to determine the effective temperatures. In contrast, {\it Gaia} DR3 uses the GSP-Phot framework\footnote{\url{https://www.mpia.de/gaia/projects/gsp} (Generalized Stellar Parametrizer)} to construct libraries (e.g., Aeneas, Priam, Support Vector Regression, and Ilium) for estimating stellar parameters through model-based analyses.

The logarithmic elemental abundances for the program stars, derived using the \texttt{MOOG} \textsc{abund} and \textsc{synth} interfaces, were normalized relative to the corresponding solar abundances, as listed in Table~\ref{tab:solarabund} and presented in Table~\ref{tab: normabund}. The sensitivity of the derived abundances to uncertainties in the stellar atmospheric parameters (i.e., $T_{\mathrm{eff}}$, $\log g$, [Fe/H], and $\xi$) is summarized in \mbox{Table~\ref{tab:modelerr}}, where the abundance variations resulting from 1$\sigma$ changes in each parameter are also shown.

\begin{table}
\tiny
    \centering
    \setlength{\tabcolsep}{0.8pt}
    \renewcommand{\arraystretch}{1.2}
    \caption{[X/Fe] abundances of the five program stars.}
    \begin{tabular}{l|ccc|ccc|ccc|ccc|ccc}
    \hline
    Species &     \multicolumn{3}{c|}{HD\,2665}       &     \multicolumn{3}{c|}{HD\,5916}       &      \multicolumn{3}{c|}{HD\,122956}      &      \multicolumn{3}{c|}{HD\,189349$^{*}$}    &      \multicolumn{3}{c}{HD\,218857}    \\
\cline{2-16}
    ~       & [X/Fe] & N & $\sigma$ & [X/Fe] & N & $\sigma$ &  [X/Fe] & N & $\sigma$  & [X/Fe] & N & $\sigma$ & [X/Fe] & N & $\sigma$ \\
    \hline
    \textbf{C\,{\sc i}}  & --    & -- & --   & --   & -- & --   & --    & -- & --   & 0.27  & 2  & 0.18 & --    & -- & --   \\
    \textbf{O\,{\sc i}}  & --    & -- & --   & --   & -- & --   & --    & -- & --   & 0.60  & 3  & 0.15 & --    & -- & --   \\
    Na\,{\sc i}          & --    & -- & --   & --   & -- & --   & --    & -- & --   & 0.22  & 3  & 0.21 & --    & -- & --   \\ 
    \textbf{Mg\,{\sc i}} & 0.23  & 2  & 0.09 & 0.19 & 2  & 0.07 & 0.33  & 2  & 0.10 & 0.12  & 3  & 0.22 & 0.16  & 2  & 0.10 \\ 
    \textbf{Al\,{\sc i}} & --    & -- & --   & 0.10 & 2  & 0.06 & 0.03  & 1  & 0.10 & 0.23  & 7  & 0.14 & --    & -- & --   \\
    Si\,{\sc i}          & --    & -- & --   & 0.17 & 7  & 0.12 & 0.31  & 1  & 0.12 & 0.27  & 18 & 0.19 & --    & -- & --   \\
    \textbf{S\,{\sc i}}  & --    & -- & --   & --   & -- & --   & --    & -- & --   & 0.03  & 1  & 0.13 & --    & -- & --   \\
    Ca\,{\sc i}          & 0.25  & 8  & 0.13 & 0.17 & 8  & 0.15 & 0.21  & 10 & 0.15 & 0.17  & 18 & 0.19 & 0.30  & 5  & 0.15 \\ 
    Sc\,{\sc ii}         & -0.01 & 4  & 0.12 & 0.22 & 5  & 0.15 & -0.02 & 4  & 0.15 & 0.28  & 10 & 0.18 & --    & -- & --   \\ 
    Ti\,{\sc i}          & 0.09  & 19 & 0.16 & 0.20 & 12 & 0.17 & 0.13  & 20 & 0.18 & 0.13  & 50 & 0.20 & 0.15  & 6  & 0.18 \\ 
    Ti\,{\sc ii}         & 0.21  & 12 & 0.17 & 0.27 & 3  & 0.13 & 0.22  & 10 & 0.16 & 0.19  & 10 & 0.21 & 0.34  & 3  & 0.18 \\ 
    \textbf{V\,{\sc i}}  & --    & -- & --   & 0.15 & 5  & 0.12 & -0.09 & 5  & 0.14 & -0.01 & 3  & 0.14 & --    & -- & --   \\ 
    Cr\,{\sc i}          & -0.20 & 9  & 0.13 & -0.11& 8  & 0.15 & -0.24 & 11 & 0.13 & -0.16 & 20 & 0.19 & --    & -- & --   \\ 
    Cr\,{\sc ii}         & --    & -- & --   & 0.03 & 3  & 0.18 & 0.05  & 2  & 0.18 & -0.05 & 4  & 0.21 & --    & -- & --   \\ 
    \textbf{Mn\,{\sc i}} & -0.50 & 4  & 0.26 & -0.26& 6  & 0.21 & -0.51 & 6  & 0.23 & -0.36 & 12 & 0.17 & -0.43 & 2  & 0.24 \\ 
    Fe\,{\sc i}          & -0.03 & 44 & 0.17 & -0.03& 62 & 0.14 & -0.06 & 75 & 0.18 & -0.02 & 188& 0.21 & 0.01  & 29 & 0.17 \\ 
    Fe\,{\sc ii}         & 0.00  & 9  & 0.13 & 0.00 & 6  & 0.08 & 0.00  & 11 & 0.14 & 0.00  & 23 & 0.19 & 0.00  & 4  & 0.14 \\ 
    \textbf{Co\,{\sc i}} & 0.12  & 3  & 0.19 & 0.14 & 6  & 0.23 & -0.16 & 5  & 0.30 & -0.03 & 5  & 0.16 & 0.09  & 1  & 0.17 \\ 
    Ni\,{\sc i}          & -0.06 & 8  & 0.16 & 0.01 & 23 & 0.15 & -0.13 & 15 & 0.19 & -0.01 & 59 & 0.21 & --    & -- & --   \\
    \textbf{Cu\,{\sc i}} & --    & -- & --   & --   & -- & --   & --    & -- & --   &-0.01  & 1  & 0.15 & --    & -- & --   \\
    \textbf{Zn\,{\sc i}} & 0.03  & 2  & 0.10 & 0.08 & 2  & 0.06 & 0.01  & 2  & 0.11 & 0.13  & 2  & 0.15 & 0.21  & 2  & 0.15 \\ 
    \textbf{Sr\,{\sc i}} & --    & -- & --   &-0.39 & 1  & 0.06 & -0.34 & 1  & 0.10 & -0.30 & 1  & 0.13 & --    & -- & --   \\ 
    \textbf{Y\,{\sc ii}} & -0.47 & 2  & 0.12 & -0.19& 2  & 0.08 & -0.27 & 2  & 0.12 & -0.40 & 2  & 0.14 & -0.44 & 2  & 0.19 \\ 
    \textbf{Zr\,{\sc ii}}& -0.22 & 1  & 0.09 & --   & -- & --   &  0.08 & 1  & 0.10 & -0.26 & 1  & 0.14 & --    & -- & --   \\ 
    \textbf{Ba\,{\sc ii}}& -0.36 & 3  & 0.12 & 0.05 & 4  & 0.10 & 0.06  & 4  & 0.14 & -0.17 & 2  & 0.14 & -0.59 & 3  & 0.12 \\ 
    \textbf{Ce\,{\sc ii}}& 0.05  & 1  & 0.09 & 0.15 & 2  & 0.09 & 0.02  & 2  & 0.10 & 0.01  & 1  & 0.14 & --    & -- & --   \\ 
    \textbf{Nd\,{\sc ii}}& 0.27  & 1  & 0.13 & 0.35 & 3  & 0.12 & 0.23  & 3  & 0.15 & -0.06 & 3  & 0.20 & 0.34  & 1  & 0.13 \\ 
    \textbf{Sm\,{\sc ii}}& --    & -- & --   & 0.41 & 1  & 0.06 & 0.24  & 1  & 0.10 & 0.26  & 1  & 0.14 & --    & -- & --   \\ 
    \hline
    \end{tabular}
    \label{tab: normabund}
$^{*}$: Espadons spectrum was used for analysis.
\end{table}

\begin{table}
\tiny
\centering
\setlength{\tabcolsep}{1.3pt}
    \renewcommand{\arraystretch}{0.8}
\caption{Sensitivity of the derived abundances to uncertainties in the model atmospheric parameters of the target stars.}
\label{tab:6}
\centering
\begin{tabular}{l|c|c|c|c|cc|c|c|c}
\hline
\multicolumn{5}{c|}{\bf Sun} & \multicolumn{5}{c}{\bf HD\,122956} \\
\cline{1-10}
 & 5790	&	4.40 	&	0.00 &  0.66 & & 4700	&	1.70 	&	-1.58 & 1.83 \\
\cline{1-10}
Species & $\Delta T_{\rm eff}$	&	$\Delta \log g$ 	&	$\Delta$[Fe/H] &  $\Delta \xi$ & & $\Delta T_{\rm eff}$	&	$\Delta \log g$ 	&	$\Delta$[Fe/H] &  $\Delta \xi$ \\

\cline{2-10}
 ~    & (+45) & (+0.09) & (+0.04) & (+0.50)&   &(+89) & (+0.27) & (+0.10) & (+0.50) \\
  \hline
Na\,{\sc i}  & +0.03 & -0.02 & +0.00 & -0.06 & ~ &   --  &   --  &   --  &   --  \\
\textbf{Mg\,{\sc i}}  & +0.04 & -0.01 & +0.00 & -0.14 & ~ & +0.19 & +0.03 & +0.03 & -0.19 \\
\textbf{Al\,{\sc i}}  & -- & -- & -- & -- & ~ & +0.05 & +0.01 & +0.02 & +0.02 \\
Si\,{\sc i}  & +0.01 & +0.01 & +0.01 & -0.03 & ~ & +0.04 & +0.02 & +0.00 & -0.01 \\
Ca\,{\sc i}  & +0.03 & -0.02 & +0.00 & -0.11 & ~ & +0.09 & -0.02 & -0.01 & -0.08 \\
Sc\,{\sc ii} & +0.01 & +0.03 & +0.02 & -0.07 & ~ & +0.02 & +0.09 & +0.03 & -0.15 \\
Ti\,{\sc i}  & +0.05 & +0.00 & +0.00 & -0.12 & ~ & +0.15 & -0.02 & -0.02 & -0.16 \\
Ti\,{\sc ii} & +0.01 & +0.03 & +0.02 & -0.06 & ~ & +0.02 & +0.09 & +0.03 & -0.15 \\
\textbf{V\,{\sc i}}   & +0.05 & +0.01 & +0.00 & -0.04 & ~ & +0.13 & +0.04 & +0.05 & -0.02 \\
Cr\,{\sc i}  & +0.04 & -0.01 & +0.00 & -0.16 & ~ & +0.14 & -0.03 & -0.02 & -0.19 \\
Cr\,{\sc ii} & -0.01 & +0.02 & +0.01 & -0.10 & ~ & -0.04 & +0.09 & +0.02 & -0.08 \\
Mn\,{\sc i}  & +0.04 & -0.01 & +0.01 & -0.15 & ~ & +0.10 & -0.02 & -0.01 & -0.09 \\
Fe\,{\sc i}  & +0.04 & -0.01 & +0.01 & -0.14 & ~ & +0.11 & -0.01 & -0.01 & -0.14 \\
Fe\,{\sc ii} & +0.00 & +0.03 & +0.02 & -0.12 & ~ & -0.03 & +0.10 & +0.02 & -0.11 \\
\textbf{Co\,{\sc i}}  & +0.03 & +0.00 & +0.00 & -0.07 & ~ & +0.09 & +0.02 & +0.00 & -0.12 \\
Ni\,{\sc i}  & +0.02 & -0.01 & +0.00 & -0.09 & ~ & +0.09 & +0.00 & -0.01 & -0.06 \\
\textbf{Zn\,{\sc i}}  & +0.01 & +0.00 & +0.01 & -0.19 & ~ & +0.01 & +0.04 & -0.05 & -0.13 \\
\textbf{Sr\,{\sc i}}  & +0.04 & +0.00 & -0.01 & -0.24 & ~ &   +0.13  & +0.00 &   +0.09  &   -0.01  \\
\textbf{Y\,{\sc ii}}  & +0.01 & +0.03 & +0.01 & -0.21 & ~ & +0.03 & +0.00 & +0.04 & -0.15 \\
\textbf{Zr\,{\sc ii}} & +0.01 & +0.03 & +0.01 & -0.19 & ~ & -0.12 & -0.07 & -0.08 & -0.28 \\
\textbf{Ba\,{\sc ii}} & +0.02 & +0.01 & +0.03 & -0.16 & ~ & -0.14 & -0.13 & +0.03 & -0.33 \\
\textbf{Ce\,{\sc ii}} & +0.01 & +0.04 & +0.01 & -0.03 & ~ & -0.01 & +0.05 & +0.03 & -0.08 \\
\textbf{Nd\,{\sc ii}} & +0.02 & +0.04 & +0.02 & -0.02 & ~ & +0.04 & +0.09 & +0.01 & -0.01 \\
\textbf{Sm\,{\sc ii}} & -- & -- & -- & -- &  ~ & +0.15 & +0.20 & +0.01 & +0.07\\ \hline
\multicolumn{10}{c}{} \\
\cline{1-10}
\multicolumn{5}{c|}{\bf HD\,2665} & \multicolumn{5}{c}{\bf HD\,189349} \\
\cline{1-10}
 & 5010	&	2.39 	& -2.05 &  1.96 & & 5000	&	2.60 	&	-0.65 & 1.41 \\
\cline{1-10}
\cline{1-10}
	Species& $\Delta T_{\rm eff}$	&	$\Delta \log g$ 	&	$\Delta$[Fe/H] &  $\Delta \xi$ & & $\Delta T_{\rm eff}$	&	$\Delta \log g$ 	&	$\Delta$[Fe/H] &  $\Delta \xi$ \\
\cline{2-10}
     & (+70) & (+0.10) & (+0.09) & (+0.50)&   &(+125) & (+0.21) & (+0.13) & (+0.50) \\
  \hline
\textbf{C\,{\sc i}}  &   --  &   --  &   --  &   --  & ~ & -0.17 & +0.09 & +0.02 & -0.03 \\
\textbf{O\,{\sc i}}  &   --  &   --  &   --  &   --  & ~ & -0.13 & +0.12 & +0.03 & +0.01 \\
Na\,{\sc i}  &   --  &   --  &   --  &   --  & ~ & +0.06 & -0.03 & +0.00 & -0.12 \\
\textbf{Mg\,{\sc i}}  & +0.07 & +0.00 & +0.00 & -0.06 & ~ & +0.18 & +0.04 & +0.02 & -0.11 \\
\textbf{Al\,{\sc i}}  &   --  &   --  &   --  &   --  & ~ & +0.10 & +0.00 & +0.00 & +0.02 \\
Si\,{\sc i}  &   --  &   --  &   --  &   --  & ~ & +0.01 & +0.01 & +0.01 & -0.07 \\
S\,{\sc i}  &   --  &   --  &   --  &   --  & ~ & -0.09 & +0.05 & +0.00 & +0.00 \\
Ca\,{\sc i}  & +0.05 & +0.00 & +0.00 & -0.02 & ~ & +0.08 & -0.02 & -0.01 & -0.19 \\
Sc\,{\sc i} &   --  &   --  &   --  &   --  & ~ & +0.20 & +0.02 & +0.02 & -0.17 \\
Sc\,{\sc ii} & +0.03 & +0.02 & +0.00 & -0.12 & ~ & -0.02 & +0.05 & +0.02 & -0.23 \\
Ti\,{\sc i}  & +0.08 & +0.00 & +0.00 & -0.04 & ~ & +0.13 & +0.00 & -0.01 & -0.21 \\
Ti\,{\sc ii} & +0.03 & +0.03 & +0.00 & -0.08 & ~ & -0.02 & +0.05 & +0.02 & -0.21 \\
\textbf{V\,{\sc i}}   & +0.08 & +0.00 & +0.00 & -0.01 & ~ & +0.19 & +0.03 & +0.03 & +0.00 \\
Cr\,{\sc i}  & +0.07 & +0.00 & +0.00 & -0.05 & ~ & +0.10 & -0.01 & -0.01 & -0.23 \\
Cr\,{\sc ii} &   --  &   --  &   --  &   --  & ~ & -0.06 & +0.05 & +0.02 & -0.19 \\
\textbf{Mn\,{\sc i}}  & +0.06 & +0.00 & +0.00 & -0.02 & ~ & +0.15 & +0.02 & +0.01 & -0.03 \\
Fe\,{\sc i}  & +0.08 & +0.00 & +0.00 & -0.06 & ~ & +0.09 & +0.00 & -0.01 & -0.21 \\
Fe\,{\sc ii} & +0.00 & +0.03 & +0.00 & -0.05 & ~ & -0.06 & +0.05 & +0.02 & -0.19 \\
\textbf{Co\,{\sc i}}  & +0.10 & -0.01 & -0.01 & -0.29 & ~ & +0.14 & +0.02 & +0.03 & -0.05 \\
Ni\,{\sc i}  & +0.06 & +0.01 & +0.00 & -0.01 & ~ & +0.06 & +0.00 & +0.00 & -0.14 \\
\textbf{Cu\,{\sc i}}  &   --  &   --  &   --  &   --  & ~ & +0.10 & -0.07 & -0.09 & -0.24 \\
\textbf{Zn\,{\sc i}}  & +0.02 & +0.02 & +0.00 & -0.03 & ~ & +0.04 & +0.07 & +0.01 & -0.25 \\
\textbf{Sr\,{\sc i}}  &   --  &   --  &   --  &   --  & ~ & +0.18 & +0.03 & +0.04 & -0.07 \\
\textbf{Y\,{\sc ii}}  & +0.04 & +0.04 & +0.01 & -0.02 & ~ & +0.11 & +0.18 & +0.10 & -0.17 \\
\textbf{Zr\,{\sc ii}} & +0.03 & +0.03 & +0.00 & -0.03 & ~ & +0.04  & +0.10& +0.04 & -0.16 \\
\textbf{Ba\,{\sc ii}} & +0.06 & +0.03 & +0.01 & -0.10 & ~ & +0.06 & +0.06 & +0.03 & -0.28 \\
\textbf{Ce\,{\sc ii}} & +0.04 & +0.03 & +0.00 & -0.01 & ~ &  +0.05  & +0.10 & +0.04 & -0.07 \\
\textbf{Nd\,{\sc ii}} & +0.05 & +0.03 & +0.00 & -0.02 & ~ & +0.06 & +0.09 & +0.04 & -0.06 \\
\textbf{Sm\,{\sc ii}} &   --  &   --  &   --  &   --  & ~ & +0.06 & +0.12 & +0.03 & +0.02 \\ 
\hline
\multicolumn{10}{c}{} \\
\cline{1-10}
\multicolumn{5}{c|}{\bf HD\,5916} & \multicolumn{5}{c}{\bf HD\,218857} \\
\cline{1-10}
 & 5200	&	3.06 	&	-0.50 &  1.63 & & 5080	&	2.40 	&	-1.98 & 2.01 \\
\cline{1-10}
	Species& $\Delta T_{\rm eff}$	&	$\Delta \log g$ 	&	$\Delta$[Fe/H] &  $\Delta \xi$ & & $\Delta T_{\rm eff}$	&	$\Delta \log g$ 	&	$\Delta$[Fe/H] &  $\Delta \xi$ \\
\cline{2-10}
     & (+95) & (+0.06) & (+0.06) & (+0.50)&   &(+95) & (+0.24) & (+0.10) & (+0.50) \\
  \hline
Na\,{\sc i}  &   --  &   --  &   --  &   --  & ~ &   --  &   --  &   --  &   --  \\
\textbf{Mg\,{\sc i}}  & +0.08 & +0.02 & +0.00 & -0.09 & ~ & -0.18 & -0.09 & -0.03 & -0.08 \\
\textbf{Al\,{\sc i}}  & +0.06 & +0.00 & +0.01 & +0.00 & ~ & -- & -- & -- & -- \\
Si\,{\sc i}  & +0.00 & +0.00 & +0.00 & -0.04 & ~ &   --  &   --  &   --  &   --  \\
Ca\,{\sc i}  & +0.03 & -0.01 & +0.00 & -0.19 & ~ & +0.06 & -0.02 & -0.01 & -0.05 \\
Sc\,{\sc ii} & -0.01 & +0.02 & +0.01 & -0.22 & ~ &   --  &   --  &   --  &   --  \\
Ti\,{\sc i}  & +0.06 & +0.00 & +0.00 & -0.18 & ~ & +0.12 & -0.01 & +0.00 & -0.07 \\
Ti\,{\sc ii} & -0.01 & +0.02 & +0.02 & -0.24 & ~ & +0.04 & +0.07 & +0.01 & -0.15 \\
\textbf{V\,{\sc i}}   & +0.06 & +0.00 & +0.01 & -0.12 & ~ &   --  &   --  &   --  &   --  \\
Cr\,{\sc i}  & +0.05 & +0.00 & +0.00 & -0.25 & ~ &   --  &   --  &   --  &   --  \\
Cr\,{\sc ii} & -0.03 & +0.03 & +0.02 & -0.13 & ~ &   --  &   --  &   --  &   --  \\
Mn\,{\sc i}  & +0.04 & -0.01 & +0.00 & -0.23 & ~ & +0.09 & +0.00 & +0.00 & -0.02 \\
Fe\,{\sc i}  & +0.04 & +0.00 & +0.00 & -0.18 & ~ & +0.10 & -0.01 & -0.01 & -0.08 \\
Fe\,{\sc ii} & -0.04 & +0.02 & +0.01 & -0.18 & ~ & +0.00 & +0.08 & +0.00 & -0.06 \\
\textbf{Co\,{\sc i}}  & +0.03 & +0.00 & +0.00 & -0.04 & ~ &   --  &   --  &   --  &   --  \\
Ni\,{\sc i}  & +0.02 & +0.00 & +0.00 & -0.10 & ~ &   --  &   --  &   --  &   --  \\
\textbf{Zn\,{\sc i}}  & -0.01 & +0.01 & +0.02 & -0.26 & ~ &   --  &   --  &   --  &   --  \\
\textbf{Sr\,{\sc i}}  &   +0.07  &   +0.01  &   -0.05  &  +0.03  & ~ &   --  &   --  &   --  &   --  \\
\textbf{Y\,{\sc ii}}  &   -0.06  &   -0.04  &   +0.04  &   -0.20  & ~ & +0.04 & +0.08 & +0.01 & -0.03 \\
\textbf{Zr\,{\sc ii}} &   --  &   --  &   --  &   --  & ~ &   --  &   --  &   --  &   --  \\
\textbf{Ba\,{\sc ii}} & +0.01 & +0.02 & +0.02 & -0.31 & ~ & +0.06 & +0.08 & +0.01 & -0.12 \\
\textbf{Ce\,{\sc ii}} & -0.06  & -0.04 & +0.05 & -0.14 & ~ &   --  &   --  &   --  &   --  \\
\textbf{Nd\,{\sc ii}} & +0.00 & +0.02 & +0.02 & -0.09 & ~ & +0.06 & +0.07 & -- & -0.01 \\
\textbf{Sm\,{\sc ii}} & -0.05 & -0.02 & +0.03 & -0.07 & ~ & -- & -- & -- & --\\
\hline
\end{tabular}
\label{tab:modelerr}
\end{table}

\subsubsection{Non-LTE Abundance Corrections}

To account for the effects of non-LTE on the Fe\,{\sc i} lines, we employed the 1D non-LTE analysis of \citet{Lind2012} using the INSPECT v1.0 interface. In contrast, the non-LTE effects on the Fe\,{\sc ii} lines were considered negligible based on previous investigations \citep{Bergemann2012, Lind2012, Bensby2014}. \citet{Lind2012} demonstrated that deviations from LTE in Fe\,{\sc ii} lines with low excitation potentials ($<$ 8 eV) are minimal for stars with [Fe/H] $>-3$ dex. We followed the methodology outlined in \citet{Lind2012}, including their Figures 4, 5, and 7, to assess the influence of non-LTE effects on the determination of atmospheric parameters using Fe\,{\sc i} and Fe\,{\sc ii} lines. For the five program stars analyzed, the non-LTE corrections to the effective temperature ranged between 85 \kelvin~and 105 \kelvin. The largest correction (105 \kelvin) was found for HD\,2665, the most metal-poor star in the sample ([Fe/H]=-2.05 dex), while the smallest correction (85 \kelvin) was obtained for HD\,5916 ([Fe/H]=-0.50 dex).

\begin{figure}
    \centering
    \includegraphics[width=0.98\linewidth]{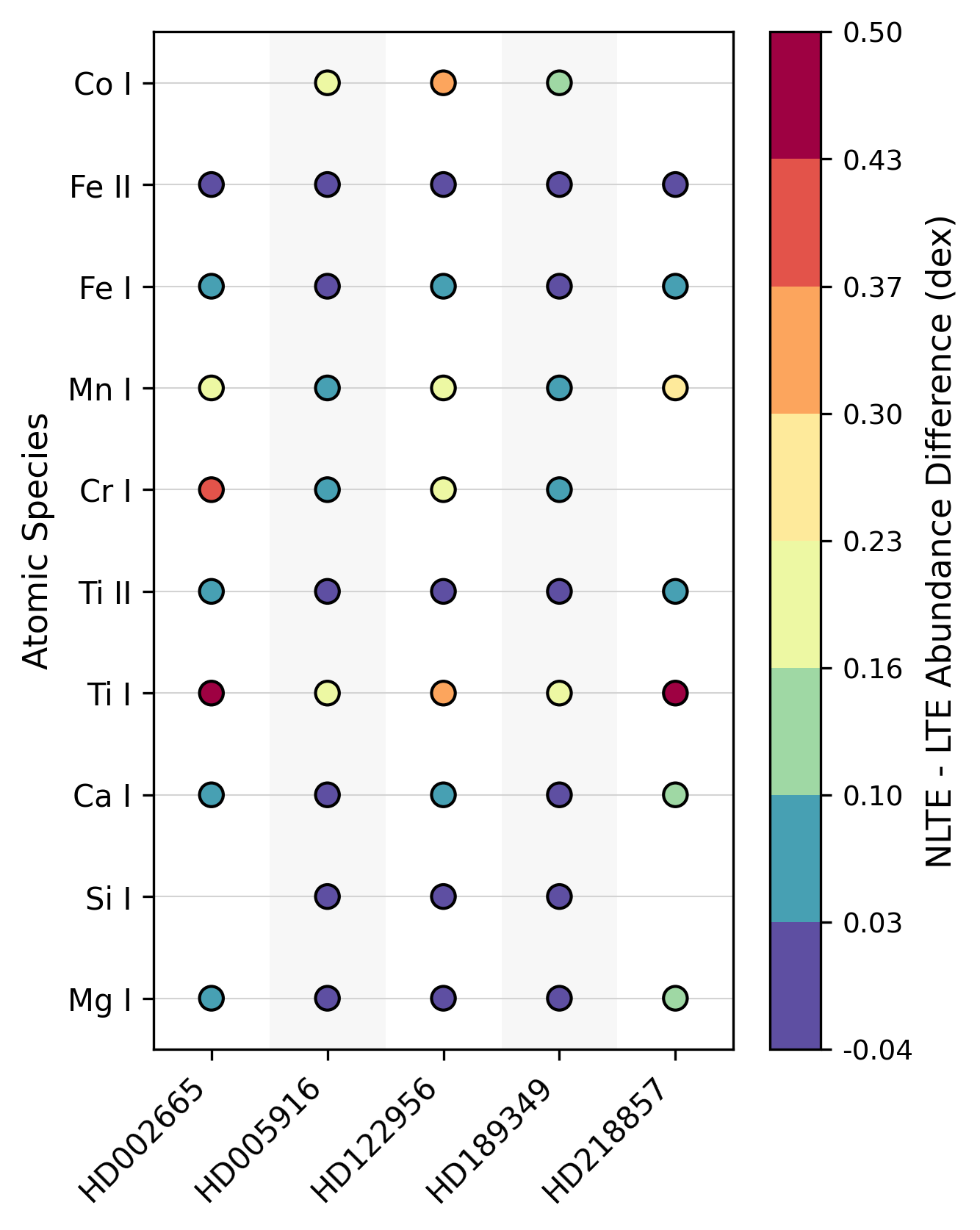}
    \caption{Non-LTE abundance corrections (in dex) for various atomic species in the five program stars. The vertical axis lists the elements and their ionization states, and the horizontal axis shows the stellar identifiers. The color indicates the magnitude of the non-LTE correction, which is defined as the abundance difference between the non-LTE and LTE calculations for each element. Corrections were computed using the available literature grids. Positive values indicate an increase in the inferred abundance when non-LTE effects are considered.}
    \label{fig:nonlte}
\end{figure}

Using the INSPECT interface, we also obtained non-LTE abundance corrections for Mg\,{\sc i}, Si\,{\sc i}, Ca\,{\sc i}, Ti\,{\sc i/ii}, Cr\,{\sc i}, Mn\,{\sc i}, Fe\,{\sc i/ii}, and Co\,{\sc i} in the program stars. These corrections were calculated for individual spectral transitions and averaged to yield a mean correction for each element. A detailed summary of the derived non-LTE corrections for these elements is provided in the subsequent part of this section and illustrated in Figure \ref{fig:nonlte}.

For Mg\,{\sc i}, HD\,122956 exhibited a mean non-LTE correction of -0.02 dex based on two transitions (4571.09\,\AA~and 5711.07\,\AA), while HD\,218857 showed a correction of +0.11 dex derived from a single line at 4571.09\,\AA. The corrections for HD\,2665, HD\,5916, and HD\,189349 ranged from 0.00 dex (HD\,5916 and HD\,189349) to +0.06 dex (HD\,2665), each based on two transitions. 

The Si\,{\sc i} abundance could not be determined for HD\,2665, and HD\,218857. For HD\,5916 (three transitions), HD\,122956 (one transition), and HD\,189349 (eight transitions), the mean non-LTE corrections ranged from -0.03 dex to -0.01 dex.  

For Ca\,{\sc i}, the applied non-LTE corrections were +0.11$\pm$0.04 dex (two transitions) for HD\,218857, +0.09$\pm$0.07 dex (eight transitions) for HD\,2665, +0.08$\pm$0.02 dex (four transitions) for HD\,122956, and +0.02 dex for HD\,5916 (three transitions) and HD\,189349 (eight transitions), respectively.

The non-LTE corrections for Ti\,{\sc i} were significant: +0.50 dex for HD\,2665 (17 transitions) and HD\,218857 (six transitions), +0.31 dex for HD\,122956 (10 transitions), and +0.17 dex for HD\,5916 (11 transitions), and HD\,189349 (31 transitions). In contrast, the Ti\,{\sc ii} corrections were minimal, ranging from -0.04 dex for HD\,5916 to +0.03  dex for both HD\,2665 and HD\,218857.  

The Cr\,{\sc i} abundance could not be determined for HD\,218857 because of the lack of reliable measurements. The non-LTE corrections applied were +0.40 $\pm$ 0.07 dex (nine transitions) for HD\,2665, +0.17 dex (10 transitions) for HD\,122956, and +0.07 dex for both HD\,5916 (nine transitions) and HD\,189349 (15 transitions).

For Mn\,{\sc i}, HD\,218857 and HD\,2665 exhibit corrections of +0.26 dex (one transition) and +0.22 dex (three transitions), respectively. HD\,122956 shows a correction of +0.17 dex (five transitions), while HD\,5916 and HD\,189349 exhibit smaller corrections of +0.07 dex, based on five and nine transitions, respectively.

The Fe\,{\sc i} corrections were small but systematic, ranging from +0.02 dex for HD\,5916 and HD\,189349 to +0.07 dex for HD\,2665 and HD\,218857, respectively. No corrections were applied to Fe\,{\sc ii}.

Co\,{\sc i} abundances have not been reported for either HD\,2665 or HD\,218857. For HD\,122956 (one transition), HD\,5916 (one transition), and HD\,189349 (five transitions), the corrections were +0.31, +0.20, and +0.15 dex, respectively.

\begin{table*}
\centering
\setlength{\tabcolsep}{5pt}
    \renewcommand{\arraystretch}{1.0}
\caption{Model atmosphere parameters ($T_{\rm eff}$, $\log g$, [Fe/H]),  heavy element ($Z$), mass ($M$), and age ($\tau$) values of the five stars analyzed in this study, obtained using spectral analysis methods. Literature median values are included for comparison.}
\label{tab:main_results}
\begin{tabular}{c|c|ccccc}
\hline
Parameter                               & Source     & HD\,2665                       & HD\,5916                      & HD\,122956                    & HD\,189349                    & HD218857 \\ 
\hline
  \multirow{2}{*}{$T_{\rm eff}$ (K)}    & This study & 5010$\pm$70                   & 5200$\pm$45                  & 4700$\pm$89                  & 5000$\pm$125                 & 5080$\pm$95\\
                                        & Literature & 5046$\pm$230                  & 4967$\pm$147                 & 4643$\pm$73                  & 5156$\pm$141                 & 5153 $\pm$182\\
\hline
\multirow{2}{*}{$\log g$ (cm s$^{-2}$)} & This study  & 2.39$\pm$0.10                & 3.06$\pm$0.06                & 1.70$\pm$0.27                & 2.60$\pm$0.21                & 2.40$\pm$0.24\\
                                        & Literature  & 2.34$\pm$0.24                & 2.25$\pm$0.38                & 1.52$\pm$0.21                & 2.57$\pm$0.26                & 2.55$\pm$0.24\\
\hline
\multirow{2}{*}{[Fe/H] (dex)}           & This study  & -2.05$\pm$0.09               & -0.50$\pm$0.06               & -1.58$\pm$0.10               & -0.65$\pm$0.13               & -1.98$\pm$0.10\\ 
                                        & Literature  & -1.88$\pm$0.30               & -0.66$\pm$0.16               & -1.76$\pm$0.20               & -0.52$\pm$0.15               & -1.86$\pm$0.28 \\ 
\hline
$Z$                                     & This study  & $0.0002_{-0.0001}^{+0.0002}$ & $0.0054_{-0.0013}^{+0.0016}$ & $0.0005_{-0.0003}^{+0.0004}$ & $0.0041_{-0.0009}^{+0.0008}$ & $0.0002_{-0.0001}^{+0.0001}$\\ 
\hline
$M$ ($M_\odot$)                         & This study  & $0.828^{+0.018}_{-0.019}$    & $1.384^{+0.180}_{-0.108}$    & $0.853^{+0.007}_{-0.008}$    & $1.341^{+0.038}_{-0.043}$    & $0.811^{+0.031}_{-0.022}$   \\ 
\hline
$\tau$ (Gyr)                            & This study  & $10.84^{+0.77}_{-0.75}$      & $2.69^{+0.47}_{-0.81}$       & $10.26^{+0.67}_{-0.47}$      & $2.77^{+0.25}_{-0.18}$       & $11.51^{+1.15}_{-1.48}$     \\ 
\hline 
\end{tabular}%
\end{table*}

Odd-Z iron-peak elements, such as V, Mn, and Co, exhibit additional spectral line broadening due to the hyperfine structure (hfs), which was accounted for in the abundance calculations via spectrum synthesis. The wavelengths and $\log gf$ values for the hfs components were taken from \citet{Lawler2014} for V, \citet{DenHartog2011} for Mn, and \citet{Lawler2015} for Co. \citet{Bergemann2019} provided new experimental transition probabilities for manganese lines and noted that some of the $gf$ values were typically 0.05--0.1 dex lower than previously adopted values. The difference between our adopted (hfs-included) $gf$ values and those from Bergemann was small, amounting to $-0.06\pm0.08$ dex. 

\subsection{Stellar Age Estimation}

In the present study, estimating the ages of the examined stars was essential for elucidating their formation history. To this end, stellar ages were derived using a Monte Carlo Markov Chain (MCMC) technique, which incorporates maximum likelihood estimation alongside an isochrone grid constructed from derived spectroscopic input parameters. The isochrone grid implemented in this approach was generated using version CMD 3.8\footnote{\url{http://stev.oapd.inaf.it/cgi-bin/cmd}} of the {\sc parsec} stellar isochrone database \citep{Bressan2012}. The grid exhibits the following characteristics: in $\log \tau$ space, it employs increments of 0.05 spanning the interval $6 \leq \log \tau \leq 10.13$, while for metallicity values $Z$, it uses steps of 0.0005 across the range $0 \leq Z \leq 0.03$. Despite its fine resolution, the grid does not form a fully continuous parameter space, which can introduce uncertainties in the age estimation process and create difficulties in the convergence of the MCMC walkers. To address this problem, interpolation between grid points is required. The Delaunay function from Python’s \texttt{Scipy} library \citep{Scipy} was used. This function performs three-dimensional interpolation over the defined parameter space, effectively transforming a discrete grid into a quasi-continuous grid. Consequently, the solution becomes independent of the original grid step sizes and allows for smoother exploration of the parameter space within the MCMC framework.

During this procedure, a function describing the target observational parameters within the isochrone grid space is generated using the Delaunay method, denoted as $I (Z, \tau, M) \xrightarrow[\text{Delaunay}]{ } \theta$. This interpolated function enables the estimation of stellar ages based on two distinct sets of $\theta$ input parameters by employing the MCMC technique in combination with the maximum-likelihood approach. The dataset is based on spectroscopic observations, incorporating stellar atmospheric model parameters such as effective temperature ($T_{\text{eff}}$), surface gravity ($\log g$), and metallicity ([Fe/H]). These parameters were utilized within the isochrone framework to identify the most probable stellar age through the application of MCMC, guided by the following maximum-likelihood function:
\begin{equation}
\log \mathcal{L} = -\frac{1}{2} \sum_{i} \left( \frac{\theta_i^{\text{obs}} - \theta_i^{\text{model}}}{\sigma_i} \right)^2  
\end{equation}
Here, $\mathcal{L}$ denotes the maximum likelihood function. The observed parameters, indicated by $\theta_i^{\text{obs}}$, are $T_{\rm eff}$, $\log g$, and [Fe/H] for the spectroscopic data set. Correspondingly, $\theta_i^{\text{model}}$ represents the theoretical parameters interpolated from the isochrone grid, and $\sigma_i$ denotes the measurement uncertainties associated with each observed quantity. This likelihood function minimizes the squared residuals between the observed and model values normalized by their respective uncertainties, thereby yielding an optimal estimation of stellar ages. The MCMC algorithm uses this log-likelihood formulation to systematically explore the parameter space and identify the most probable stellar age.

\begin{table*}
\setlength{\tabcolsep}{3.0pt}
\caption{Input parameters required for kinematic and dynamic orbit analysis of five stars (upper table): equatorial coordinates, radial velocities, proper-motion components, trigonometric parallaxes, and estimated distances from the Sun for five G-type HPM stars. Kinematic and dynamic orbit parameters of the five stars (middle and lower tables). The calculated space velocity components ($U, V, W$) and total space velocities ($S$), apogalactic ($R_{\rm a}$) and perigalactic ($R_{\rm p}$) distance, maximum distance from the Galactic plane ($Z_{\rm max}$), planar eccentricities ($e_{\rm p}$), azimuthal ($T_{\rm P}$) and radial ($T_{\rm R}$) periods, total energy of the star on the orbit ($E$) and angular momentum of the star on the three axes ($L_{\rm X}, L_{\rm Y}, L_{\rm Z}$).}
\centering
\begin{tabular}{lccccccc}
\hline
\multicolumn{8}{c}{Input Parameters}\\
\hline
Star& $\alpha$ & $\delta$ &  $V_{\rm rad}$ & $\mu_{\rm \alpha} \cos\delta$ & $\mu_{\rm \delta}$ & $\varpi$ & $d$ \\
    &  (hh~mm~ss) &  (dd~mm~ss) & (km s$^{\rm -1}$) & (mas yr$^{-1}$)  &     (mas yr$^{-1}$)   &  (mas)   &(pc)\\	 
\hline
HD\,002665 & 00 30 45.45 &   +57 03 53.63 & $-$382.04$\pm$0.12 &   +41.669$\pm$0.014 & $-$64.713$\pm$0.017 & 3.7263$\pm$0.0181 & 268$\pm$1.30\\
HD\,005916 & 01 01 19.02 &   +45 27 07.39 & ~$-$68.03$\pm$0.12 &   +98.719$\pm$0.023 & $-$25.564$\pm$0.016 & 6.0024$\pm$0.0244 & 167$\pm$0.68\\
HD\,122956 & 14 05 13.02 & $-$14 51 25.46 &   +165.81$\pm$0.14 & $-$94.040$\pm$0.024 & $-$44.033$\pm$0.019 & 2.7070$\pm$0.0210 & 369$\pm$2.86\\
HD\,189349 & 19 58 02.38 &   +40 55 36.63 & $-$104.21$\pm$0.12 &   +07.280$\pm$0.015 &   ~16.114$\pm$0.016 & 5.0153$\pm$0.0141 & 199$\pm$0.56\\
HD\,218857 & 23 11 24.59 & $-$16 15 04.03 & $-$169.40$\pm$0.19 & $-$50.883$\pm$0.020 & $-$95.475$\pm$0.018 & 2.8529$\pm$0.0191 & 351$\pm$2.35\\
\hline 
\multicolumn{8}{c}{Output Kinematic and Galactic Orbit Parameters}\\
\hline
 Star	  & $U_{\rm LSR}$         &  $V_{\rm LSR}$          & $W_{\rm LSR}$           & $S_{\rm LSR}$              & $R_{\rm a}$ & $R_{\rm p}$ & $Z_{\rm max}$ \\
\cline{2-8}
            & \multicolumn{4}{c}{(km s$^{-1}$)}                          &   \multicolumn{3}{c}{(kpc)}         \\
\hline
HD\,002665 & 157.59$\pm$0.18  &  -344.94$\pm$0.18 &  ~-41.44$\pm$0.42 & 381.49$\pm$0.49 & 12.13$\pm$0.01 & 2.55$\pm$0.01 &  1.18$\pm$0.01 \\
HD\,005916 & -19.49$\pm$0.26  &  ~-87.98$\pm$0.22 &  ~10.63$\pm$0.08  & ~90.74$\pm$0.35 & ~8.14$\pm$0.01 & 3.20$\pm$0.01 &  0.16$\pm$0.01 \\
HD\,122956 &  ~21.38$\pm$0.72 &  -205.50$\pm$1.22 &   118.19$\pm$0.11 & 238.03$\pm$1.42 & ~7.96$\pm$0.01 & 0.46$\pm$0.01 &  6.29$\pm$0.02 \\
HD\,189349 & -37.27$\pm$0.06  &  ~-82.80$\pm$0.12 &  ~-2.56$\pm$0.02  & ~90.84$\pm$0.14 & ~8.12$\pm$0.01 & 3.29$\pm$0.01 &  0.06$\pm$0.01 \\
HD\,218857 & 102.52$\pm$0.95  &  -155.80$\pm$0.75 &   157.53$\pm$0.17 & 244.13$\pm$1.22 & 10.18$\pm$0.02 & 2.86$\pm$0.01 &  8.30$\pm$0.03 \\
\hline 
\multicolumn{8}{c}{Output Galactic Orbit Parameters}\\
\hline
 Star	  & $e_{\rm P}$         &  $T_{\rm P}$          & $T_{\rm R}$     & $E$              & $L_{\rm X}$ & $L_{\rm Y}$ & $L_{\rm Z}$ \\
           &     &     \multicolumn{2}{c}{(Gyr)} & ($10^5\times$~km$^2$ s$^{-2})$ & \multicolumn{3}{c}{(kpc km s$^{-1}$)}\\
\hline
HD\,002665 & 0.65$\pm$0.01 &  0.23$\pm$0.01  & 0.15$\pm$0.01 &  -0.43$\pm$0.01 & -10.37$\pm$1.21 & ~~341.82$\pm$1.05 & -987.65$\pm$1.46 \\
HD\,005916 & 0.44$\pm$0.01 &  0.16$\pm$0.01  & 0.11$\pm$0.01 &  -0.57$\pm$0.01 & ~~5.15$\pm$0.52 & ~~-86.40$\pm$1.55 & 1057.64$\pm$1.75 \\
HD\,122956 & 0.89$\pm$0.01 &  0.16$\pm$0.01  & 0.09$\pm$0.01 &  -0.61$\pm$0.01 & -20.37$\pm$0.30 & ~-922.96$\pm$2.52 & ~108.23$\pm$3.51 \\
HD\,189349 & 0.42$\pm$0.01 &  0.16$\pm$0.01  & 0.11$\pm$0.01 &  -0.58$\pm$0.01 & ~-6.20$\pm$0.05 & ~~~21.48$\pm$0.23 & 1079.26$\pm$2.75 \\
HD\,218857 & 0.56$\pm$0.01 &  0.18$\pm$0.01  & 0.13$\pm$0.01 &  -0.46$\pm$0.01 & ~37.79$\pm$0.41 & -1212.02$\pm$1.36 & ~515.97$\pm$1.75 \\
\hline 
\end{tabular}
\label{tab:uvw}
\end{table*}

\begin{figure*}
\centering 
\includegraphics[width=0.96\linewidth]{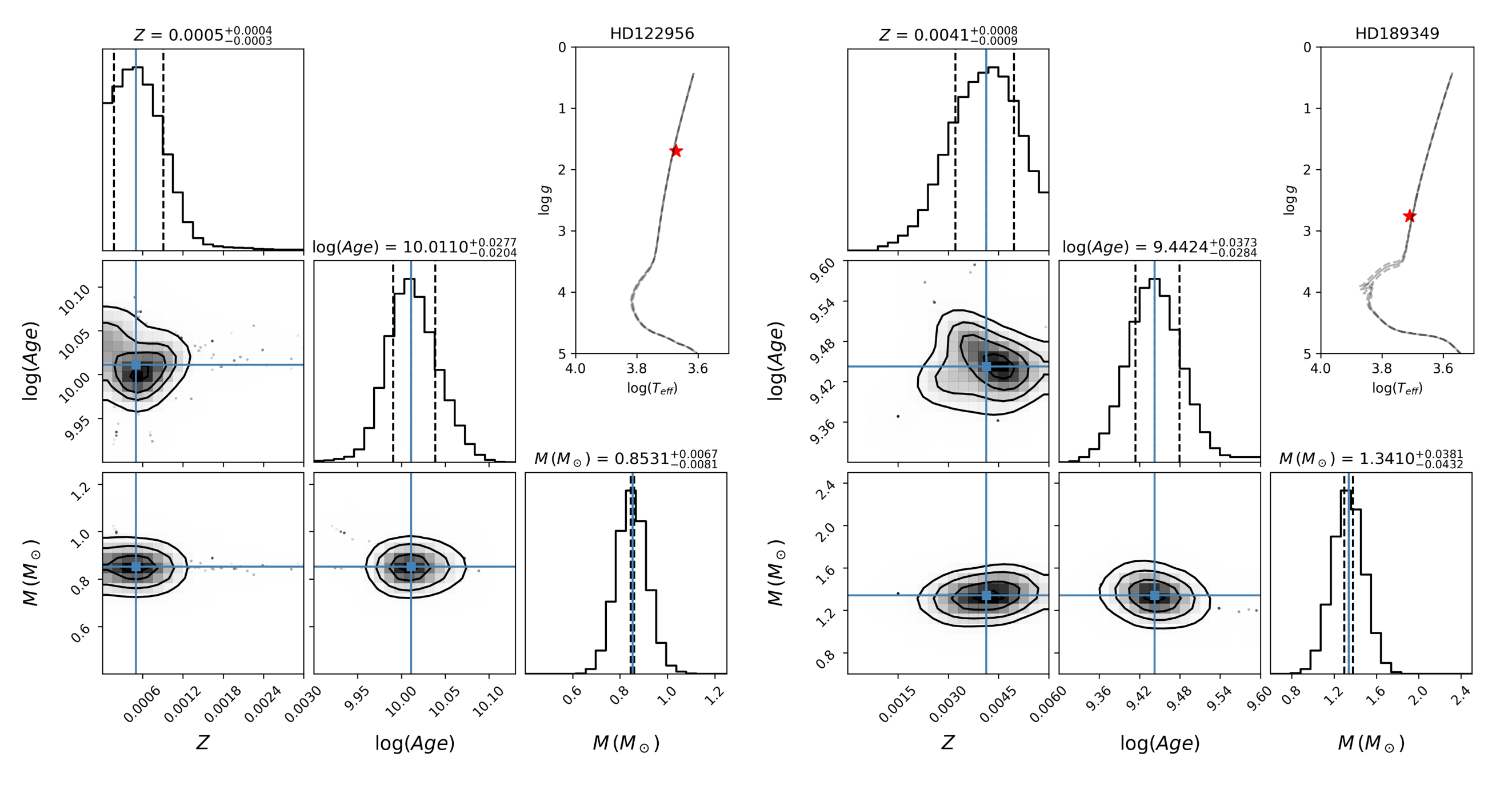}
\caption{Corner plot displaying the posterior probability distributions for HD\,122956 and HD\,189349, with confidence levels marked at 68\%, 90\%, and 95\%. The one-dimensional marginal distributions indicate the median values, along with the 16th and 84th percentiles. The right side of each panel shows the corresponding positions of the stars on the Kiel diagram. Corner plots for the remaining three stars (HD\,2665, HD\,5916, and HD\,218857) are provided in Appendix Figure \ref{fig:age-corner}.}
\label{fig:age}
\end{figure*}

For the final age determination, each star was analyzed using 22 independent walkers over 5,000 iterations of the MCMC algorithm. The resulting posterior distributions were employed to extract the median stellar age, with uncertainties defined by the interval spanning one standard deviation above and below the median value. This methodology offers a statistically reliable approach for age estimation that fully incorporates the inherent uncertainties in observational data.

A summary of these findings is presented in Table~\ref{tab:main_results}. Figure~\ref{fig:age} and Figure~\ref{fig:age-corner} present the posterior distributions and corresponding best-fitting stellar ages derived from spectroscopic atmospheric parameters. The additional parameters listed in Table~\ref{tab:main_results} were inferred through interpolation on the isochrone grid derived from most probable solutions.

\subsection{Kinematic and Orbital Dynamic Analyses of the HPM stars} \label{sec:dynamics}
The {\tt galpy} code was used to calculate the space velocity components of the five program stars and Galactic orbital parameters. {\tt galpy} is a powerful Python library designed for Galactic dynamics that enables researchers to analyze and simulate the motion of stars in the Milky Way \citep{Bovy2015}. We assume that the Milky Way is well represented by the {\tt galpy} potential {\it MilkywayPotential2014}, which is composed of three potentials that constitute the gravitational fields of the disk, halo, and bulge components of the Galaxy. We used the potential proposed by \citet{Miyamoto1975} for the Galactic disk components. The potential of the halo component was determined according to the method described by \citet{Navarro1996}. Finally, we prefer a spherical power-law density profile \citep{Bovy2013} for the bulge component. In the calculations, the parameter sets given by \citet{Bovy2015} (see their Table 1) were used for parameters related to the potentials and structure of the Galaxy.

Astrometric ($\alpha$, $\delta$, $\mu_{\alpha}\cos \delta$, $\mu_{\delta}$, $\varpi$) and spectroscopic data ($\gamma$) provided in {\it Gaia} DR3 \citep{Gaia2023} were used for  $U, V, W$ space velocity component calculations. The input kinematic parameters are presented in Table \ref{tab:uvw}. For the calculations, epoch J2000 was adopted, as described in the International Celestial Reference System (ICRS). The transformation matrix uses the notation of a right-handed system. Hence, $U$, $V$, and $W$ are the components of a velocity vector of a star concerning the Sun, where $U$ is positive towards the Galactic center ($l=0^{\circ}$), $V$ is positive in the direction of Galactic rotation ($l=90^{\circ}$) and $W$ is positive towards the North Galactic Pole ($b=90^{\circ}$).

Correction for differential Galactic rotation is necessary for the accurate determination of the space velocity components. This effect is proportional to the projection of the distance to the stars onto the Galactic plane; that is, the $W$ velocity component is unaffected by Galactic differential rotation \citep{Mihalas1981}. We applied the procedure of \cite{Mihalas1981} to the distribution of the five program stars and calculated the first-order Galactic differential rotation corrections for $U$ and $V$ velocity components of the five HPM stars. The ranges of these corrections were $-3.8<dU~({\rm km~ s}^{-1})<6.3$ and $-0.4<dV~({\rm km~ s}^{-1})<0.6$ for $U$ and $V$, respectively. As expected, $U$ was affected more than $V$. The space velocity components were also reduced to the local standard of rest (LSR) by adopting the solar LSR velocities in \citet{Coskunoglu2011}, that is, $(U,V,W)_{\odot}$=(8.83$\pm$0.24, 14.19$\pm$0.34, 6.57$\pm$0.21) km s$^{-1}$. The total space velocities and their errors for the five-star programs are listed in Table \ref{tab:uvw}.

The uncertainties of the space velocity components ($U_{\rm err}$, $V_{\rm err}$, $W_{\rm err}$) were computed by propagating the uncertainties of the proper motion components, radial velocity, and trigonometric parallaxes using the algorithm proposed by \cite{Johnson1987}. The uncertainty of the total space velocity ($S_{\rm err}$) of the program stars was calculated using the following equation: $S_{\rm err}=\sqrt{U_{\rm err}^{2}+V_{\rm err}^{2}+W_{\rm err}^{2}}$.

The Galactic orbits of the program stars were calculated with 0.05 Myr time steps and a 13 Gyr integration time. We used the same input data used in the calculations of the space-velocity components to estimate the orbital parameters, as shown in Table~\ref{tab:uvw}. The apogalactic ($R_{\rm a}$) and perigalactic distances ($R_{\rm p}$), maximum height above the Galactic plane ($Z_{\rm max}$), planar eccentricity ($e_{\rm p}$), azimuthal ($T_{\rm P}$), and radial ($T_{\rm R}$) periods, total energy of the star on the orbit ($E$), and angular momentum of the star on the three axes ($L_{\rm X}, L_{\rm Y}, L_{\rm Z}$) were calculated for each star. The calculated orbital parameters of the program stars using the {\tt galpy} code are listed in Table \ref{tab:uvw}.

\begin{figure}
\centering 
\includegraphics[width=\columnwidth]{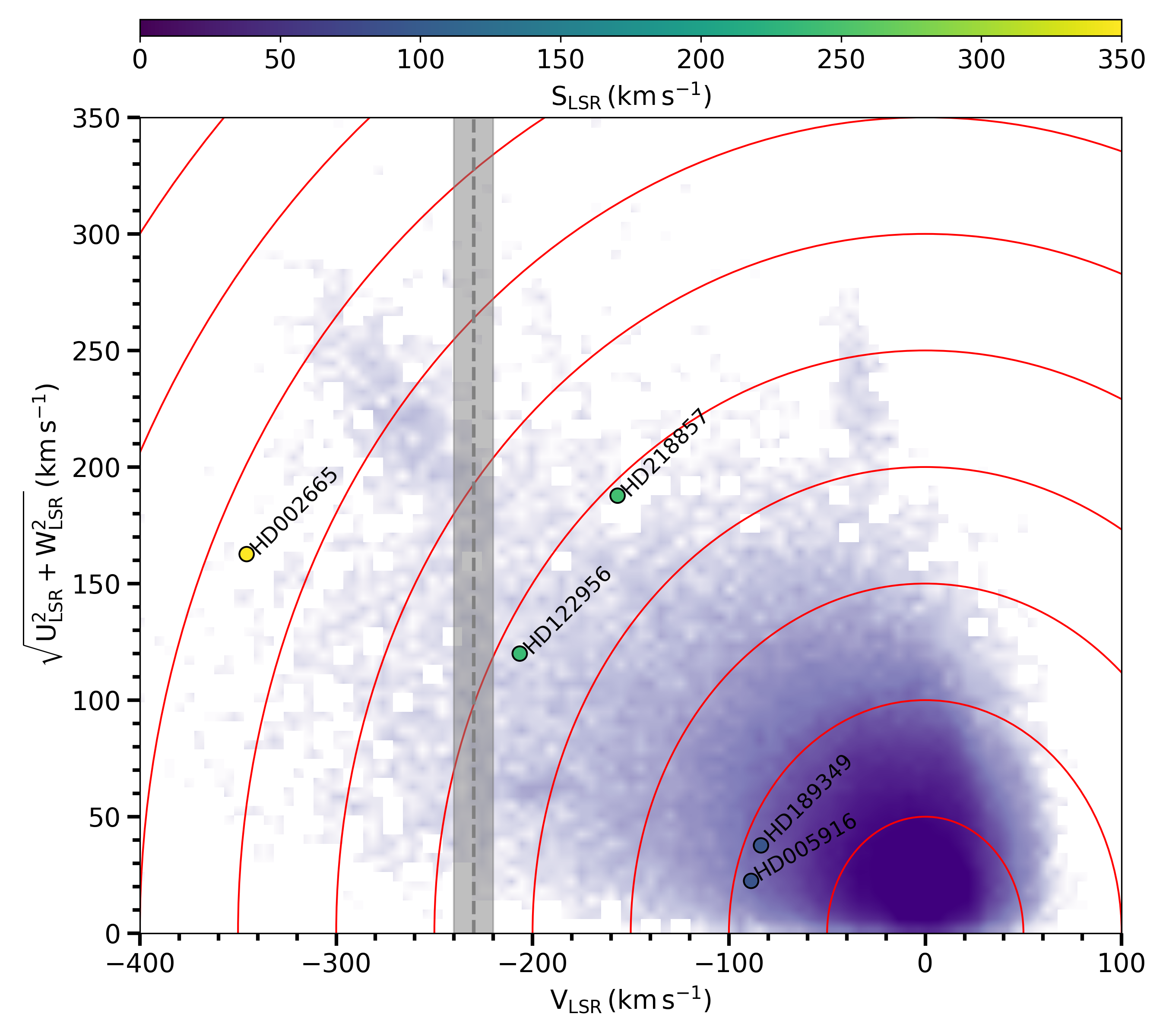}
\caption{Toomre diagram for five HPM stars. The solid red iso-velocity curves represent constant total space velocities from 50 to 500 steps of 50 km s$^{-1}$. The vertical grey band corresponds to a velocity of $S_{\rm LSR}$ = -230$\pm$10 km s$^{-1}$ \citep{Necib2022}. Stars are classified as retrograde if their velocity relative to the LSR satisfies $V_{\rm LSR}<-230$ km s$^{-1}$, whereas stars with $V_{\rm LSR}\geq -230$ km s$^{-1}$ are considered prograde in the Galaxy. Background blue points represent APOGEE DR17 stars selected using quality cuts as described in the text.}
\label{fig:Toomre}
\end{figure}

\begin{figure}
\centering 
\includegraphics[width=\columnwidth]{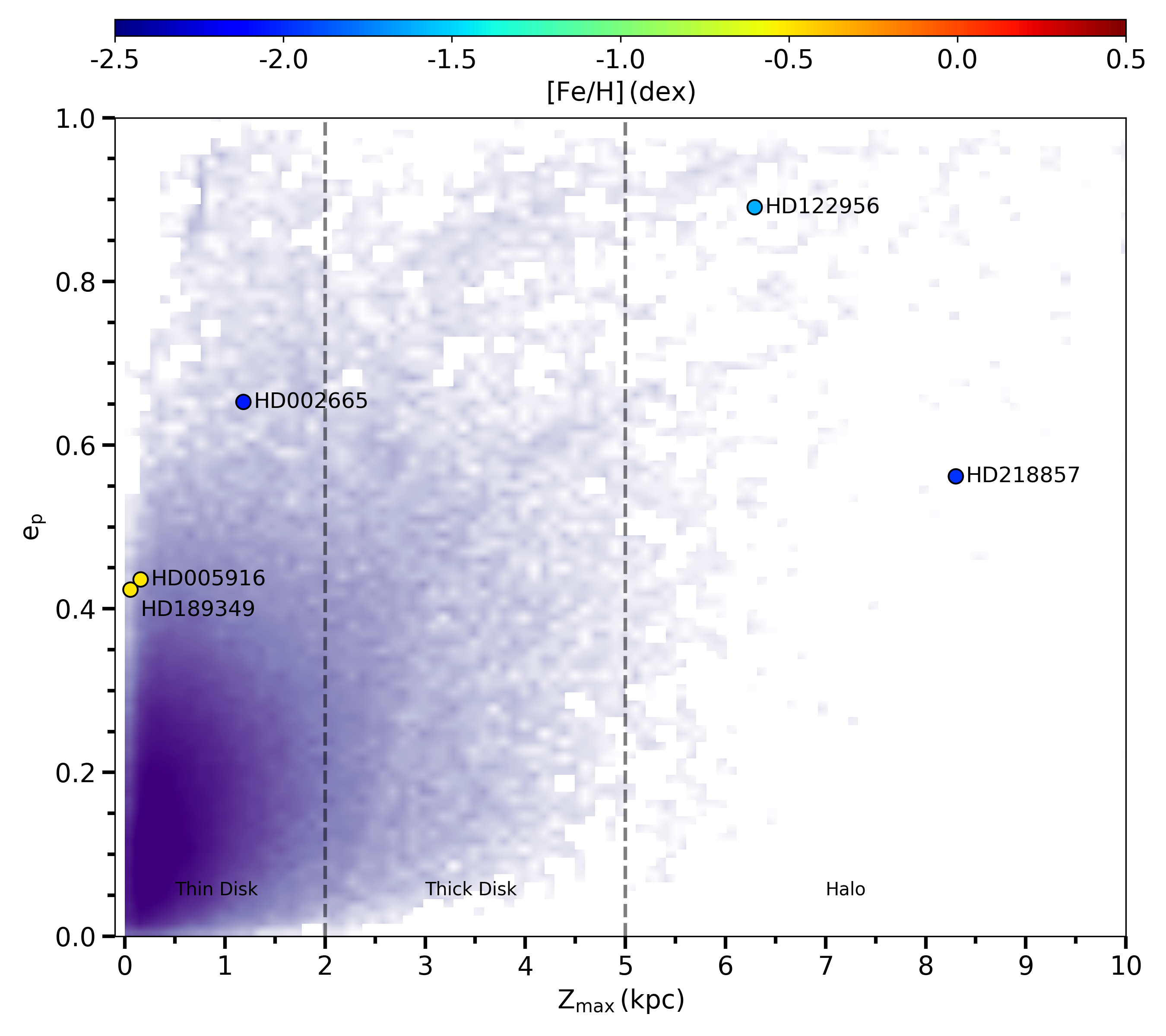}
\caption{$Z_{\rm max} \times e_{\rm p}$ diagram for five stars. The stars are color-coded according to their metallicities. Background blue points represent APOGEE DR17 stars selected using quality cuts as described in the text.}
\label{fig:Zmax-ep}
\end{figure}

\begin{figure*}
    \centering
    \includegraphics[width=0.99\linewidth]{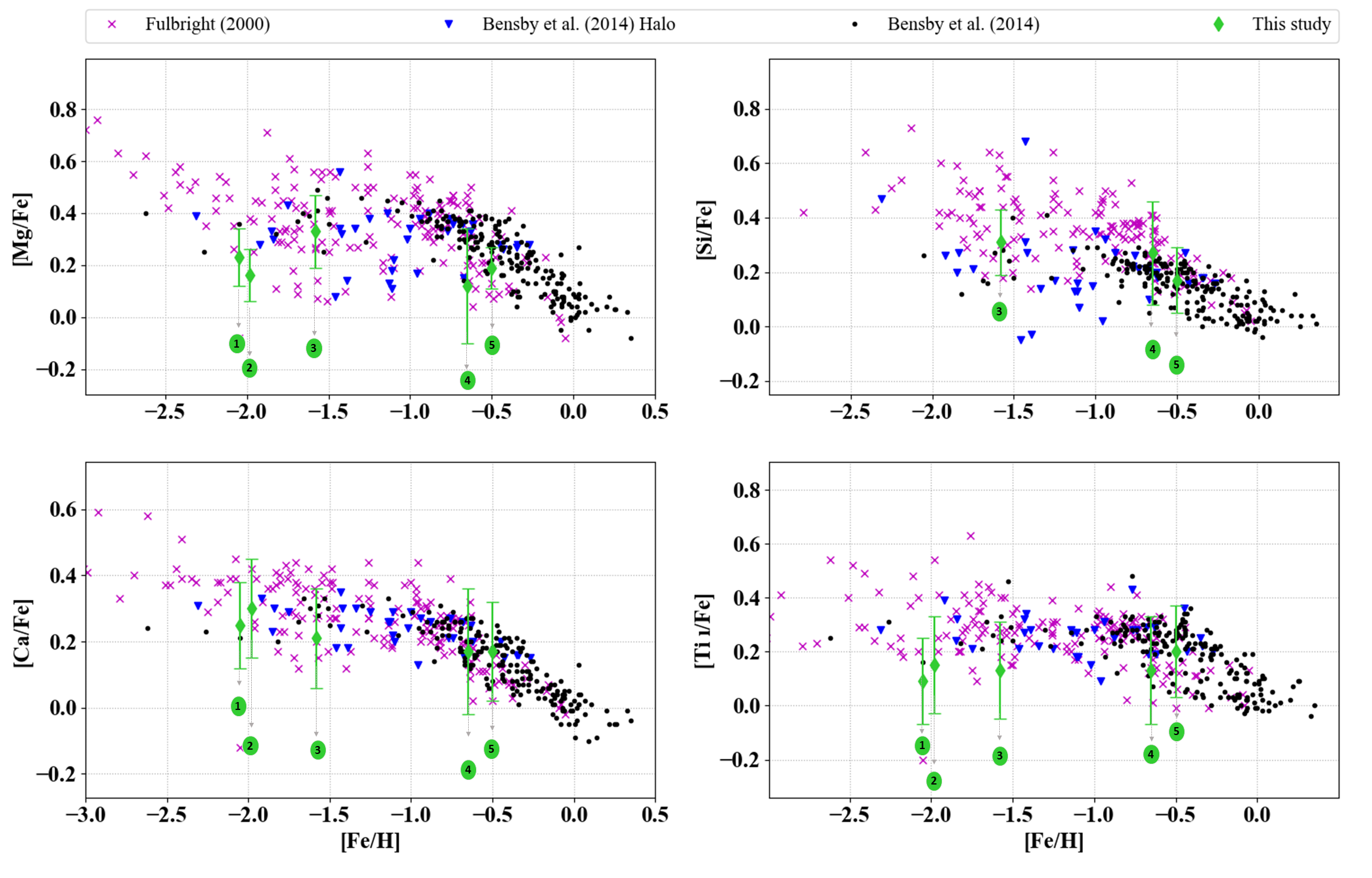}
    \caption{Abundance ratios of $\alpha$-elements ([Mg/Fe], [Si/Fe], [Ca/Fe], and [Ti\,{\sc i}/Fe]) as a function of metallicity ([Fe/H]) for stars analyzed in this study (green diamonds with error bars and numbered from 1 to 5, in order of increasing metallicity: HD 2665 (1), HD 218857 (2), HD 122956 (3), HD 189349 (4), and HD 5916 (5)) compared to literature data. The data from \citet{Fulbright2000} are shown as pink crosses, and the magenta squares indicate the mean values for metal-poor dwarfs from \citet{Fulbright2000}. The blue downward triangles represent halo stars from \citet{Bensby2014}, and their mean values are shown as larger blue symbols. Black dots represent thick-disk stars from \citet{Bensby2014}, and large black dots indicate the average abundances in 0.25 dex intervals of [Fe/H] for thick-disk stars. Individual error bars in the [X/Fe] axis for the metal-poor HPM stars are also indicated.}
    \label{fig:alfa_abund}
\end{figure*}

\begin{figure*}
\centering 
\includegraphics[width=1.02\textwidth]{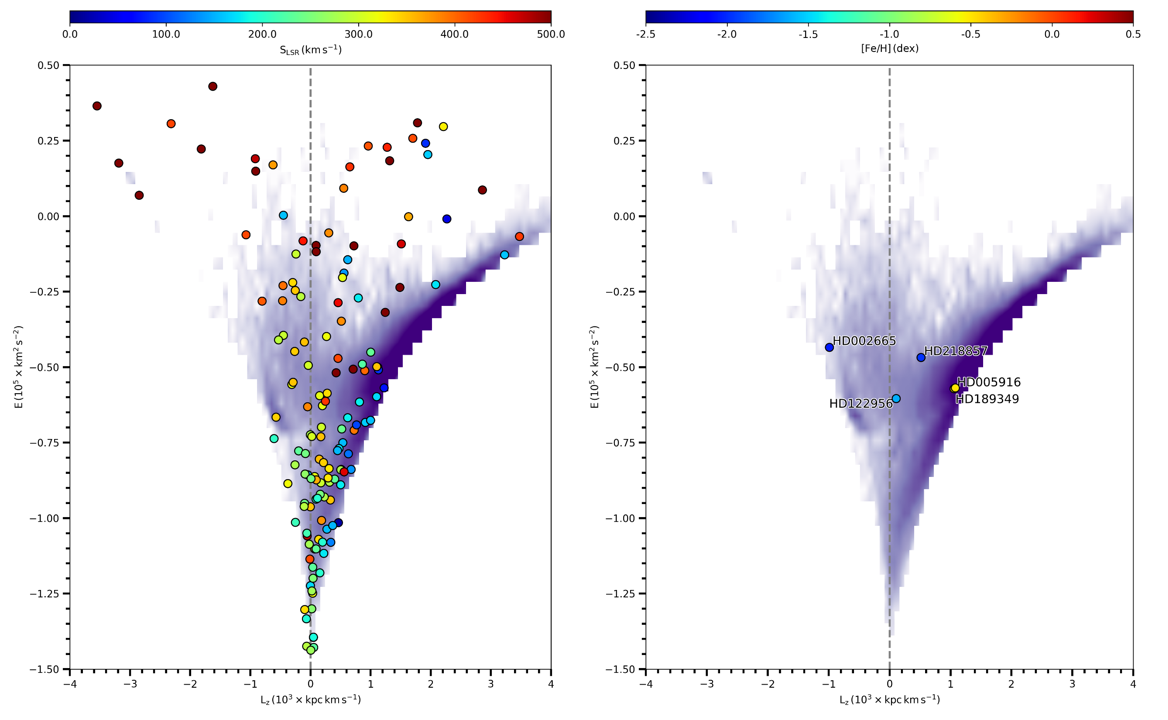}
\caption{Orbital properties of five stars in the Lindblad diagram. Color-coded distribution of stellar angular momentum ($L_{\rm tot}$) over the Milky Way stellar density map, with colored circles marking GC positions in the Galaxy (left panel). The same diagram is color-coded according to metallicity (right panel). The overplotted labeled markers indicate the positions of the selected stars (HD\,122956, HD\,189349, HD\,005916, HD\,122956, and HD\,218857) used in this study. The vertical dashed line corresponds to \mbox{$L_{\rm z}$ = 0}, which separates prograde (right) and retrograde (left) orbits.}
\label{fig:lindband}
\end{figure*}

To study the Galactic population types of the program stars, the stars were plotted in a Toomre energy diagram calibrated according to their iso-velocity curves (Figure \ref{fig:Toomre}). According to \citet{Nissen2004}, thin-disk stars have a total space velocity of $S_{\rm LSR}<60$ km s$^{\rm -1}$, whereas the total space velocities of thick-disk stars have been reported to exhibit a larger interval in velocity, that is, $80<S_{\rm LSR}~({\rm km~s}^{-1})<180$. The total space velocity of the halo stars in the Solar neighborhood was greater than $S_{\rm LSR}$=180 km s$^{\rm -1}$. \citet{Nissen2004}'s kinematic criteria indicate that HD\,5916 and HD\,189349 are members of the thick-disk population, and HD\,2665, HD\,122956, and HD\,218857 stars are members of the halo population. It also shows that HD\,2665 rotates in the opposite direction of the Galaxy's rotation from its position in the Toomre diagram (see Figure~\ref{fig:Toomre}), which indicates that the star could have come from satellite galaxies interacting with the Milky Way. To provide a broader comparison, we used the APOGEE-2 program's 17th data release DR17 \citep{ApogeeDR17} for the Galactic star sample as background stars in Figure~\ref{fig:Toomre}. These stars were selected based on standard quality cuts: \emph{starflag}=0, \emph{ascapflag}=0, \emph{fe\_h\_flag}=0, and a signal-to-noise ratio (SNR) greater than 100. The space velocities of these stars were computed using the same methods described in Section~\ref{sec:dynamics}.

The $e_{\rm p} \times Z_{\rm max}$ diagram composed according to the Galactic orbital parameters of the five stars is shown in Figure \ref{fig:Zmax-ep} where we also include background stars from the APOGEE DR17 survey. The orbits of these stars were computed using the same Galactic potential and methods described in Section~\ref{sec:dynamics}. In addition, the efficiency ranges in $Z_{\rm max}$ of the three Galactic populations (thin disk, thick disk, and halo) are shown in this diagram \citep{TuncelGuctekin2019}. Considering the maximum distances the stars can reach from the Galactic plane on the nuclear timescale, HD\,2665, HD\,5916, and HD\,189349 thin disks, and HD\,122956 and HD\,218857 halo populations were also seen as members. As shown in Figure \ref{fig:Zmax-ep}, the orbital eccentricities of HD\,2665, belonging to the thin-disk population, are similar to those of the thick disk and halo stars. The flattened orbital motion of HD\,2665 in the Galactic disk could be caused by gravitational interactions within the Galactic disk, as shown by \citet{OnalTas2018}.

To refine the Galactic population classification of metal-poor HPM stars, Figure \ref{fig:alfa_abund}  presents the [$\alpha$/Fe] abundance ratio as a function of [Fe/H]. For comparison, we included representative samples of thick-disk and halo stars from  \citet{Bensby2014} and halo and disk stars, including metal-poor dwarfs, from \citet{Fulbright2000}. This comparison clearly illustrates the chemical evolution trends among different Galactic components and contextualizes the $\alpha$-element abundances of the HPM stars relative to previously studied populations. As shown in Figure \ref{fig:alfa_abund}, all HPM stars exhibit enhanced $\alpha$-element abundances relative to iron, which is consistent with their membership in the thick disk or halo. Moreover, the agreement among the silicon, calcium, and titanium abundances was satisfactory, reinforcing their classification as $\alpha$-rich stars.

The left panel of Figure~\ref{fig:lindband} displays the APOGEE DR17 sample as the background distribution as in as in Figure \ref{fig:Toomre} and Figure \ref{fig:Zmax-ep}. In this Lindblad ($E$–$L_{\rm z}$) diagram\footnote{The Lindblad diagram delineates the two-dimensional (2D) projection of the orbital phase space.}, the colored circles marking positions of GCs from \citet{Baumgardt2019, Vasiliev2021} and they provide additional context for assessing the dynamical properties of the program stars relative to known cluster populations.

The Lindblad diagram in Figure~\ref{fig:lindband} reveals distinct dynamic histories for the five program stars, contextualizing their origins within the structural components of the MW. HD\,2665, with $L_{\rm z} = -987.65 \pm 1.46$ kpc km s$^{\rm -1}$ and $E=(-0.43 \pm 0.01)\times 10^5$ km$^{-2}$s$^{-2}$, occupies the retrograde halo region ($L_{\rm z} < 0$ kpc km s$^{-1}$). Although there is no certain way to identify stars of accreted origin, there are some clues, such as high space velocities and/or low metallicity accompanied by a low [$\alpha$/Fe] ratio \citep{Majewski1992, Gratton2003}. They may also exhibit high eccentricities \citep{Naidu2020}. For instance, stars in the GSE can have a range of eccentricity values \citep{Perottoni2022}. The high planar eccentricity of HD\,2665 ($e_{\rm p}=0.65\pm 0.01$) and its moderate vertical excursion ($Z_{\rm max}=1.18\pm 0.01$ kpc) may suggest an accreted origin; however, its spectrum shows moderate enrichment in [Mg/Fe]. A high [Mg/Fe] ratio has been reported for retrograde stars \citep[for example,][]{helmi2018, Feuillet2021}. In contrast, HD\,5916 ($L_{\rm z} = 1057.64 \pm 1.75$ kpc km s$^{\rm -1}$, $E = (-0.57\pm 0.01)\times 10^5$ km$^{-2}$s$^{-2}$) lies in the high-$L_{\rm z}$, which is the tightly bound energy regime of the thick disk. Its low eccentricity ($e_{\rm p}=0.44\pm 0.01$) and minimal vertical motion ($Z_{\rm max} = 0.16\pm 0.01$ kpc) reflect a stable in-situ disk orbit, consistent with stars that are undisturbed by major mergers. 

HD\,122956 ($L_{\rm z}=108.23\pm 3.51$ kpc km s$^{\rm -1}$, $E=(-0.61 \pm 0.01)\times 10^5$ km$^{-2}$s$^{-2}$) lies near the disk-halo transition and is characterized by extreme eccentricity ($e_{\rm p}=0.89\pm 0.01$) and a large vertical excursion ($Z_{\rm max}=6.29\pm 0.02$ kpc). This combination implies a halo star with a highly eccentric orbit, possibly dynamically heated or accreted from a progenitor with a residual angular momentum. Conversely, HD\,189349 ($L_{\rm z} = 1079.26 \pm 2.75$ kpc km s$^{\rm -1}$, $E=(-0.58 \pm 0.01)\times 10^5$ km$^{-2}$s$^{-2}$) clusters with HD\,5916 in the high-$L_{\rm z}$ regime, indicative of the thin/thick disk. Its moderately flattened orbit ($e_{\rm p}=0.42\pm 0.01$) and negligible vertical motion (\(Z_{\rm max} = 0.06\pm 0.01\) kpc) underline its origin in an undisturbed MW disk.

HD\,218857 ($L_{\rm z}=515.97\pm 1.75$ kpc km s$^{\rm -1}$, $E=(-0.46\pm 0.01)\times 10^5$ km$^{-2}$s$^{-2}$) presents an ambiguous case, straddling the metal-poor thick disk and inner halo. Its moderate eccentricity ($e_{\rm p}=0.56\pm 0.01$) and extreme vertical excursion ($Z_{\rm max} = 8.30\pm 0.03$ kpc) suggest a halo star with a prograde motion, possibly heated by mergers or originating from a disk-like satellite. The prograde $L_{\rm z}$ of the star contrasts with its halo-like kinematics, highlighting the complexity of disentangling in situ and accreted populations. 

The Lindblad diagram emphasizes the hierarchical structure of MW. Retrograde stars such as HD\,2665 trace ancient accretion events, while high-$L_{\rm z}$ disk stars (HD\,5916 and HD\, 189349) map the secular evolution of the Galaxy. Transitional objects, such as HD\,122956 and HD\,218857, exemplify the dynamical continuum between the disk and halo, likely shaped by early mergers or heating processes. These results demonstrate the benefit of the diagram in probing Galactic history, with future studies focusing on resolving ambiguities through detailed chemical tagging.  

A comparison of Galactic orbital parameters computed for five stars using the \citet{McMillan2017} and {\sc MilkyWayPotential2014} potentials \citep{Bovy2015} reveals no significant differences apart from the total energy. The mean differences in apogalactic and perigalactic distances were -0.02 kpc and 0.23 kpc, respectively, while the mean difference in $Z_{\rm max}$ was -0.10 kpc, and the orbital eccentricity difference averaged at -0.03. The differences in the three components of angular momentum were generally less than 10\%. These results indicate that, despite differences in the gravitational potential models, the morphology and orientation of stellar orbits remain largely consistent. However, the total energy values differ significantly, with the energies computed in the \citet{McMillan2017}  potential being nearly twice as negative as those obtained using the {\sc MilkyWayPotential2014} \citep{Bovy2015}. This discrepancy arises from the deeper mass distribution represented in the \citet{McMillan2017} potentials, which is calibrated with more recent observational constraints and includes a more massive dark matter halo, denser disk component, and more detailed baryonic structure, including gas. While such differences are expected in quantities sensitive to the depth of the potential, the close agreement in other orbital parameters suggests that both models yield dynamically consistent results overall.

\section{Summary and Discussion}

\subsection{Summary of Spectral Analysis Results}

\begin{figure}
    \centering
    \includegraphics[width=1\linewidth]{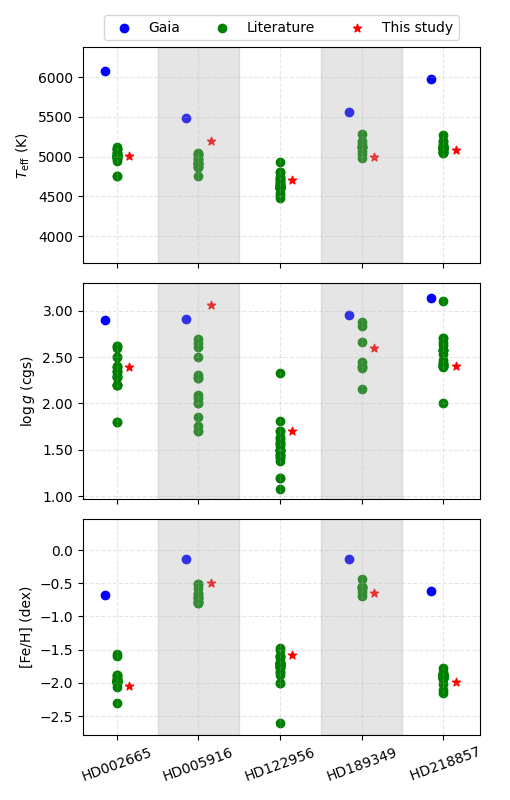}
    \caption{Comparison of atmospheric parameters for the five program stars derived in this study (red stars) with values from the literature (green circles) and the {\it Gaia} DR3 catalog (blue circles). From top to bottom, the panels show the effective temperature ($T_{\mathrm{eff}}$), surface gravity ($\log g$), and metallicity ([Fe/H]). The shaded background bands alternate for clarity between adjacent stars.}
    \label{fig:literature}
\end{figure}

\begin{figure*}
    \centering
    \includegraphics[width=0.97\linewidth]{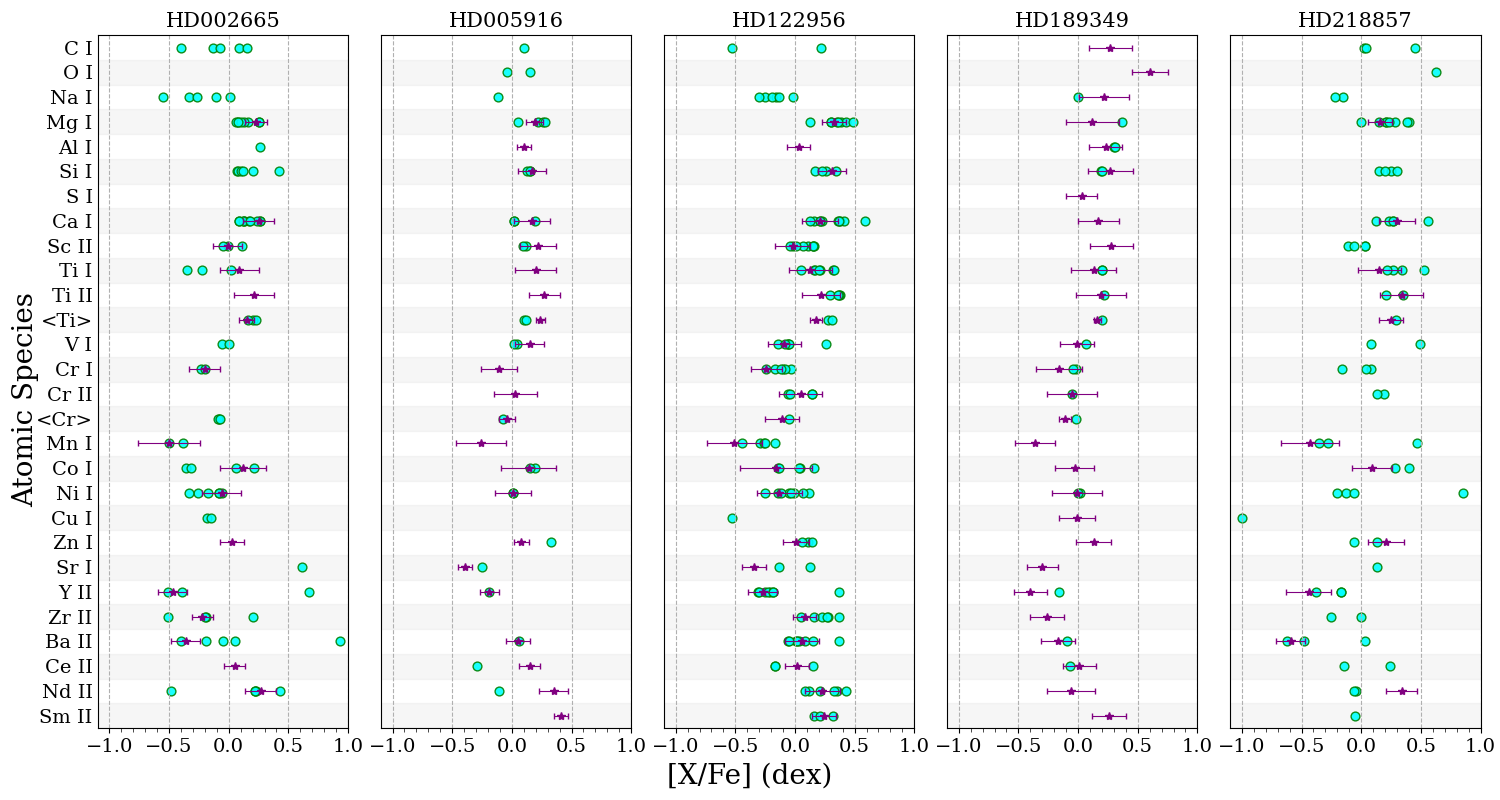}
    \caption{Comparison of elemental abundance ratios [X/Fe] for the five program stars analyzed in this study (purple stars with error bars) against values compiled from the literature (cyan circles). Each panel corresponds to a different star, with the vertical axis listing the atomic species and the horizontal axis representing the [X/Fe] abundance ratio in dex unit. Error bars indicate the $\pm 1\sigma$ uncertainties of the measurements derived in this study. For elements for which both neutral and ionized species are available (e.g., Ti\,{\sc i}, Ti\,{\sc ii}, Cr\,{\sc i}, and Cr\,{\sc ii}), we included all individual measurements as well as their averaged abundances. The combined [X/Fe] values are shown to facilitate comparison with literature studies that report only averaged quantities.}
    \label{fig:lit-abund}
\end{figure*}

The model atmospheric parameters of the program stars reported in the literature are listed in Table \ref{tab:lit1}. To complement this, Figure \ref{fig:literature} provides a visual comparison of the atmospheric parameters derived in this study with values from both the literature and the {\it Gaia} DR3 catalog, allowing a clearer assessment of the agreement and systematic differences among the sources. Furthermore, \mbox{Figure \ref{fig:lit-abund}} presents a comparison of the elemental abundances determined for the five program stars with those reported in the literature, offering additional validation of our abundance analysis. The mean model atmospheric parameters derived by averaging the reported values are presented in Table \ref{tab:main_results}. In the following subsections, we provide a summary of the spectral analysis results for each of the five program stars, including comparisons with literature values and discussions of their individual spectroscopic characteristics.

\subsubsection{HD\,2665}

Compared to the mean literature values, the spectroscopically derived effective temperature for HD\,2665 is 36 K cooler, exhibits 0.05 dex higher surface gravity, and is 0.17 dex more metal-poor.

\citet{2004venn}, who classified Galactic components using a Bayesian method based on Gaussian velocity ellipsoids, assigned HD\,2665 to the halo group. Their study compiled space velocities and abundances from \citet{Fulbright2000,2002Fulbright}. \citet{Fulbright2000} analyzed the high-resolution ($R \sim 50\,000$) spectrum of HD\,2665 obtained with the 3m Shane Telescope and Hamilton spectrograph and reported the atmospheric parameters of 5050 $\pm$ 40 K, $\log g = 2.20 \pm 0.06$ cgs, [Fe/H] $= -1.80 \pm 0.04$ dex, and microturbulent velocity $v_{\text{micro}} = 1.60 \pm 0.11$ km s$^{-1}$ for the star across the wavelength range $\lambda\lambda$4\,190–9\,900. The uncertainties represent the error estimates calculated via the Monte Carlo methodology applied to multi-epoch observational data for 18 stars. The atmospheric model used to determine the abundances in their work was 40 K hotter, exhibited 0.19 dex lower surface gravity, and was 0.25 dex more metal-rich than our parameters. Kurucz models were adopted for the spectrum synthesis, with non-LTE effects corrected for Fe lines. When comparing the elemental abundances jointly reported in their study and ours, [Mg\,{\sc i}/Fe], [Ca\,{\sc i}/Fe], $\langle$[Ti/Fe]$\rangle$, [Ni\,{\sc i}/Fe], [Y\,{\sc ii}/Fe], and [Zr\,{\sc ii}/Fe] showed discrepancies of $<$ 0.05 dex, whereas $\langle$[Cr/Fe]$\rangle$ and [Ba\,{\sc ii}/Fe] agreed within $<$ 0.20 dex.  

\begin{figure*}
    \centering
    \includegraphics[width=0.9\linewidth]{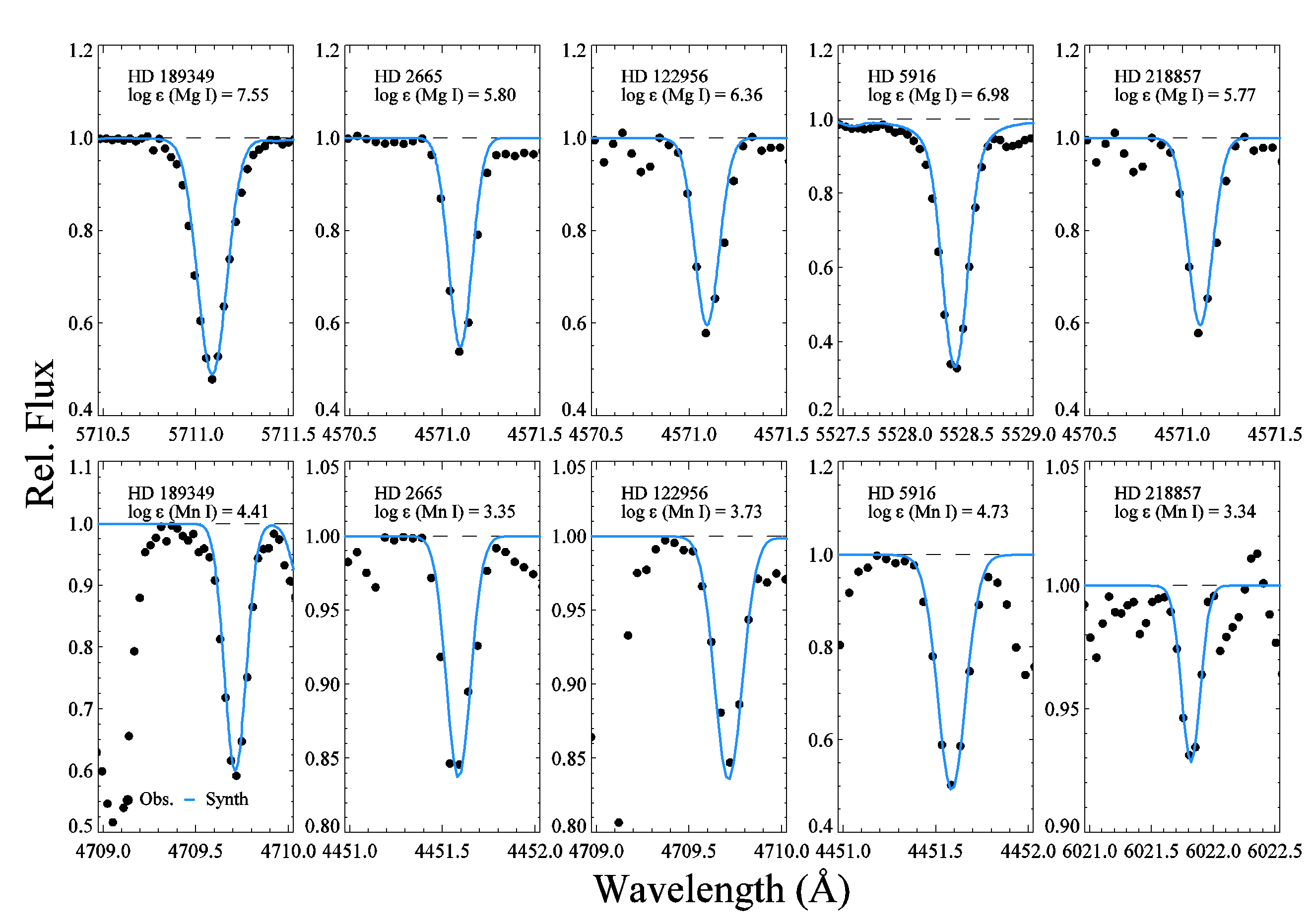}
    \caption{Comparison of observed (black dotted) and synthetic (blue) spectra for Mg\,{\sc i} and Mn\,{\sc i} lines in the program stars HD\,189349, HD\,2665, HD\,122956, HD\,5916, and HD\,218857. The spectral regions shown include the Mg\,{\sc i} and Mn\,{\sc i} lines. Derived logarithmic abundances ($\log~\epsilon$) for Mg\,{\sc i} and Mn\,{\sc i} are also indicated.}
    \label{fig:synth-mg-mn}
\end{figure*}

\citet{2006Sobeck}, which reports data for HD~2665 in the CDS but is not cited in SIMBAD's references, compares the [Mn\,{\sc i}/Fe] abundances of over 200 field stars and 19 GCs in the metallicity range of $-2.7<\mathrm{[Fe/H]~(dex)}<-0.7$. This study aimed to investigate the nucleosynthetic contribution of massive stars undergoing supernovae (SN) to Mn. The Mn transitions identified by \citet{2006Sobeck} are 6013.51 \AA, 6016.64 \AA, and 6021.82 \AA. The abundance of these hyperfine structure (HFS)-affected transitions was computed using the spectrum synthesis method. They adopted stellar atmospheric parameters from \citet{Fulbright2000}. The reported [Mn/Fe] ratio for the star was -0.38 dex and differed $\approx$0.1 dex from the Mn abundance reported in this study for the star. Notably, the [Mn/Fe] ratio by \citet{2006Sobeck} was not calculated relative to model metallicity. Instead, it was determined locally using Fe transitions satisfying the following criteria: no non-LTE contributions, the presence of the same spectral order as the Mn lines, and excitation potentials comparable to the Mn triplet. Two such transitions, (6024.06 \AA~ and 6027.05 \AA), were used, yielding [Fe/H] = -2.01 dex for HD\,2665. This value is 0.21 dex more metal-poor than the model metallicity reported by \citet{Fulbright2000}, but agrees with the metallicity (-2.05 dex) reported in this study.  

In our study, the [Mn/Fe] = -0.50 $\pm$ 0.26 dex value was derived using the spectrum synthesis of Mn\,{\sc i} lines at the 4082.9 \AA, 4451.5 \AA,  4765.8 \AA, and 4783.4 \AA\ transitions (Figure \ref{fig:synth-mg-mn}). The laboratory oscillator strengths of these Mn lines include the HFS.

\citet{Heiter2015}, aiming to obtain reliable, spectroscopy-independent atmospheric parameters using {\it Gaia} data, identified HD\,2665 as a candidate {\it Gaia} benchmark star. Based on this rationale, \citet{2020Karovicova} determined the radius of a star via interferometric measurements and derived an effective temperature of $4883 \pm 95$ K using the Stefan-Boltzmann relation combined with bolometric flux calculations. Surface gravity and metallicity were determined via isochrone fitting and spectroscopy, yielding $\log g = 2.21 \pm 0.03$ cgs and [Fe/H] $= -2.10 \pm 0.10$ dex, respectively. Their model is a 167 K cooler, exhibits a 0.18 dex lower surface gravity, and is 0.05 dex more metal-poor than our model. 

\citet{Boeche2016}, who created an EW library for atmospheric parameter grids by calibrating $\log gf$ values for 4643 absorption lines across $\lambda\lambda$5212–6860 and $\lambda\lambda$8400–8924\footnote{A cross-entropy algorithm was applied to high-resolution spectra of five well-known stars to ensure consistent measurements}, aimed to derive precise atmospheric parameters and abundances from low- to medium-resolution spectra ($R \sim 2\,000$–20\,000). Their Fortran-based SP-ACE code calculates the EWs, generates synthetic spectra, and performs $\chi^2$ optimization to identify the optimal atmospheric parameters. The line list in the EW library includes transitions observed in the spectra of the Sun (G), Arcturus (K), and Procyon (F), with atomic data compiled from VALD and EWs computed using the MOOG. Applying the SP-ACE code to the high-resolution ELODIE spectrum of HD\,2665, they reported the atmospheric parameters of $T_{\text{eff}} = 5046^{+26}_{-134}$ K, $\log g = 2.39^{+0.04}_{-0.69}$ cgs, and [Fe/H] $= -1.96^{+0.01}_{-0.14}$ dex, describing an atmosphere 4 K cooler with identical surface gravity and 0.09 dex higher metallicity compared to our model. For commonly reported elemental abundances,  [Sc\,{\sc ii}/Fe] (11), $\langle$[Ti/Fe]$\rangle$ (32), [Co\,{\sc i}/Fe] (12), and [Ni\,{\sc i}/Fe] (44) agree within < 0.05 dex, whereas Mg\,{\sc i} two lines), Ca\,{\sc i} (26), and $\langle$[Cr/Fe]$\rangle$ (28) show discrepancies of < 0.15 dex.  

The medium-resolution INT library of empirical spectra (MILES), an empirical stellar library comprising low-/medium-resolution ($R \sim 2\,000$) spectra from the Isaac Newton Telescope, was expanded by \citet{2021garcia} to include new observations, including that of HD\,2665. Automated spectrum synthesis yielded atmospheric parameters of $T_{\text{eff}} = 5109.83 \pm 2.72$ K, $\log g = 1.24 \pm 0.02$ cgs, [Fe/H] $= -2.03 \pm 0.01$ dex, and $v_{\text{micro}} = 0.58 \pm 0.00$ km s$^{-1}$, describing an atmosphere 60 K hotter, 1.15 cgs lower in surface gravity, and 0.02 dex more metal-rich than our model. The optimization used the FERRE library \citep{AllendePrieto_2006} with parameter steps of 250 K ($T_{\rm eff}$), 0.5 ($\log g$), 0.25 ([Fe/H]), and 0.3 ($\xi$). Their reported $\alpha$-abundance averaged over [O/Fe], [Ne/Fe], [Mg/Fe], [Ti/Fe], [Si/Fe], [S/Fe], and [Ca/Fe] was $0.27 \pm 0.02$ dex, whereas our $\alpha$-abundance derived from [Mg\,{\sc i}/Fe], [Ca\,{\sc i}/Fe], and $\langle$[Ti/Fe]$\rangle$ was $0.21 \pm 0.05$ dex, differing by 0.06 dex. Although the study does not specify a line list for synthesis, it acknowledges the use of theoretical Kurucz line lists. In our study, the $\alpha$-abundance for HD\,2665 was based on 41 transitions, with atomic data updated using laboratory results.  

Investigating the origin of neutron-capture elements, \citet{Burris2000} calculated the abundance of metal-poor giants using the blue ($\lambda\lambda$4070–4710, $R \sim 20\,000$) and red ($\lambda\lambda$6100–6180, $R \sim 22\,000$) spectral regions. For HD\,2665, they adopted atmospheric parameters from \citet{Pilachowski1996}, who derived $T_{\text{eff}} = 5000$ K from Strömgren photometry and $\log g = 2.20$ cgs via color-magnitude diagram placement and ionization balance (with a 0.14 dex discrepancy). Their model, which had a 50 K cooler and 0.19 dex lower surface gravity than ours, reported [Fe/H] = -3 dex, although the methodology is unclear, preventing an explanation of the 0.95 dex discrepancy with our metallicity. However, \citet{Burris2000} used [Fe/H] $=-1.97$ dex (based on eight Fe\,{\sc i} and one Fe\,{\sc ii} line) instead of the nominal -3 dex, resulting in a 0.08 dex difference from our metallicity estimate. Jointly reported [Y\,{\sc ii}/Fe], [Zr\,{\sc ii}/Fe], and [Nd\,{\sc ii}/Fe] abundances agree within < 0.10 dex, whereas [Ba\,{\sc ii}/Fe] shows a 0.31 dex discrepancy. Their Ba analysis used 4554 \AA~ and 6141 \AA, whereas we employed 5853 \AA, 6141 \AA, and 6496 \AA. The shared 6141 \AA~ line shows a 0.03 dex $\log gf$ difference. The exclusion of 4554 \AA~ in our work is due to its strong profile, which reduces the reliability under microturbulence variations. The 0.6 km s$^{-1}$ difference in $v_{\text{micro}}$ between studies may explain the Ba discrepancy if 4554 \AA~ is prioritized.  

No Ce abundance in HD\,2665 has been reported in the literature for the past 25 years. This paper presents the first detection of the Ce {\sc ii} line at 4628.16 \AA. The atomic data adopted for this transition include an excitation potential of 0.52 eV and an oscillator strength ($\log gf$) of 0.14 dex \citep{Lawler2009}.  

\subsubsection{HD\,189349}

This star, with 36 publication records since 1850, was reported as an HB-type red giant \citep{2014Skiff, Afsar2018}.  

The methodology initiated by \citet{Heiter2015} for metal-poor regimes in the {\it Gaia} Benchmark Star survey to obtain precise model parameters independent of spectroscopy \citet{2024Soubiran} contributed to the analysis of an additional 200 stars. Using interferometric measurements from \citet{2020Karovicova,2022Karovicova}, the limb-darkened angular diameter was combined with the bolometric flux and effective temperature to determine the effective temperature of the star as 5175$\pm$31 K. The radius was calculated as $8.931 \pm 0.109 R_\odot$ via the Stefan-Boltzmann relation, whereas the mass was derived as $0.723 \pm 0.057 M_\odot$ using BASTI evolutionary tracks \citep[e.g.][]{Hidalgo2018, Pietrinferni2021, Salaris2022} 
 based on the temperature-luminosity relationship. The surface gravity, computed from $M$ and $R$ using Newton's law of gravitation and {\it Gaia} DR3 parallax, was consistent; that is, 2.39$\pm$0.04 cgs. The metallicity ([Fe/H]) obtained from the PASTEL database \citep[c.f.][]{Soubiran2010, Soubiran2016}, was -0.59$\pm$0.03 dex. Thus, their model parameters are 175 K hotter, 0.21 dex lower in surface gravity, and 0.06 dex more metal-rich than our spectroscopically determined model atmospheric parameters. \citet{2016Takeda} noted that stars have a 68$\%$ probability of being RGB stars, whereas seismic analysis suggests HB characteristics.

Using high-resolution Subaru telescope HDS ($R\sim~$80\,000) data, \citet{Takeda2015} and \citet{2016Takeda} investigated the atmospheric parameters and age-metallicity relation for selected giant Kepler field stars and determined atmospheric parameters of 5026 K / 2.45 cgs / -0.63 dex / 1.46 km s$^{-1}$ for HD\,189349. Compared to our model, these values are 26 K hotter, 0.15 dex lower in surface gravity, and 0.02 dex more metal-rich. Spectroscopic excitation and ionization equilibria yielded a $\log g$ value consistent with seismic determinations, thus supporting the high-reliability mass estimates. The PARSEC evolutionary model \citep{Bressan2012} provides a mass of $0.78M_{\odot}$, which differs from $0.56M_{\odot}$ in this study. This discrepancy can be attributed to the constraints of the adopted evolutionary tracks. The reported age of 16.33 Gyr\footnote{\(\log(\text{age})=10.213\) years} contrasts with the derived age of 2.77 Gyr.  

\citet{2019Liu} reported the elemental abundances of this star using the atmospheric parameters of \citet{Takeda2015}. Relative to our abundances, differences are $<$ 0.1 dex for [Al\,{\sc i}/Fe], [Si\,{\sc i}/Fe], [Ca\,{\sc i}/Fe], [Ti/Fe], [V\,{\sc i}/Fe], [Cr/Fe], [Ni\,{\sc i}/Fe], Ba\,{\sc ii}/Fe], and [Ce\,{\sc ii}/Fe], and $<$ 0.25 dex for [Na\,{\sc i}/Fe], [Mg\,{\sc i}/Fe], and [Y\,{\sc ii}/Fe]. For elements with 0.1 dex differences, the uncertainty arises from variations in metallicity. The 0.25 dex difference suggests additional factors beyond metallicity. Our spectroscopic comparison benchmarks were: (i) atmospheric parameters, (ii) atomic data cross-validation, and (iii) reference solar abundances. The 1$\sigma$ agreement between \citet{Takeda2015} and our atmospheric determinations rules out parameter differences as a significant source of error. This is evident from Table \ref{tab:modelerr}, which shows the abundance variations under 1$\sigma$ parameter shifts for each element. However, \citet{Li2019} did not share their line lists, thus precluding atomic data-level comparisons. Finally, the reference solar abundances were derived directly from the solar spectrum analysis using published values. \citet{Li2019} adopted lunar-based reference abundances but did not report specific values. \citet{Li2019} also identified the star as belonging to the thick disk using {\it Gaia} DR2 dynamical data via \citet{Bensby2003}'s methodology. No population analysis using {\it Gaia} DR3 data is available in the literature. This study provides the first assessment using {\it Gaia} DR3 data. 

\citet{Afsar2018} spectroscopically determined atmospheric parameters of 4978 K / 2.16 cgs / -0.69 dex/ 1.58 km s$^{-1}$ for HD\,189349 using high-resolution spectra from McDonald Observatory's HJS telescope and the 2.7m/Tull spectrograph. Compared to our model, these values are 22 K cooler, 0.44 dex lower in surface gravity, and 0.04 dex more metal-poor. \citet{Afsar2018} noted that their surface gravity determination was 0.29 dex lower than the literature values, but did not investigate the cause. Their $\log g$ determination via Fe and Ti ionization equilibria utilized up to 70 Fe\,{\sc i}, 12 Fe\,{\sc ii}, 11 Ti\,{\sc i}, and six Ti\,{\sc ii} lines. Our analysis employed more extensive line samples: 188 Fe\,{\sc i}, 23 Fe\,{\sc ii}, 50 Ti\,{\sc i}, and 10 Ti\,{\sc ii} lines for the ionization balance. Thus, our results benefit from a more robust line analysis than that of \citet{Afsar2018}. The reported \(\alpha\)-element abundances ([Si\,{\sc i}/Fe]: 0.31 $\pm$ 0.05 [15 lines], [Ca\,{\sc i}/Fe]: 0.20 $\pm$ 0.04 [10], [Ti\,{\sc i}/Fe]: 0.20 $\pm$ 0.05 [10], and [Ti\,{\sc ii}/Fe]: 0.22 $\pm$ 0.06 [6]) differed from ours by <0.1 dex. The Fe-peak elements show differences of 0.12 dex for [Cr\,{\sc i}/Fe] (20 lines), 0.01 dex for [Ni\,{\sc i}/Fe] (59 lines), and 0.05 dex for [Cr\,{\sc ii}/Fe] (5 lines) between their reported abundances ([Cr\,{\sc i}/Fe]: -0.04 $\pm$ 0.07 [17], [Cr\,{\sc ii}/Fe]: -0.05 $\pm$ 0.08 [5], [Ni\,{\sc i}/Fe]: 0.00 $\pm$ 0.06 [27]) and ours.  

We present the first abundance measurements for C\,{\sc i} (8335.19, 9111.85), O\,{\sc i} (7771.96, 7774.18, 7775.40), S\,{\sc i} (8694.70), Sc\,{\sc ii} (4246.84, 5239.82, 5526.82, 5640.99, 5657.88, 5667.15, 5669.04, 6245.63, 6320.85, 6604.60), Mn\,{\sc i} (4055.55, 4082.94, 4451.59, 4470.14, 4709.72, 4739.11, 4765.86, 4766.42, 4783.42, 5432.55, 6013.50, 6021.80), Co\,{\sc i} (4792.86, 4813.48, 5352.05, 5647.23, 6093.15), Cu\,{\sc i} (5105.54), Zn\,{\sc i} (4722.16, 4810.54), Sr\,{\sc i} (4607.34), Zr\,{\sc ii} (4208.98), Nd\,{\sc ii} (4446.40, 5092.80, 5293.17), and Sm\,{\sc ii} (4577.69) in HD\,189349, with atomic data provided in \citet{Sahin2024}. The transitions of C\,{\sc i} (8335.19, 9111.85) and O\,{\sc i} (7771.96, 7774.18, 7775.40) were outside the ELODIE data range. Our comprehensive spectroscopic analysis, synergistically combined with high-precision photometric data from {\it TESS} and {\it Kepler}, will enable detailed asteroseismic investigations to reveal the internal structure and evolutionary history of the stars.

\subsubsection{HD\,5916}

The star, classified as a giant in the HPM luminosity class, is listed in the {\it Hubble Space Telescope} (HST) guide star catalog \citep{2008Lasker}. Belonging to the Arcturus group\footnote{stars with ages $\sim$14 Gyr and [Fe/H]$\sim$ -0.65 dex}, a population defining the old disk \citep[see][]{1998Eggen}, its literature-reported atmospheric parameters are presented in Table \ref{tab:lit1}. Averaging these values revealed that our spectroscopically determined atmospheric parameters are 233 K cooler, 0.81 dex lower in surface gravity, and 0.16 dex more metal poor. To better understand this discrepancy, the following paragraphs detail the methodologies used in spectroscopic studies.

\citet{Niedzielski2016} derived atmospheric parameters of 4915$\pm$10 K / 2.27$\pm$0.04 cgs / -0.72$\pm$0.01 dex / 1.39$\pm$0.05 km s$^{-1}$ for HD\,5916 via automated EW measurements of Fe I-II lines from high-resolution HET ($R\sim$60\,000) spectra. Their model has a 285 K cooler, 0.79 dex lower surface gravity, and 0.22 dex more metal-poor than ours. While their analysis relied solely on the Fe ionization equilibrium, our study incorporated Ti and Cr ionization equilibria to improve precision. 

\citet{Mishenina2001} determined atmospheric parameters of 4863$\pm$100 K / 1.7$\pm$0.3 cgs / -0.51$\pm$0.25 dex / 1.2$\pm$0.2 km s$^{-1}$ by fitting H$_\alpha$ wings and Fe ionization equilibrium in ELODIE spectra, using WIDTH-9 for abundance calculations. Compared to our model, this is a 337 K cooler, 1.36 dex lower in surface gravity, and 0.01 dex more metal-poor. Although \citet{Mishenina2001} focused solely on the Fe equilibrium, we achieved a Ti{\sc i/ii} ionization balance within 0.1 dex. The chromospheric activity reported for H$_\alpha$ by \citet{2018Adamov} suggests that the temperature determination from line-wing fitting may be less reliable than full spectrum analyses. Despite the parameter differences, the abundance discrepancies for the common elements were small: [Mg\,{\sc i}/Fe], [Si\,{\sc i}/Fe], [Ca\,{\sc i}/Fe], [Y\,{\sc ii}/Fe], [Ba\,{\sc ii}/Fe]  $<$ 0.07 dex; [Sr\,{\sc i}/Fe] $ <$ 0.15 dex; [Ce\,{\sc ii}/Fe], and [Nd\,{\sc ii}/Fe]  $<$ 0.5 dex. Our Nd abundances from the 4446 \AA, 5092 \AA, and 5293 \AA\ transitions differ from their unlisted lines. Ce\,{\sc ii} transitions at 4562.37 and 4628.16 \AA\, show $\sim$0.04 dex $\log gf$ differences versus \citet{Mishenina2001}, while the single shared Sr\,{\sc i} line (4607.34 \AA) exhibits a 0.11 dex log $gf$ difference, explaining the 0.13 dex abundance discrepancy. [Zn\,{\sc i}/Fe] abundances differ by 0.22 dex (0.33$\pm$0.26 in \citet{2002Mishenina} versus our work).  

\citet{Boeche2016} derived parameters of 5050$_{-40}^{+8}$ K / 2.66$_{-0.07}^{+0.05}$ cgs/ -0.64$_{-0.04}^{+0.01}$ dex using the SP-ACE code on the ELODIE spectra. Their model had a 150 K cooler, 0.4 dex lower surface gravity, and 0.14 dex more metal-poor than ours. The abundance differences remain modest: $<$ 0.05 dex for [Mg\,{\sc i}/Fe], $\langle$[Cr/Fe]$\rangle$, [Co\,{\sc i}/Fe], and [Ni\,{\sc i}/Fe]$<$ 0.15 dex for [Ca\,{\sc i}/Fe], [Sc\,{\sc ii}/Fe], $\langle$[Ti/Fe]$\rangle$. Despite the surface gravity discrepancies, the minimal abundance variations suggest stable thermonuclear and $\alpha$ processes in the core/envelope. The ionization equilibria of Fe, Cr, and Ti surpassed the unspecified methodology of SP-ACE in terms of precision.  

\citet{2021garcia} obtained parameters of $T_{\text{eff}}$=4899$\pm$2.89 K, $\log g$=1.92$\pm$0.02 cgs, [Fe/H]=$-0.54\pm0.01$ dex, and $\xi=0.05\pm0.01$ km s$^{-1}$ via spectral synthesis of MILES library data. This model is 301 K cooler and 1.14 dex lower in surface gravity than ours, yet yields comparable [$\alpha$/Fe]=0.12$\pm$0.01 (from O, Ne, Mg, Si, S, Ca, Ti) versus 0.19$\pm$0.03 dex ([Mg\,{\sc i}/Fe], [Si\,{\sc i}/Fe], [Ca\,{\sc i}/Fe], and $\langle$[Ti/Fe]$\rangle$). The $\alpha$-abundance consistency across parameter differences suggests an intrinsic $\alpha$-anomaly characteristic of this particular star.  

Notably, Al\,{\sc i} (6695.97, 6698.63), V\,{\sc i} (4437.84, 4577.18, 5727.06, 6119.53, 6243.11) Sm\,{\sc ii} (4577.69), and Mn\,{\sc i} (4055.55, 4451.59, 4709.72, 4739.11, 4783.42, 6021.80) abundances were reported for the first time for this star using spectral synthesis (see Figure \ref{fig:alsynth} for Al lines). 


\subsubsection{HD\,122956}

A star was identified as an inner halo member \citet{Ishigaki2013}. The infrared spectrum shows strong evidence of the 10830 \AA\, He\,{\sc i} transition. The absorption profile lacks P-Cygni features, indicating chromospheric rather than wind-driven origins \citep{2009Dupree}. The sensitivity of this transition to magnetic fields explains its prominence in interstellar polarization studies \citep{2019Gontcharov}. The HST spectra further revealed chromospheric outflows via asymmetric 2800 \AA~ Mg{\sc ii} emission \citep{2009Dupree}.  

Our spectroscopically determined model is 57 K hotter, 0.18 dex higher in surface gravity, and 0.18 dex more metal-rich than the average atmospheric parameters reported for HD\,122956 in the literature (Table \ref{tab:lit1}). 

\citet{2022Lombordo} derived the atmospheric parameters $T_{\text{eff}} =$ 4642 K, $\log g = 1.58$ cgs, [Fe/H] = -1.87 dex, $\xi =$ 1.81 km s$^{-1}$ for HD\,122956 using {\it Gaia} DR3 photometry/parallax and ATLAS9 grids. This model is 58 K cooler, 0.12 dex lower in gravity, and $\approx$0.3 dex more metal-poor than our model. The abundance differences were $<$0.05 dex for [Mg\,{\sc i}/Fe], [Si\,{\sc i}/Fe], [V\,{\sc i}/Fe], [Y\,{\sc ii}/Fe], $\langle$[Cr/Fe]$\rangle$, and $<$ 0.20 dex for [Ca\,{\sc i}/Fe], [Sc\,{\sc ii}/Fe], $\langle$[Ti/Fe]$\rangle$, [Mn\,{\sc i}/Fe], [Co\,{\sc i}/Fe], [Ni\,{\sc i}/Fe], [Zn\,{\sc i}/Fe],  $<$ 0.30 dex for [Zr\,{\sc ii}/Fe]. 

\citet{2021garcia} reported $T_{\text{eff}} = 4717\pm6.68$ K, $\log g = 0.64\pm0.02$ cgs, [Fe/H] = -1.69$\pm$0.01 dex, $\xi = 0.31\pm0.01$ km s$^{-1}$ from MILES spectra, yielding [$\alpha$/Fe] = 0.48$\pm$0.02 versus our 0.25$\pm$0.08 dex. The 0.23 dex $\alpha$-abundance difference likely stems from the Kurucz line lists and microturbulence (1.5~km~s$^{-1}$ higher here). Table \ref{tab:modelerr} shows that the Mg and Ti abundances vary by $\sim$0.2 dex per 0.5 kms$^{-1}$ in $\xi$, supporting this hypothesis.  

{\citet{Boeche2016}'s SP-ACE analysis of the ELODIE spectra yielded 4728$_{-107}^{+5}$K / 1.81$_{-0.22}^{+0.07}$ cgs / -1.67$_{-0.05}^{+0.01}$ dex, 28 K hotter but 0.11 dex lower in \(\log g\) than in our model. Abundance differences remain $<$0.05 dex for [Mg\,{\sc i}/Fe], [Ca\,{\sc i}/Fe], [Sc\,{\sc ii}/Fe], $\langle$[Ti/Fe]$\rangle$, [V\,{\sc i}/Fe], $\langle$[Cr/Fe]$\rangle$; <0.15 dex for [Si\,{\sc i}/Fe], [Co\,{\sc i}/Fe], [Ni\,{\sc i}/Fe].

\citet{2013Frabel} derived \(4600~{\rm K} / 1.0~{\rm cgs} / -1.8 ~{\rm dex} /~2.0 ~{\rm km~s}^{-1}\) from $R~\sim 22\,000$ MIKE spectra, 100 K cooler and 0.7 dex lower in \(\log g\) than our values. Abundance comparisons show <0.05 dex differences for [Cr\,{\sc i}/Fe]; $<$0.10 dex for [Mg\,{\sc i}/Fe], $\langle$[Ti/Fe]$\rangle$; <0.20 dex for [Ca\,{\sc i}/Fe], [Sc\,{\sc ii}/Fe], [Zn\,{\sc i}/Fe]; <0.35 dex for [Mn\,{\sc i}/Fe], [Ni\,{\sc i}/Fe], [Ba\,{\sc ii}/Fe]. Their results align with those of \citet{Fulbright2000} within 0.15 dex, which is consistent with our $<$0.10 dex difference for most elements.  

\citet{Ishigaki2013} obtained 4609 K / 1.6 cgs / -1.71 dex / 1.7 km s$^{-1}$ from Subaru/HDS ($R\sim$55\,000) spectra, 91 K cooler and 0.13 dex more metal-poor than our model. Abundances agree within $<$0.05 dex for [V\,{\sc i}/Fe], [Co\,{\sc i}/Fe], [Ni\,{\sc i}/Fe], [Zn\,{\sc i}/Fe], [Y\,{\sc ii}/Fe], [Ba\,{\sc ii}/Fe], [Nd\,{\sc ii}/Fe]; $<$0.15 dex for [Sc\,{\sc ii}/Fe], $\langle$[Cr/Fe]$\rangle$, [Sm\,{\sc ii}/Fe]; $<$0.25 dex for [Mn\,{\sc i}/Fe], [Zr\,{\sc ii}/Fe]. \(\alpha\)-abundances match \citet{Ishigaki2012} within $<$ 0.05 dex ([Mg\,{\sc i}/Fe], [Si\,{\sc i}/Fe], [Ca\,{\sc i}/Fe]) and $<$ 0.10 dex ($\langle$[Ti/Fe]$\rangle$).  

\citet{Mishenina2001}'s ELODIE analysis yielded 4635 K / 1.5 cgs / -1.60 dex / 1.5 km s$^{-1}$, 65 K cooler and 0.2 dex lower in $\log g$ than our model.The abundances differed by $<$0.10 dex for [Si\,{\sc i}/Fe], [Ca\,{\sc i}/Fe], [Y\,{\sc ii}/Fe], [Ba\,{\sc ii}/Fe], and [Nd\,{\sc ii}/Fe], and $<$0.25 dex for [Mg\,{\sc i}/Fe], [Sr\,{\sc i}/Fe], and [Ce\,{\sc ii}/Fe].  

This study is the first to report the 6695.97 \AA~ Al\,{\sc i} transition abundance, expanding the chemical profile of HD\,122956. 


\subsubsection{HD\,218857}

\citet{Hawkins2016} classified this star among metal-poor benchmark candidates. \citet{Ishigaki2013} identified stars as inner halo members based on elemental abundance trends.  

\citet{2023Santos} analyzed the IR Ca\,{\sc i} triplet (8498.02, 8542.09, 8662.14 \AA\,) and Mg\,{\sc i}b triplet (5167.3, 5172.7, 5183.6 \AA) using X-Shooter VIS/NIR
spectra, adopting atmospheric parameters $T_{\text{eff}} =$ 5107 K, $\log g =$ 2.58 cgs, [Fe/H] = -1.90 dex from \citet{Arentsen2019}, who employed ULySS spectral fitting to MILES templates ($R\sim$10\,000). This model is 27 K hotter, 0.18 dex higher in \(\log g\), and 0.08 dex more metal-rich than ours. The reported abundances [Mg\,{\sc i}/Fe] = 0.28$\pm$0.03 and [Ca\,{\sc i}/Fe] = 0.27$\pm$0.03 agree with our results within 0.12 and 0.03 dex, respectively.  

\citet{2021garcia} derived $T_{\text{eff}} = 5130\pm3.90$ K, $\log g = 1.58\pm0.03$ cgs, [Fe/H] = -2.03$\pm$0.01 dex, $\xi = 0.29\pm0.01$ km s$^{-1}$ from MILES spectra, yielding [$\alpha$/Fe] = 0.44$\pm$0.03. Our [$\alpha$/Fe] = 0.24$\pm$0.07 ([Mg\,{\sc i}/Fe], [Ca\,{\sc i}/Fe], $\langle$[Ti/Fe]$\rangle$) shows a 0.15 dex discrepancy, which is attributed to systematic differences in $\log g$ and $\xi$ determination methodologies, consistent with the HD\,5916 and HD\,122956 analyses.  

\citet{Mashonkina2017} reported high-precision parameters for metal-poor giants in dwarf galaxies and the MW, by combining photometric $T_{\text{eff}}$, astrometric $\log g$, and VLT/UVES ($R\sim$56\,000) spectroscopy. For HD\,218857, they derived (5060 K / 2.53 cgs / -1.92 dex / 1.4 km s$^{-1}$), which is 20 K cooler, 0.13 dex higher in $\log g$, and 0.06 dex more metal-rich than our model. non-LTE abundances $\log \langle {\rm Ti} \rangle = 3.26\pm0.02$, $\log \langle {\rm Fe} \rangle = 5.56\pm0.02$ yield [Ti/Fe] = 0.27 dex, consistent with our LTE results within $<$0.03 dex.  

\citet{Boeche2016} obtained 5071$_{-168}^{+17}$ K, 2.63$_{-0.48}^{+0.19}$ cgs, -2.00$_{-0.23}^{+0.02}$ dex via SP-ACE analysis of ELODIE spectra, 9 K cooler and 0.23 dex higher in \(\log g\) than our model. Abundance differences are <0.05 dex ([Mg\,{\sc i}/Fe]), <0.20 dex ([Ca\,{\sc i}/Fe], [Co\,{\sc i}/Fe]), and $<$0.30 dex ($\langle$[Ti/Fe]$\rangle$). The potential causes of Ti discrepancies include the following:  (i) overestimated ionized Ti due to \(\log g\) differences affecting the ionization balance, (ii) isotopic contributions skewing neutral Ti abundances, and (iii) exclusive use of Ti\,{\sc ii} lines. Considering (iii), comparing only Ti\,{\sc ii} reduces the discrepancy to 0.2 dex.  

\citet{Ishigaki2013} derived 5107 K / 2.71 cgs / -1.94 dex / 1.70 km s$^{-1}$ from Subaru/HDS ($R\sim$50\,000) spectra, 27 K hotter and 0.31 dex higher in \(\log g\) than our model. The abundances agree within $<$0.10 dex ([Mn\,{\sc i}/Fe], [Zn\,{\sc i}/Fe], [Y\,{\sc ii}/Fe], [Ba\,{\sc ii}/Fe]), and $<$0.40 dex ([Nd\,{\sc ii}/Fe]). The EW measurements for six Nd\,{\sc ii} lines (five $<$10 m\AA ~ and one 15 mÅ at 4061.08 \AA) contrast our spectrum synthesis results (3 m\AA\, EW at 4446.38 \AA). \citet{Ishigaki2012} reports [$\alpha$/Fe]) values ([Mg\,{\sc i}/Fe], [Ca\,{\sc i}/Fe], $\langle$[Ti/Fe]$\rangle$) consistent within $<$ 0.10 dex. 

\subsection{[Mg\,{\sc i}/Mn\,{\sc i}] vs [Al\,{\sc i}/Fe] diagram}

Figure~\ref{fig:chemicalplane} shows the distribution of program stars in the [Mg/Mn]–[Al/Fe] abundance plane, which is a diagnostic tool for distinguishing between in-situ and ex-situ stellar populations. A background comparison sample drawn from the APOGEE DR17 survey was also included, following the same quality cuts used in Figures \ref{fig:Toomre}, \ref{fig:Zmax-ep}, and \ref{fig:lindband}. The three-star HD\,122956, HD\,5916, and HD\,189349 exhibit distinct chemical signatures consistent with their dynamical properties derived from the Lindblad diagram (Figure \ref{fig:lindband}), providing a coherent narrative of their Galactic origins. Colored error bars reflect propagated uncertainties\footnote{Errors in [Mg/Mn] are calculated as $\sqrt{\sigma^2_{\rm [Mg/Fe]} + \sigma^2_{\rm [Mn/Fe]}}$, while errors in [Al/Fe] are determined as $\sqrt{\sigma^2_{\rm [Al/H]} + \sigma^2_{\rm [Fe/H]}}$}.

HD\,122956, located in the high-$\alpha$ region ([Mg/Mn] $>$ 0.3 dex) in Figure~\ref{fig:chemicalplane}, occupies the ex situ region of the chemical plane, which is consistent with its retrograde-like kinematics ($L_{\rm z}=108.23\pm 3.51$ kpc km s$^{\rm-1}$) and highly eccentric orbit ($e_{\rm p} = 0.89\pm 0.01$). The extreme vertical excursion ($Z_{\rm max}=6.29\pm 0.02$ kpc) and low orbital energy ($E=(-0.61\pm 0.01)\times 10^5$ km$^{2}$ s$^{-2}$) further support its classification as a metal-poor halo star, likely accreted from a disrupted satellite galaxy.

\begin{figure}
    \centering
    \includegraphics[width=1\linewidth]{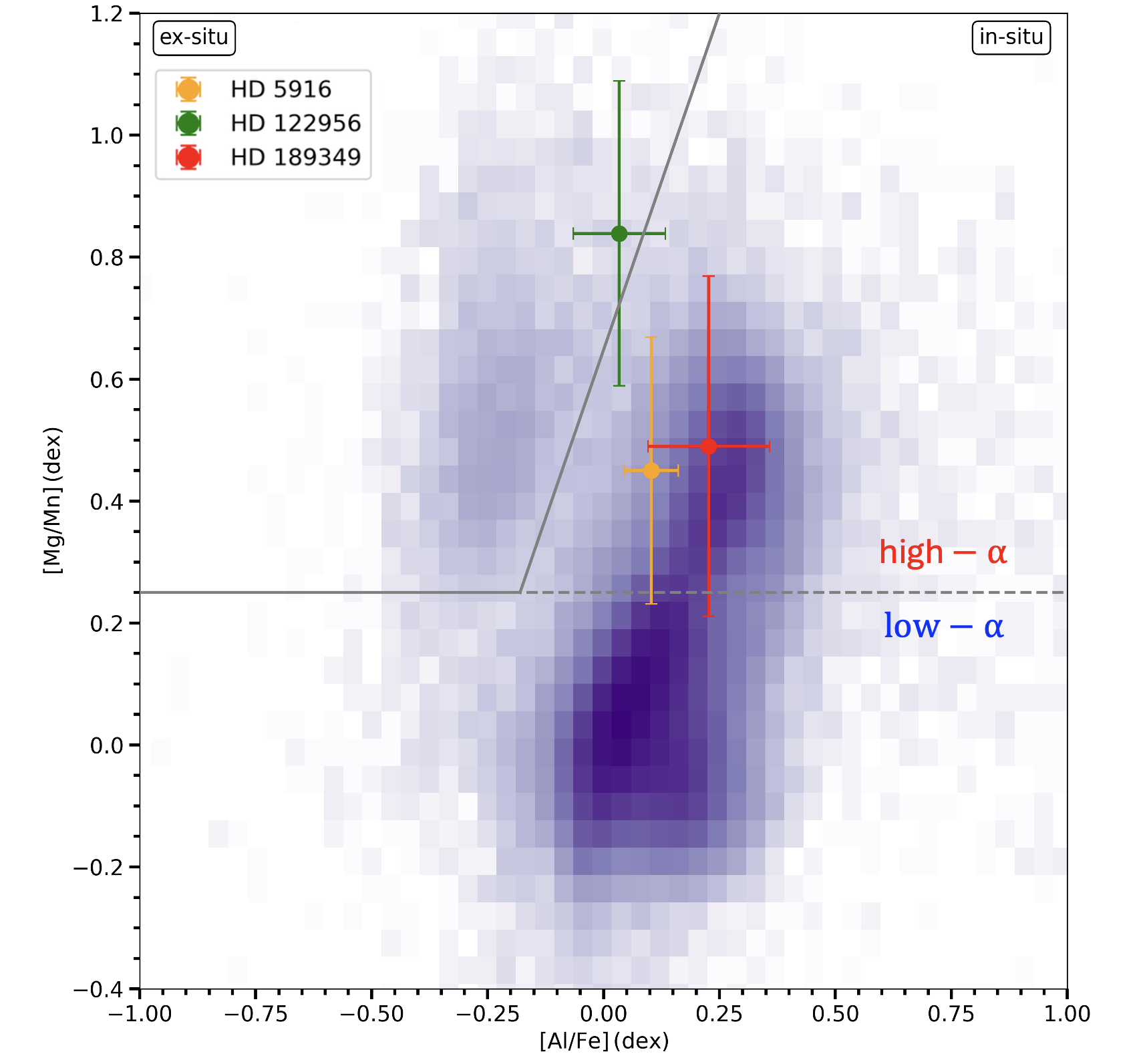}
    \caption{Distribution of selected stars in the [Mg/Mn] vs. [Al/Fe] abundance plane. The solid black lines indicate the empirical separation between the in-situ and ex-situ stellar populations, as proposed in recent chemical tagging studies. The colored error bars denote individual stars, with labels corresponding to their HD catalog numbers. The dashed horizontal line at [Mg/Mn] = 0.3 dex separates high-$\alpha$ and low-$\alpha$ populations, highlighted in red and blue, respectively. Background blue points represent APOGEE DR17 stars selected using quality cuts as described in the text.}
    \label{fig:chemicalplane}
\end{figure}

In Figure~\ref{fig:chemicalplane}, HD\,5916 lies close to the empirical boundary separating the in-situ and ex-situ populations, reflecting its intermediate chemical properties. This ambiguity parallels its kinematic behavior: positioned near $L_{\rm z} \approx 0$ with moderate orbital energy ($E=(-0.57 \pm 0.01)\times 10^5$ km$^{2}$ s$^{-2}$), HD\,5916 exhibits a ``splash'' orbit ($Z_{\rm max}=0.16\pm 0.01$ kpc) with slight eccentricity ($e_{\rm p}=0.44\pm 0.01$). Such transitional characteristics suggest that it may belong to a heated disk population or a merger remnant that retains partial angular momentum, bridging the disk-halo interface in both chemical and dynamical spaces.

HD\,189349, which is firmly situated in the high-$\alpha$, in-situ region of the chemical plane, aligns with its strongly prograding motion ($L_{\rm z} = 1079.26 \pm 2.75$ kpc km s$^{\rm-1}$) and tightly bound orbit ($E=(-0.58\pm 0.01)\times 10^{5}$ km$^{2}$ s$^{-2}$) in the Lindblad diagram. Its nearly circular orbit ($e_{\rm p}=0.42\pm 0.01$) and minimal vertical motion ($Z_{\rm max} = 0.06\pm 0.01$ kpc) are hallmarks of a disk-born star, which likely originates from a chemically enriched thin or thick disk. The elevated [Al/Fe] ratio of the star further supports its in-situ origin, reflecting rapid star formation in the early Galactic disk.}

The synergy between chemical abundance and orbital dynamics emphasizes the multifaceted nature of Galactic archaeology. HD\,122956’s ex-situ chemical signature and retrograde kinematics exemplify accreted halo populations, while HD\,189349’s in situ chemistry and prograde disk kinematics map the MW’s internal evolution. The transitional properties of HD\,5916 highlight the complex interplay between accretion events and dynamical heating and emphasize the need for combined chemical-kinematic analyses to resolve ambiguous populations. These results reinforce the utility of the [Mg/Mn]–[Al/Fe] plane as a complement to the Lindblad diagram for disentangling the assembly history of the MW. 

\begin{table*}
\centering
\caption{The Galactic coordinates ($l$, $b$), position ($P(\theta$)), velocity ($P(\nu)$) and their combined probability values ($P$(origin\textbar{}$\Theta$,$\nu$)) maximum distance from the Galactic plane ($Z_{\rm max}$), iron ($\rm [Fe/H]$) abundances, ages ($\tau$), and magnesium ($\rm [Mg/Fe]$) abundances under the scenario of the four stars leaving the GCs. The last column of the table contains the reference for [Fe/H] and [Mg/Fe] abundances and ages ($\tau$).}
\setlength{\tabcolsep}{9pt}
\footnotesize
\begin{tabular}{c|c|c|c|c|c|c|c|c|c}
\hline
Cluster & \textit{l} & \textit{b} & \textit{P}($\theta$) & \textit{P}($\nu$) & \textit{P}(origin|$\theta$,$\nu$) & [Fe/H] & $\tau$ & [Mg/Fe] & References \\	
        & $(^\circ)$ & $(^\circ)$ &        $\%$       &         $\%$      &              $\%$                & (dex) &  (Gyr) & (dex)   &            \\
\cline{1-10}
\textbf{NGC\,5139}	   &	309.10	&   ~14.97	&	96	&	84	&	81	&	-2.03$\pm$0.21	        &	13.0     	    &	0.43$\pm$0.22	&	18, 01, 02	\\
NGC\,6388	   &	345.56	&	~-6.74	&	96	&	70	&	67	&	-0.44$\pm$0.07	&	11.7	        &	0.06$\pm$0.17	&	03, 03, 03	\\
NGC\,6441	   &	353.53	&	~~5.01	&	76	&	72	&	54	&	-0.44$\pm$0.07	&   11.2$\pm$2.4    &	0.11$\pm$0.06	&	04, 05, 06	\\
NGC\,2808	   &	282.19	&	-11.25	&	48	&	93	&	45	&	-0.92$\pm$0.07	&	11.2	        &	0.22$\pm$0.04	&	07, 07, 06	\\
NGC\,6544	   &	~~5.84	&	~-2.20	&	90	&	31	&	28	&	-1.46$\pm$0.02	&	12.0	        &	0.24$\pm$0.03	&	08, 08, 08	\\
\hline
\bf HD\,2665   &\bf 120.11	&\bf ~-5.69	&\bf --	&\bf --	&\bf --	&\bf -2.05$\pm$0.09	&\bf 10.62$\pm$1.11 &\bf  0.23$\pm$0.09	&\bf	This Study	    \\
\hline
\hline
\textbf{NGC\,6441}	   &	353.53	&	~~5.01	&	87	&	90	&	79	&	-0.44$\pm$0.07	&	11.2$\pm$2.4   &	0.11$\pm$0.06	&	04, 05, 06	\\
NGC\,7078	   &	~64.01	&	-27.31	&	66	&	81	&	53	&	-2.22$\pm$0.14	&	13.6	        &	0.41$\pm$0.03	&	07, 07, 06	\\
NGC\,5927	   &	326.60	&	~~4.86	&	74	&	63	&	47	&	-0.48$\pm$0.05	&	12	            &	0.39$\pm$0.04	&	07, 07, 07	\\
NGC\,1851	   &	244.51	&	-35.04	&	67	&	61	&	41	&	-1.11$\pm$0.04	&	11.5	        &	0.22$\pm$0.08	&	07, 07, 07	\\
NGC\,2808	   &	282.19	&	-11.25	&	57	&	64	&	36	&	-0.92$\pm$0.07	&	11.2	        &	0.22$\pm$0.04	&	07, 07, 06	\\
\hline
\bf HD\,5916   &\bf 124.75	&\bf -17.39	&\bf --	&\bf --	&\bf --	&\bf -0.50$\pm$0.06	&\bf 2.72$\pm$1.88	&\bf 0.19$\pm$0.07	&\bf 	This Study      \\
\hline
\hline
\textbf{NGC\,6864}	   &	~20.30	&	-25.75	&	75	&	84	&	64	&	-1.29$\pm$0.14	&	11.25	        &	0.35$\pm$0.05	&	06, 09, 06	\\
NGC\,6517	   &	~19.22	&	~~6.76	&	75	&	83	&	62	&	-1.58	        &	12.5$\pm$2.0    &	0.24$\pm$0.05     &	10, 11,	16  \\
NGC\,6388	   &	345.56	&	~-6.74	&	69	&	82	&	57	&	-0.44$\pm$0.07	&	11.7            &	0.06$\pm$0.17   &	03, 03, 03	\\
NGC\,6093	   &	352.67	&	~19.46	&	86	&	65	&	56	&	-1.79$\pm$0.08	&	13.7$\pm$0.9    &	0.45$\pm$0.10	&	12, 11, 12	\\
NGC\,104	   &	305.90	&	-44.89	&	60	&	89	&	54	&	-0.74$\pm$0.03	&	12.5	        &	0.38$\pm$0.05	&	07, 07, 07	\\
\hline
\bf HD\,122956 &\bf 328.25	&\bf 44.36	&\bf --	&\bf --	&\bf --	&\bf -1.58$\pm$0.10	&\bf 10.24$\pm$1.59	&\bf 0.33$\pm$0.10	&\bf      This Study	\\
\hline
\hline
\textbf{NGC\,5927}	   &	326.60	&	~~4.86	&	90	&	70	&	63	&	-0.48$\pm$0.05	&	12	        &	0.27$\pm$0.02	&	07, 07, 15	\\
Pal\,10	       &	~52.44	&	~~2.72	&	77	&	56	&	43	&	-0.53$\pm$0.05	&	--	            &	0.12$\pm$0.01	&	06,   -, 06	\\
NGC\,7078	   &	~65.01	&	-27.31	&	42	&	89	&	37	&	-2.22$\pm$0.14	&	13.6	        &	0.41$\pm$0.03	&	07, 07, 06	\\
NGC\,6656	   &	~~6.72	&	~10.21	&	42	&	82	&	35	&	-1.52$\pm$0.09	&	12.7	        &	0.50$\pm$0.01	&	07, 07, 06	\\
NGC\,5139	   &	309.10	&	~14.97	&	71	&	46	&	33	&	-2.03$\pm$0.21  &	13	            &	0.43$\pm$0.22*	&	18, 01, 02	\\
\hline
\bf HD\,189349 &\bf	76.22	&\bf  6.09	&\bf --	&\bf --	&\bf --	&\bf -0.65$\pm$0.13	&\bf 2.60$\pm$1.43	&\bf 0.12$\pm$0.22  &\bf 	This Study	    \\
\hline
\hline
NGC\,5139      &	309.10	&	~14.97	&	97	&	68	&	66	&	-2.03$\pm$0.21	        &	13            	&	0.43$\pm$0.22	&	18, 01, 02	\\
{\bf NGC\,5634}&	342.21	&	~49.26	&	69	&	79	&	55	&	-1.94$\pm$0.10	&	11.2            &	0.52$\pm$0.03	&	13, 13, 17	\\
NGC\,2808	   &	282.19	&	-11.25	&	78	&	62	&	49	&	-0.92$\pm$0.07	&	11.2	        &	0.22$\pm$0.04	&	07, 07, 06	\\
NGC\,6356	   &	~~6.72	&	~10.21	&	57	&	79	&	45	&	-0.35$\pm$0.14	&	11.35$\pm$0.41	&	0.12$\pm$0.04	&	14, 14, 06	\\
NGC\,7078	   &	65.01	&	-27.31	&	61	&	69	&	42	&	-2.22$\pm$0.14	&	13.6	        &	0.31$\pm$0.10	&	07, 07, 06	\\
\hline
\bf HD\,218857   &\bf 52.93	&\bf -64.41	&\bf --	&\bf --	&\bf --	&\bf -1.98$\pm$0.10	&\bf 11.35$\pm$1.12	&\bf 0.16$\pm$0.10	&\bf This Study	\\
\hline 
\end{tabular}
\\
\label{tab:origin}
(1) \cite{Villanova2014}; (2) \cite{Magurno2019}; (3) \cite{Meszaros2020}; (4) \cite{Carretta2009}; (5) \cite{Marin-Franch2009}; (6) \citet{Dias2016}; (7) \cite{Kovalev2019}; (8) \cite{Gran2021}; (9) \cite{Usher2019}; (10) \cite{Schiavon2024}; (11) \cite{Santos2004}; (12) \cite{Carretta2015}; (13) \cite{Bellazzini2002}, (14) \cite{Koleva2008}, (15) \cite{2017Guzman}, (16) APOGEE DR17 Holtzman et al. (2025, in preparation), (17) \cite{carretta2017}, (18) \cite{Nitschai2024}
\end{table*}

\begin{figure*}
\centering
\includegraphics[width=18cm,height=14.cm,angle=0]{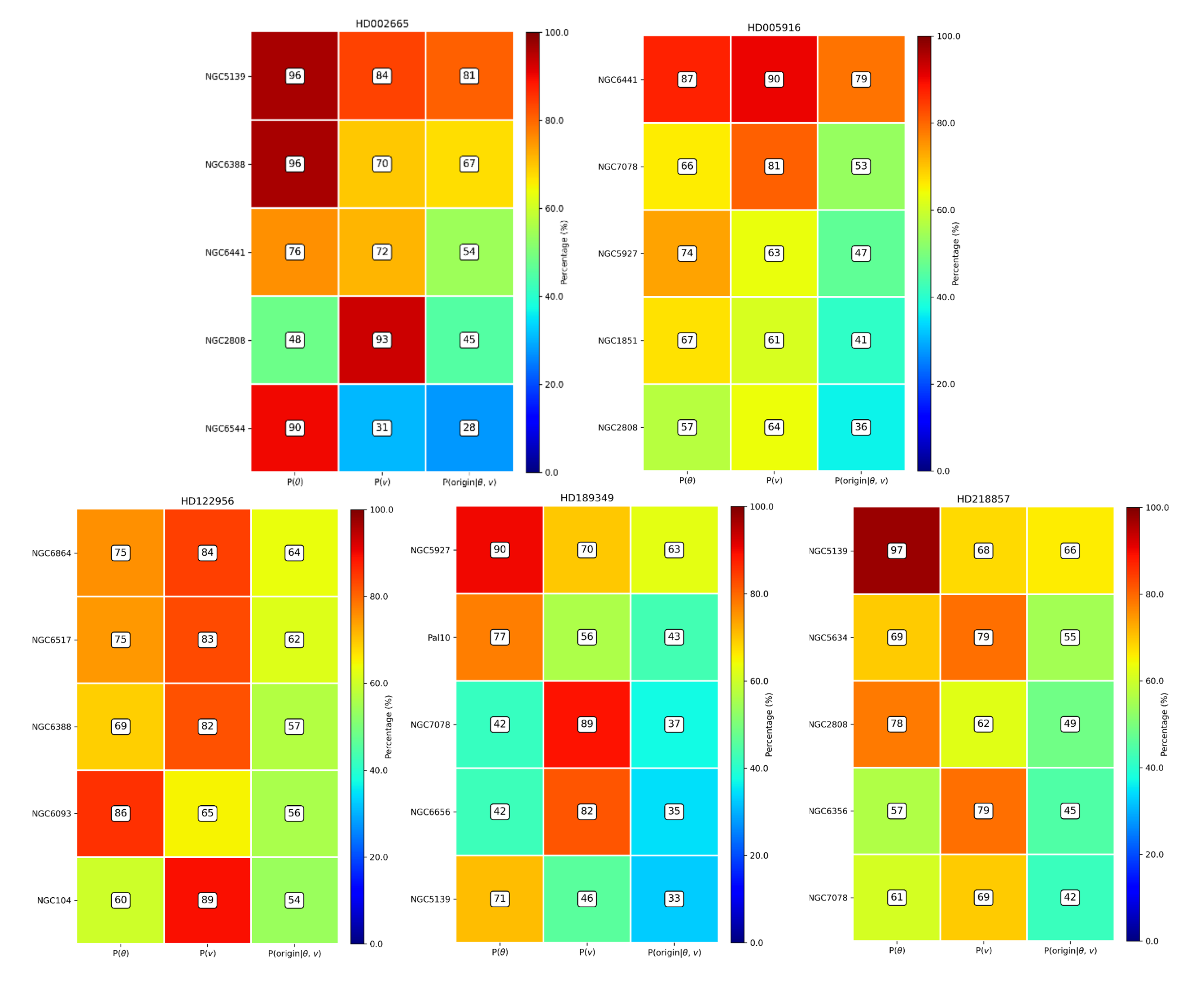}
\caption{Likelihood of stars encountering various GCs, based on spatial and velocity parameters. The matrix displays three key probability values: spatial probability ($P(\theta)$), velocity probability ($P(\nu)$), and combined probability ($P({\rm origin | \theta, \nu})$) for each cluster. Colors in the matrix reflect the probability values, as indicated by the scale.}
\label{fig:origin}
\end{figure*}

\subsection{Galactic Possible Origins} \label{sec:origin}

The detection of nearby stars with HPM and low metallicity is considered indicative of a possible origin involving ejection from GCs. Metal-poor HPM stars in the solar neighborhood exhibit distinct chemical and kinematic properties when compared to nearby Population I stars. Their presence in the local Galactic volume is difficult to reconcile with in-situ star formation scenarios and is more plausibly explained by an origin in a different Galactic component. Since most stars are thought to form within stellar clusters and can be dynamically ejected due to internal interactions or external tidal forces, probably, the stars examined in this study were originally associated with GCs. Motivated by this hypothesis, the present study investigates the possible origins of these stars by analyzing their kinematics, spatial distribution, and elemental abundance patterns in the context of the known properties of Galactic GCs. To investigate the possible Galactic origins of the five program stars, their orbital parameters were reconstructed and compared with those of known Galactic GCs. This comparison was based on observed properties including equatorial positions, distances, proper motions, and radial velocities, compiled from \citet{Baumgardt2019} and \citet{Vasiliev2021}. A comprehensive kinematic and dynamical analysis was then conducted for a total of 170 identified Galactic GCs.

To determine the Galactic orbital parameters of the GCs, symmetric Galactic potential models were employed using the {\tt galpy} Python library \citep{Bovy2015}, specifically incorporating models such as the {\sc MWPotential2014}. These calculations, extending from the present time back to 13 Gyr ago, involved simulating orbital motions using 20 million data-point. A time step of 650 years between successive points was selected after several iterations to ensure an optimal balance between the computational efficiency and accuracy. This resolution was found to be adequate for capturing dynamic interactions between stars and GCs. The use of multicore processors significantly accelerated the computational process. The resulting Galactic orbital parameters of the stars under investigation are listed in Table \ref{tab:uvw}.

Equatorial coordinates, proper motion components, trigonometric parallaxes, and radial velocities for the stars HD\,2665, HD\,5916, HD\,122956, HD\,189349, and HD\,218857 were obtained from the {\it Gaia} DR3 catalog \citep{GaiaDR3}. The orbital trajectories of these stars were modeled using the same computational framework applied to GC stars, involving backward integration over a 13 Gyr time. However, in calculating the likelihood of past encounters, the time intervals preceding the formation of stars were excluded to ensure that the analysis focused only on dynamically meaningful periods.

\begin{table*}
\setlength{\tabcolsep}{3.8pt}
    \renewcommand{\arraystretch}{1}
\caption{Chemical abundance ratios with uncertainties for GCs (NGC 5139, NGC 6441, NGC 6864, NGC 6517, NGC 5927, and NGC 5634).}
\footnotesize
    \centering
    \begin{tabular}{l|c|c|c|c|c|c|c|c|c|l}
    \hline
    Cluster 	&[Mg\,{\sc i}/Fe]& [Si\,{\sc i}/Fe]& [Ca\,{\sc i}/Fe]& [Sc\,{\sc ii}/Fe]& [Ti\,{\sc i}/Fe]& [Cr\,{\sc i}/Fe]& [Ni\,{\sc i}/Fe]& [Zn\,{\sc i}/Fe]& [Y\,{\sc ii}/Fe]&Ref\\
    \hline
    \hline

NGC 5139 	&0.43$\pm$0.22&  -- &  0.29$\pm$0.12&  0.11$\pm$0.21&  0.17$\pm$0.16&  0.09$\pm$0.18&  0.06$\pm$0.16&  0.30$\pm$0.11&  0.25$\pm$0.31& 1\\
NGC 6441 	&0.11$^{*}$$\pm$0.06&  0.36$\pm$0.18&  0.20$\pm$0.17&  0.15$\pm$0.15&  0.32$\pm$0.17& -0.06$\pm$0.20&  0.13$\pm$0.07&  --& -0.05$\pm$0.24& 2, 3\\
NGC 6864 	&0.35$^{*}$$\pm$0.05& 0.28$\pm$0.10&  0.17$\pm$0.13&  0.17$\pm$0.10&  0.16$\pm$0.11& -0.13$\pm$0.14& -0.11$\pm$0.09& -0.03$\pm$0.16&  0.06$\pm$0.20& 4\\
NGC 6517 	&0.24$\pm$0.05&  0.25$\pm$0.05&  0.33$\pm$0.07&  0.14$\pm$0.24&  0.57$\pm$0.28& -0.25$\pm$0.06&  --&  --&  --& DR17\\
NGC 5927 	&0.27$\pm$0.02&  0.2
4$\pm$0.03&  0.15$\pm$0.01&  0.32$\pm$0.02&  0.32$\pm$0.02&  0.03$\pm$0.02&  0.17$\pm$0.02& -0.04$\pm$0.04& -0.15$\pm$0.04& 5\\
NGC 5634 	&0.52$\pm$0.03&  --&  0.30$\pm$0.02&  --&  0.14$\pm$0.02&  --&  --&  0.01$\pm$0.05& -0.08$\pm$0.11& 6\\
\hline
    \end{tabular}
    
    \label{tab:cluster_abund}
    (*)\citet{Dias2016}, (1) \citet{Magurno2019}, (2) \citet{roediger2013}, (3) \citet{gratton2006}, (4)\citet{2013Kacharov}, (5) \citet{2017Guzman}, (6) \citet{carretta2017}
\end{table*}

In this study, the co-evolving orbital paths of five selected stars were examined to evaluate their spatial proximity to 170 GC centers \citep{Vasiliev2021} within the MW. The analysis focused on identifying potential encounters within a boundary defined as five times the tidal radius of each cluster. At every time step, the distance between each star and the GC center was calculated, and an encounter was recorded when the distance fell below the specified threshold. Additionally, the relative velocity ($\Delta \nu$) and angular separation ($\Delta \theta$) between the objects were determined using the following expressions:
\begin{equation}
    \Delta \theta = \sqrt{(X_{\rm s}-X_{\rm GC})^2+(Y_{\rm s}-Y_{\rm GC})^2+(Z_{\rm s}-Z_{\rm GC})^2},
\end{equation}
\begin{equation}
    \Delta \nu = \sqrt{(U_{\rm s}-U_{\rm GC})^2+(V_{\rm s}-V_{\rm GC})^2+(W_{\rm s}-W_{\rm GC})^2},
\end{equation}
Here, $\Delta \theta$ denotes the three-dimensional Cartesian distance between a star and the center of a GC at a given time $\tau$, and $\Delta \nu$ indicates the relative velocity. All spatial ($X, Y, Z$) and velocity ($U, V, W$) components are expressed in the Galactic coordinate system. In cases where multiple encounters occurred between a star and the same cluster over consecutive time steps, the corresponding probabilities were aggregated to obtain cumulative values. For encounters distributed across distinct epochs, the highest probability associated with each cluster was used for comparison. The probabilities related to the position ($P(\theta)$) and velocity ($P(\nu)$) were derived from Gaussian distributions extended over cosmological time intervals \citep[see also][]{Marismak2024}.
\begin{equation}
    P({\rm \theta}) = \frac{1}{\sqrt{2\pi R_{\rm tidal}}}\exp\left(-\frac{(\Delta \theta)^2}{2 R_{\rm tidal}}\right),
\end{equation}
\begin{equation}
    P({\rm \nu})= \frac{1}{\sqrt{2\pi V_{\rm escape}}}\exp\left(-\frac{(\Delta \nu)^2}{2 V_{\rm escape}}\right),
\end{equation}
with $R_{\rm tidal} = 5 \times R_{t}$ (five tidal radii) and $V_{\rm escape}$ values adopted from version 4 of the MW GC Database\footnote{\href{https://people.smp.uq.edu.au/HolgerBaumgardt/globular/}{{\url https://people.smp.uq.edu.au/HolgerBaumgardt/globular/}}}. The joint origin probability $P({\rm origin | \theta, \nu})$ is derived as follows:
\begin{equation}
    P({\rm origin | \theta, \nu})= P({\rm \theta}) \times P({\rm \nu}),
\end{equation}

Figure~\ref{fig:origin} shows the probability matrices that compare stars HD\,2665, HD\,5916, HD\,122956, HD\,189349, and HD\,218857 with 170 GCs within the MW \citep{Vasiliev2021}, emphasizing the five GCs with the highest likelihood of association. Complementing this visual representation, Table \ref{tab:origin} consolidates the calculated encounter probabilities, offering a structured basis for exploring the potential dynamic origins of these stars. The findings are grounded in detailed kinematic evaluations that incorporate both positional alignment and relative velocity consistency between the stars and GCs.

\begin{figure*}
    \centering
    \includegraphics[width=\linewidth]{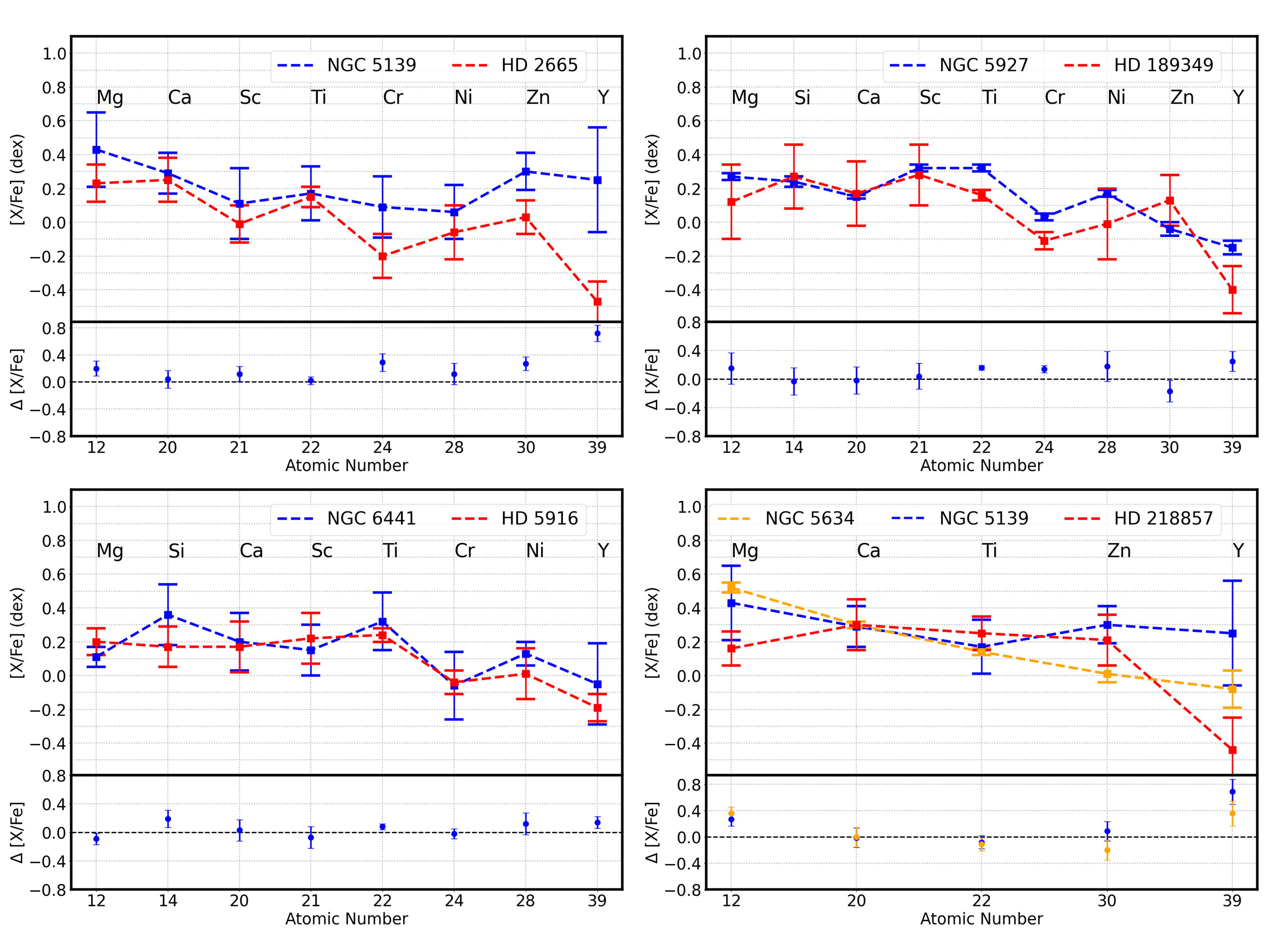}
    \caption{Elemental abundance ratios [X/Fe] as a function of atomic number for stars in the GCs NGC 5139 (top-left), NGC 6441 \citep[bottom-left,][]{roediger2013, gratton2006},  NGC 5927 \citep[top-right,][]{2017Guzman} and NGC 5634 \citep[bottom-right,][]{carretta2017},  compared with HD\,2665, HD\,5916, HD\,189349 and HD\,218857, respectively. The dashed blue lines represent the mean abundance patterns of GCs, and the red dashed lines indicate the field star abundance. The colored symbols with error bars denote individual elemental abundances, and the panels below each plot show the residuals $\Delta$[X/Fe] between the clusters and their corresponding comparison stars. Element symbols are labeled above their respective atomic numbers.}
    \label{fig:cluster_ab}
\end{figure*}

The program stars and top five GCs (out of 170) exhibiting the highest cumulative probability matches, along with their literature-derived metallicities, ages, and [Mg\,{\sc i}/Fe] abundances, are listed in Table \ref{tab:origin}. For HD\,2665, the dynamic analysis identified NGC\,5139, NGC\,6388, NGC\,6441, NGC\,2808, and NGC\,6544 as the most probable clusters of origin. NGC\,5139 exhibited the strongest association, with an 81$\%$ probability of spatial and kinematic similarity. The metallicity and Mg abundance of HD\,2665 align closely with the literature values for NGC\,5139. While an apparent age discrepancy exists, it remains consistent with the 2 Gyr age variation observed among its subpopulations, which span a metallicity range of [Fe/H] = -1.83 to -0.42 dex, as reported by \citet{Villanova2014} for 172 subgiant stars. The metallicity spread of the cluster was further exemplified by \citet{Johnson2020}, who identified very metal-poor members ([Fe/H] $<$ -2.5 dex) among the 395 SGB and RGB stars. Notably, \citet{Nitschai2024} identified a distinct subgroup within NGC\,5139 with [Fe/H] = -2.03 dex (see Table 2), further emphasizing the chemical complexity of the cluster. A detailed comparison of the cluster and stellar abundances, presented in Figure \ref{fig:cluster_ab} and Table \ref{tab:cluster_abund}, reveals an exceptional agreement ($<$0.1 dex) for Ca\,{\sc i}, Sc\,{\sc ii}, Ti\,{\sc i}, and Ni\,{\sc i}. For Mg\,{\sc i}, Cr\,{\sc i}, and Zn\,{\sc i}, the differences remained below 0.3 dex, whereas Y\,{\sc ii} exhibited a marginally larger deviation. The robust dynamical, metallicity, and chemical coherence, coupled with an interpretative framework accommodating age variations in NGC\,5139, supports the conclusion that HD\,2665 likely originated from this globular cluster. 

To place these findings in broader context, we computed the total number of close encounters between old program stars and GCs of comparable ages ($\approx$13 Gyr), using a maximum separation criterion of five tidal radii. The results suggest that approximately 50\,000–60\,000 such encounters occurred over the 13 Gyr integration time (see also \citealt{Cinar2025}), corresponding to an average of 3\,800-4\,600 encounters per Gyr. However, it is important to emphasize that the Galactic potential used in this analysis is static and does not evolve with time. Given that the integration spans a period as long as 13 Gyr, during which the mass distribution and structural properties of the Galaxy may have undergone significant evolution. As such, our proximity estimates should be interpreted within a probabilistic and statistical framework rather than as definitive timing determinations.

\begin{figure}
    \centering
        \includegraphics[width=0.92\linewidth]{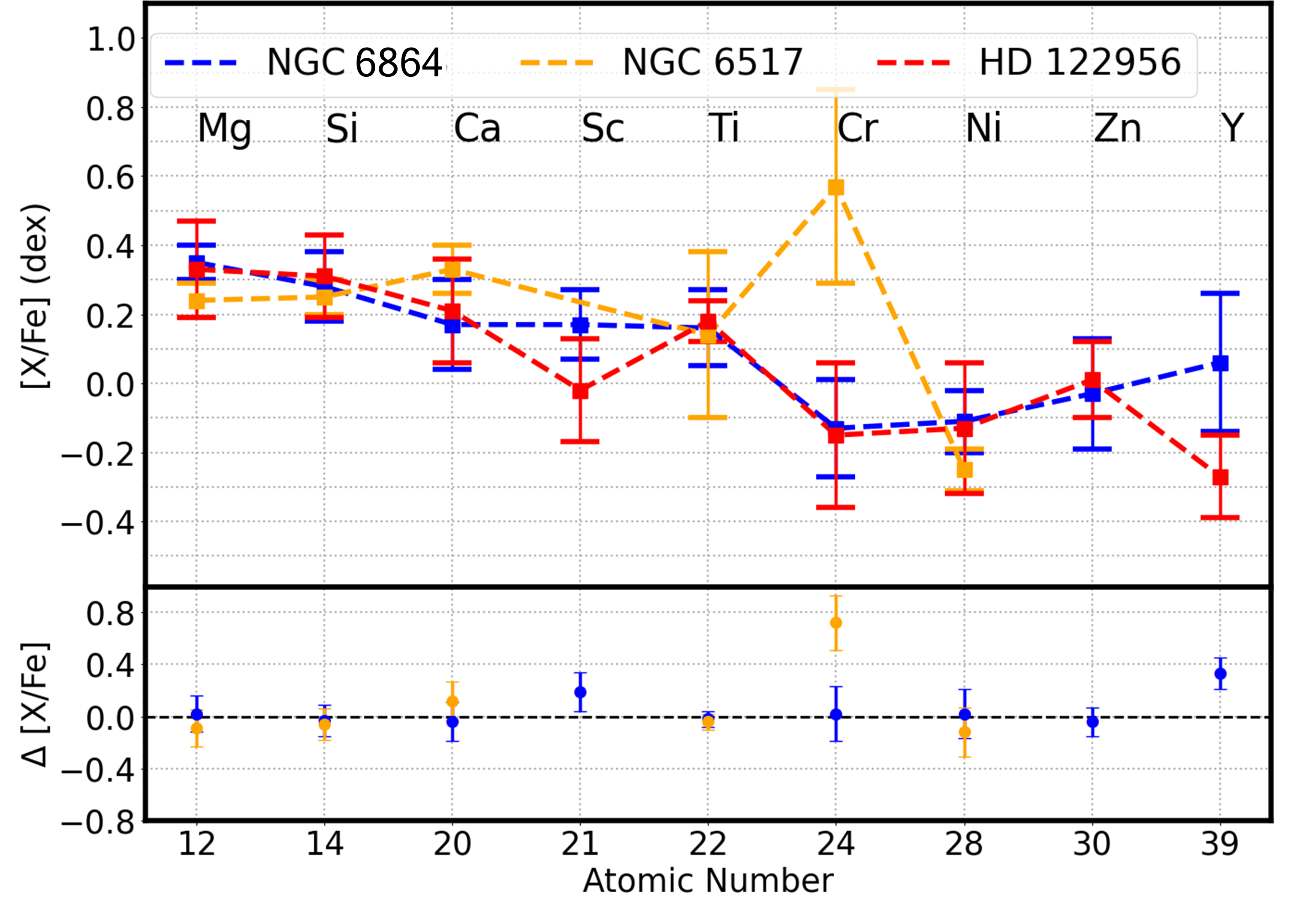}
    \caption{Comparative chemical abundance patterns of elements (Mg to Y) relative to Fe in the stars of GCs NGC 6517 (orange dashed line) and NGC 6864 (blue dashed line) versus field star HD\,122956. The upper panels show individual abundance ratios [X/Fe] with associated uncertainties for each element, while the lower panel displays the differences 
$\Delta$[X/Fe] between NGC 6517 - HD\,122956 (orange) and NGC 6864 - HD\,122956 (blue). The colored markers represent the observed data, with error bars indicating measurement uncertainties. }
\label{fig: hd122ab}
\end{figure}

For HD\,5916, the top five candidate clusters of dynamic origin were NGC 6441, NGC 7078, NGC 5927, NGC 1851, and NGC 2808. The star shows a 79$\%$ dynamical match (combined probability $P(\text{origin}|\theta, \nu$)) with NGC 6441, and its metallicity and Mg abundance closely align with those reported for this cluster. However, a significant age discrepancy persists between HD\,5916 and NGC\,6441, complicating definitive association. Numerous studies on NGC\,6441 include measurements of Mg\,{\sc i}, Si\,{\sc i}, Ca\,{\sc i}, Sc\,{\sc ii}, Ti\,{\sc i}, Cr\,{\sc i}, Ni\,{\sc i}, and Y\,{\sc ii} \citep{gratton2006, 2008Origlia, roediger2013}. These studies reported metallicities of -0.50$\pm$0.02, -0.46$\pm$0.06, and -0.43$\pm$0.08 dex, respectively. Notably, \citet{roediger2013} focused exclusively on $\alpha$-element abundance, whereas \citet{gratton2006} extended their analysis to include both $\alpha$- and Fe-peak elements. In Table \ref{tab:cluster_abund}, the Si\,{\sc i}, Ca\,{\sc i}, Ti\,{\sc i}, and Cr\,{\sc i} abundances were compiled from \citet{roediger2013}, while Sc\,{\sc ii}, Ni\,{\sc i}, and Y\,{\sc ii} values were taken from \citet{gratton2006}. The Mg abundance was further updated using high-precision measurements from \citet{Dias2016}. Despite the star’s chemical consistency with NGC\,6441 (abundance differences $<$0.2 dex across Mg\,{\sc i}, Si\,{\sc i}, Ca\,{\sc i}, Sc\,{\sc ii}, Ti\,{\sc i}, Cr\,{\sc i}, Ni\,{\sc i}, and Y\,{\sc ii}), the unresolved age disparity challenges the origin of GC. This inconsistency, coupled with the absence of corroborating dynamical evidence, strongly supports the classification of HD\,5916 as a field star rather than a tidally stripped cluster member. 

HD\,122956 shows 64$\%$ dynamical compatibility (combined probability $P(\text{origin}|\theta, \nu$)) with GC NGC\,6864. When metallicity, Mg abundance, and age criteria were considered, 62$\%$ agreement was observed with NGC\,6517. For NGC\,6864, elemental abundances derived from 16 giant stars were reported by \citet{2013Kacharov}. A systematic search of the APOGEE DR17 archive identified one confirmed member of NGC\,6517; however, no elemental abundance values for this cluster have been reported in other spectral archives or in the literature. The abundance ratios of HD\,122956 (Mg\,{\sc i}, Si\,{\sc i}, Ca\,{\sc i}, Ti\,{\sc i}, Cr\,{\sc i}, Ni\,{\sc i}, and Zn\,{\sc i}) exhibit discrepancies below 0.1 dex relative to NGC 6864. For Sc\,{\sc ii} and Y\,{\sc ii}, the differences remained under 0.2 dex and 0.3 dex, respectively. In contrast, comparisons with NGC 6517 reveal abundance differences below 0.1 dex for Mg\,{\sc i}, Si\,{\sc i}, Ca\,{\sc i}, and Cr\,{\sc i}, while Sc\,{\sc ii} shows a 0.2 dex offset (Figure \ref{fig: hd122ab}), and Ti\,{\sc i} diverges by approximately 0.4 dex. Although the metallicity of HD\,122956 is more consistent with that of NGC\,6517, the limited sample size (comprising only one member star) and the significant Cr\,{\sc i} discrepancy (0.4 dex) suggest that the origin associated with this GC is improbable. This conclusion may warrant reassessment pending future high-precision abundance studies of NGC\,6517.

For HD\,189349, a systematic dynamical analysis of 170 GCs identified NGC\,5927, Palomar 10, NGC\,7078, NGC\,6656, and NGC\,5139 as potential candidates with the highest posterior probabilities ($P(\text{origin}|\theta, \nu$)). Among them, NGC\,5927 emerged as the most probable progenitor, exhibiting the highest combined probability of 63$\%$. However, the remaining four clusters demonstrated significant discrepancies in metallicity, age, and [Mg/Fe] abundance ratios relative to HD\,189349. Chemically, the abundance ratios of Mg\,{\sc i}, Si\,{\sc i}, Ca\,{\sc i}, Sc\,{\sc ii}, Ti\,{\sc i}, Cr\,{\sc i}, Ni\,{\sc i}, Zn\,{\sc i}, and Y\,{\sc ii} in NGC\,5927—reported by \citet{2017Guzman} based on seven member stars—exhibit differences below 0.2 dex compared to those in HD\,189349. Notably, the abundance uncertainties reported by \citet{2017Guzman} reflect standard errors rather than standard deviations, which leads to comparatively small formal uncertainties in the present study. Nevertheless, the pronounced Y\,{\sc ii} discrepancy (+0.25 dex) between NGC\,5927 and HD\,189349, combined with the dynamical mismatches and chemical inconsistencies in other elements, strongly supports the classification of HD\,189349 as a field star rather than a tidally stripped cluster member. This interpretation aligns with the absence of conclusive chemical or dynamical evidence of the GC origin.

NGC\,5139 ($\omega$\,Cen) emerged as the most probable progenitor cluster for HD\,218857, with a 66$\%$ dynamical association probability. NGC\,5634 remains a plausible alternative, exhibiting a 55$\%$ probability when metallicity and age constraints are considered. Figure \ref{fig:cluster_ab} compares the abundance patterns of HD\,218857 with those of NGC\,5139 \citep{Magurno2019} and NGC\,5634 \citep{carretta2017}. For both clusters, Mg\,{\sc i}, Ca\,{\sc i}, and Ti\,{\sc i} abundances agree with HD\,218857 to within $<$0.1 dex. However, Zn\,{\sc i} and Y\,{\sc ii} exhibit larger discrepancies of $\approx$0.3 dex. While NGC\,5634 exhibits a marginal Mg\,{\sc i} and Y\,{\sc ii} offset (0.36 dex), its Ca\,{\sc i} and Ti\,{\sc i} abundances are closely aligned with those of the star. The strong dynamical and chemical coherence with NGC\,5139 supports its candidacy as the primary progenitor, although NGC\,5634 cannot be ruled out, pending future abundance studies on its members.
\subsection{Future works}
Although this study provides robust evidence for the origins of HD\,2665, HD\,5916, HD\,122956, HD\,189349, and HD\,218857, several avenues require further exploration. The subtle discrepancies in the Y\,{\sc ii} and Zn\,{\sc i} abundances (e.g., for HD\,2665 and HD\,218857) highlight the need for high-resolution, multi-epoch spectroscopic follow-up to refine the abundances of neutron-capture elements (e.g., Y, Ce, Nd), which are critical for distinguishing between GC ejecta and accreted halo stars. Improved N-body simulations of GC tidal interactions, coupled with {\it Gaia} DR4 astrometry, could further constrain the dynamical pathways linking field stars such as HD\,122956 to proposed progenitor clusters such as NGC\,6864 and test scenarios of tidal stripping versus early cluster dissolution. For chemically complex GCs such as NGC\,5139, systematic abundance studies of low-mass stars (e.g., turn-off and subgiants) are needed to clarify whether field stars such as HD\,2665 originate from specific metallicity subpopulations or represent stripped intracluster debris from previous galactic interactions. Additionally, for GCs such as NGC\,6517, where current abundance data rely on a single APOGEE ($H$-band) measurement, complementary optical spectroscopy of member stars is essential to validate chemical homogeneity and resolve discrepancies in key elements such as Ti and Cr.

\section*{Acknowledgments}
This study was supported by the Scientific and Technological Research Council of Türkiye (TÜBİTAK) under project number MFAG-121F265. The research presented here is part of the Ph.D. thesis of Gizay Yolalan. We gratefully acknowledge Mahmut Marışmak, Nur Çınar, Ferhat Güney, Sena A. Şentürk, and Deniz C. Çınar for their valuable assistance in the preparation of the figures and partial contributions to the analysis. The participants were graduate bursary recipients of the MFAG-121F265 project. This research used NASA's (National Aeronautics and Space Administration) Astrophysics Data System and the SIMBAD Astronomical Database, operated at CDS, Strasbourg, France, and the NASA/IPAC Infrared Science Archive, which is operated by the Jet Propulsion Laboratory, California Institute of Technology, under contract with the National Aeronautics and Space Administration. This study used data from the European Space Agency (ESA) mission {\it Gaia} (\mbox{https://www.cosmos.esa.int/gaia}) and, processed by the {\it Gaia} Data Processing and Analysis Consortium (DPAC, \mbox{https://www.cosmos.esa.int/web/gaia/dpac/consortium}). Funding for the DPAC was provided by national institutions, particularly those participating in the {\it Gaia} Multilateral Agreement.

\software{{\tt galpy} \citep{Bovy2015}, INSPECT program v1.0 \citep{Lind2012}, LIME \citep{Sahin2017}, MOOG \citep{Sneden1973}, SPECTRE \citep{Sneden1973}.
}

\bibliography{yolalan}
\bibliographystyle{aasjournal}

\begin{appendix}

\setcounter{table}{0}
\renewcommand{\thetable}{A\arabic{table}}

\setcounter{figure}{0}
\renewcommand{\thefigure}{A\arabic{figure}}

\begin{table*}
\setlength{\tabcolsep}{3pt}
\scriptsize
\caption{Literature-derived model atmosphere parameters for the five program stars, including effective temperature ($T_{\rm eff}$) surface gravity ($\log g$), and metallicity ([Fe/H]).}
\centering
\begin{tabular}{|l|ccc|ccc|ccc|ccc|ccc|}
\hline
 & \multicolumn{3}{c|}{HD\,002665} & \multicolumn{3}{c|}{HD\,005916} & \multicolumn{3}{c|}{HD\,122956} & \multicolumn{3}{c|}{HD\,189349} &\multicolumn{3}{c|}{HD\,218857} \\	
\cline{2-16}
References & $T_{\rm eff}$ & $\log g$ & [Fe/H] & $T_{\rm eff}$ & $\log g$ & [Fe/H] & $T_{\rm eff}$ & $\log g$ & [Fe/H] & $T_{\rm eff}$ & $\log g$ & [Fe/H] & $T_{\rm eff}$ & $\log g$ & [Fe/H] \\
\cline{2-16}
           &	(K)	       &	(cgs)  & (dex)  &	    (K)	    &  (cgs)   & (dex)	&	(K) &(cgs)  &(dex) &	(K) &(cgs)  &(dex) &	(K) &(cgs)  & (dex)\\
\cline{1-16}
TS (Spectroscopy)               & 5010 &	2.39	&	-2.05	&	5200	&	3.06	&	-0.50	&	4700	&	1.70	&	-1.58	&	5000	&	2.60	&	-0.65	&	5080	&	2.40	&	-1.98	\\
\citet{Gaia2022}	    & 6073 &	2.90	&	-0.67	&	5488	&	2.91	&	-0.13	&	--	&	--	&	--	&	5557	&	2.95&	-0.13&	5973	&	3.14	&	-0.61	\\
\citet{Gaia2018}	    & 4986 &	--	&	--	&	5029&	--	&	--	&	4712	&	--	&	--	&	5114&	--	&	--	&	5194	&	--	&	--	\\
\citet{Koelbloed1967}   & 4755 &	1.80	&	-1.56	&	--	&	--	&	--	&	--	&	--	&	--	&	--	&	--	&	--	&	--	&	--	&	--	\\
\citet{Sneden1974}	    & 4950&	2.60	&	-2.30	&	--	&	--	&	--	&	--	&	--	&	--	&	--	&	--	&	--	&	--	&	--	&	--	\\
\citet{Spite1978}	    & 4755 &	1.80	&	-1.60	&	--	&	--	&	--	&	--	&	--	&	--	&	--	&	--	&	--	&	--	&	--	&	--	\\
\citet{Luck1981}	    & --   &	--	&	--	&	--	&	--	&	--	&	--	&	--	&	--	&	--	&	--	&	--	&	5200&	2.00	&	-2.11	\\
\citet{Gratton1984}	    & --   &	--	&	--	&	--	&	--	&	--	&	4630&	1.70&	-2.00&	--	&	--	&	--	&	--	&	--	&	--	\\
\citet{Cottrell1986}    & --   &	--	&	--	&	4750&	2.00	&	-0.80	&	--	&	--	&	--	&	--	&	--	&	--	&		&		&		\\
\citet{Gratton1986}	    & --   &	--	&	--	&	--	&	--	&	--	&	4630&	1.70	&	-2.00&	--	&	--	&	--	&	--	&	--	&	--	\\
\citet{Gilroy1988}	    & 5000 &	2.50	&	-2.00	&	--	&	--	&	--	&	4800	&	1.50	&	-1.70&	--	&	--	&	--	&	--	&	--	&	--	\\
\citet{Gratton1991}	    & --   &	--	&	--	&	--	&	--	&	--	&	4609	&	1.38	&	-1.50&	--	&	--	&	--	&	--	&	--	&	--	\\
\citet{Sneden1991}	    & --   &	--	&	--	&	--	&	--	&	--	&	4650&	1.20&	-2.61&	--	&	--	&	--	&	--	&	--	&	--	\\
\citet{Francois1993}    & 5000 &	2.50	&	-2.00	&	--	&	--	&	--	&	4800	&	1.50	&	-1.70&	--	&	--	&	--	&	--	&	--	&	--	\\
\citet{Pilachowski1993} & 5100 &	2.20	&	-1.89	&	--	&	--	&	--	&	--	&	--	&	--	&	--	&	--	&	--	&	5100	&	--	&	-1.87	\\
\citet{Axer1994}	    & --   &	--	&	--	&	--	&	--	&	--	&	--	&	--	&	--	&	--	&	--	&	--	&	5130	&	3.10	&	-1.90	\\
\citet{Gratton1994}  	& --   &	--	&	--	&	--	&	--	&	--	&	4609	&	1.56	&	-1.70&	--	&	--	&	--	&	--	&	--	&	-1.90	\\
\citet{Gratton1996}	    & --   &	--	&	--	&	--	&	--	&	--	&	4485	&	1.08	&	-1.76	&	--	&	--	&	--	&	--	&	--	&	--	\\
\citet{Pilachowski1996}	& 5000 &	2.20	&	-1.97	&	--	&	--	&	--	&	4600	&	1.50	&	-1.78	&	--	&	--	&	--	&	5125	&	2.40	&	-1.86	\\
\citet{Alonso1999}	    & 4990 &	2.40	& -1.92 &	4944	&	2.00&	-0.80&	--	&	--	&	--	&	--	&	--	&	--	&	--	&	--	&	--	\\
\citet{Tomkin1999}	    & --   &	--	&	--	&	--	&	--	&	--	&	4600	&	1.20	&	-1.88	&	--	&	--	&	--	&	--	&	--	&	--	\\
\citet{Burris2000}	    & 5000 &	2.20	&	-1.97	&	--	&	--	&	--	&	4600	&	1.50	&	-1.78	&	--	&	--	&	--	&	5125	&	2.40	&	-1.86	\\
\citet{Fulbright2000}	& 5050 &	2.20	&	-1.96	&	--	&	--	&	--	&	&	&	&	--	&	--	&	--	&	--	&	--	&	--	\\
\citet{Gratton2000}	    & 5061 &	2.35	&	-1.94	&	--	&	--	&	--	&	4670	&	1.63	&	-1.63	&	--	&	--	&	--	&	--	&	--	&	--	\\
\citet{Mishenina2001}	& --   &	--	&	--	&	4863	&	1.70	&	-0.51	&	4635	&	1.50	&	-1.60	&	--	&	--	&	--	&	5050	&	2.40	&	-1.84	\\
\citet{Melendez2002}	& --   &	--	&	--	&	--	&	--	&	--	&	4540	&	1.60	&	-1.60	&	--	&	--	&	--	&	--	&	--	&	--	\\
\citet{Mishenina2003}	& --   &	--	&	--	&	4863	&	1.70	&	-0.51	&	4635	&	1.50	&	-1.60	&	--	&	--	&	--	&	--	&	--	&	--	\\
\citet{Cenarro2007}	    & 5013 &	2.35	&	-1.96	&	4863	&	2.50	&	-0.80	&	4635	&	1.49	&	-1.75	&	--	&	--	&	--	&	5082&	2.41	&	-1.93	\\
\citet{Carney2008}	    & 5000  &	2.34	&	-1.99	&	--	&	--	&	--	&	--	&	--	&	--	&	--	&	--	&	--	&	5040&	2.42	&	-2.15	\\
\citet{Soubiran2008}	& 5021 &	2.28	&	-1.98	&	4912	&	1.76	&	-0.57	&	4613	&	1.41	&	-1.74	&	--	&	--	&	--	&	5109&	2.47	&	-1.88	\\
\citet{Gonzalez2009}	& 5123 &	2.31	&	-1.98	&	5047	&	1.85	&	-0.65	&	--	&	--	&	--	&	--	&	--	&	--	&	--	&	--	&	--	\\
\citet{Prugniel2011}	& 4986 &	2.28	&	-1.96	&	4954	&	2.31	&	-0.75	&	4709	&	1.54	&	-1.68	&	--	&	--	&	--	&	5057&	2.43	&	-1.93	\\
\citet{Ishigaki2012}	& --   &	--	&	--	&	--	&	--	&	--	&	4609	&	1.57	&	-1.72	&	--	&	--	&	--	&	5107	&	2.71	&	-1.92\\
\citet{Koleva2012}	    & 5100 &	2.62	&	-1.88	&	4977	&	2.69	&	-0.78	&	4932	&	2.33	&	-1.47	&	--	&	--	&	--	&	--	&	--	&	--	\\
\citet{Ishigaki2013}	& --   &	--	&	--	&	--	&	--	&	--	&	4609	&	1.57	&	-1.70	&	--	&	--	&	--	&	5107&	2.71&	-1.94	\\
\citet{Matrozis2013}	& --   &	--	&	--	&	--	&	--	&	--	&	4698	&	1.44	&	-1.74	&	--	&	--	&	--	&	--	&	--	&	--	\\
\citet{Molenda2013}	    & --   &	--	&	--	&	--	&	--	&	--	&	--	&	--	&	--	&	5163	&	2.88	&	-0.44	&	--	&	--	&	--	\\
\citet{Molenda2013}	    & --   &	--	&	--	&	--	&	--	&	--	&	--	&	--	&	--	&	5121	&	2.83	&	-0.56	&	--	&	--	&	--	\\
\citet{Beers2014}	    & --   &	--	&	--	&	--	&	--	&	--	&	4521	&	1.50	&	-1.83	&	--	&	--	&	--	&	5134&	2.43	&	-1.89	\\
\citet{Beers2014}$^{\rm*}$ & --&	--	&	--	&	--	&	--	&	--	&	4617	&	1.42	&	-1.73	&	--	&	--	&	--	&	5106&	2.58	&	-1.87	\\
\citet{Beers2014}	    & --   &	--	&	--	&	--	&	--	&	--	&	4748	&	1.64	&	-1.83	&	--	&	--	&	--	&	5269&	2.39	&	-1.88	\\
\citet{Huang2015}	    & --   &	--	&	--	&	--	&	--	&	--	&	--	&	--	&	--	&	5282	&	--	&	-0.56&	--	&	--	&	--	\\
\citet{Takeda2015}	    & --   &	--	&	--	&	--	&	--	&	--	&	--	&	--	&	--	&	5026	&	2.45	&	-0.63	&	--	&	--	&	--	\\
\citet{Boeche2016}	    & 4972 &	2.39	&	-2.07	&	5034	&	2.61	&	-0.71	&	4728	&	1.81	&	-1.70	&	--	&	--	&	--	&	5071&	2.63	&	-2.01	\\
\citet{Boeche2016}	    & 5046 &	2.39	&	-1.98	&	5050	&	2.65	&	-0.69	&	--	&	--	&	--	&	--	&	--	&	--	&	--	&	--	&	--	\\
\citet{Hawkins2016}	    & --   &	--	&	--	&	--	&	--	&	--	&	--	&	--	&	--	&	--	&	--	&	--	&	5162	&	2.66	&	-1.78	\\
\citet{Niedzielski2016}	& --   &	--	&	--	&	4915	&	2.27	&	-0.72	&	--	&	--	&	--	&	--	&	--	&	--	&	--	&	--	&	--	\\
\citet{Mashonkina2017}	& --   &	--	&	--	&	--	&	--	&	--	&	--	&	--	&	--	&	--	&	--	&	--	&	5060	&	2.53	&	-1.92	\\
\citet{Afsar2018}	    & --   &	--	&	--	&	--	&	--	&	--	&	--	&	--	&	--	&	4978	&	2.16	&	-0.69	&	--	&	--	&	--	\\
\citet{Deka2018}	    & --   &	--	&	--	&	4915	&	2.27	&	-0.72	&	--	&	--	&	--	&	--	&	--	&	--	&	--	&	--	&	--	\\
\citet{Yu2018}	        & --   &	--	&	--	&	--	&	--	&	--	&	--	&	--	&	--	&	5121	&	2.39&	-0.56	&	--	&	--	&	--	\\
\citet{Arentsen2019}	& --   &	--	&	--	&	--	&	--	&	--	&	4669	&	1.44	&	-1.74	&	--	&	--	&	--	&	5107	&	2.58	&	-1.90	\\
\citet{Gaulme2020}	    & --   &	--	&	--	&	--	&	--	&	--	&	--	&	--	&	--	&	5121	&	2.40	&	-	&	--	&	--	&	--	\\
\citet{Hon2021}	        & 5015   & --	&	--	&	4905	&	--	&	--	&	--	&	--	&	--	&	5193	&	--	&	--	&	5106&	--	&	--	\\
\citet{Soubiran2022}	& 5020   &	2.28	&	-1.97	&	4919	&	2.06	&	-0.66	&	4625	&	1.45	&	-1.72	&	5076	&	2.66	&	-0.59	&	5120&	2.58&	-1.89\\
\citet{Pal2023}	        & 5100  & 2.62	& -1.88	&	4977	&	2.69	&	-0.78	&	--	&	--	&	--	&	5121	&	2.38	&	-0.56	&	5107&	2.58&	-1.90\\

\hline 
\end{tabular}
\label{tab:lit1}
\end{table*}

\begin{figure}
\centering
\includegraphics[width=1\linewidth]{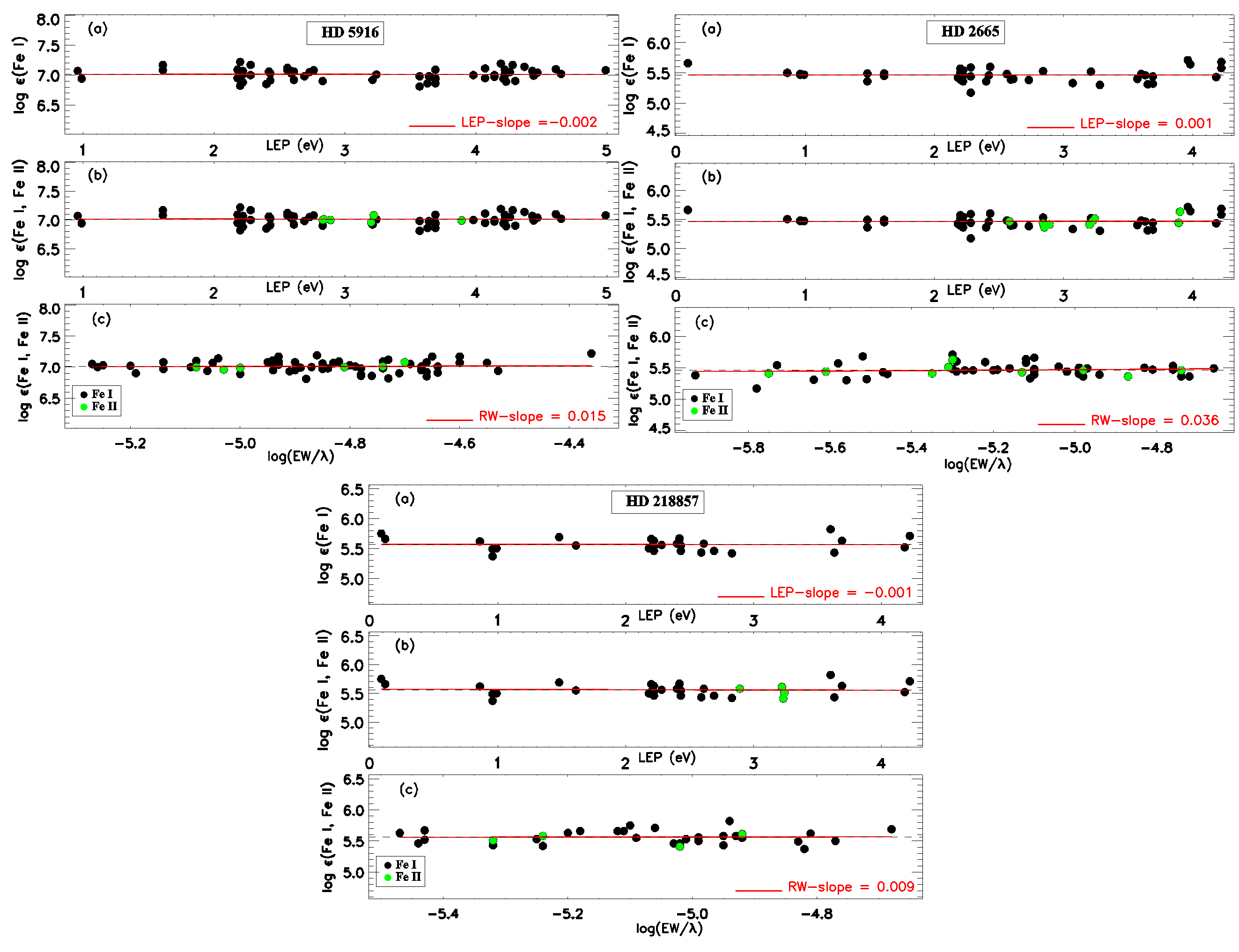}
\caption{An example for the determination of atmospheric parameters $T_{\rm eff}$ and $\xi$ using abundance (log$\epsilon$) as a function of both lower level excitation potential (LEP, panels a and b) and reduced EW (REW; log (EW/$\lambda$), panels c) for HD\,5916, HD\,2665, and HD\,218857. The solid line in all panels is the least-squares fit to the data.}
\label{fig:moog_results_2}
\end{figure}

\begin{figure}
    \centering
    \includegraphics[width=1\linewidth]{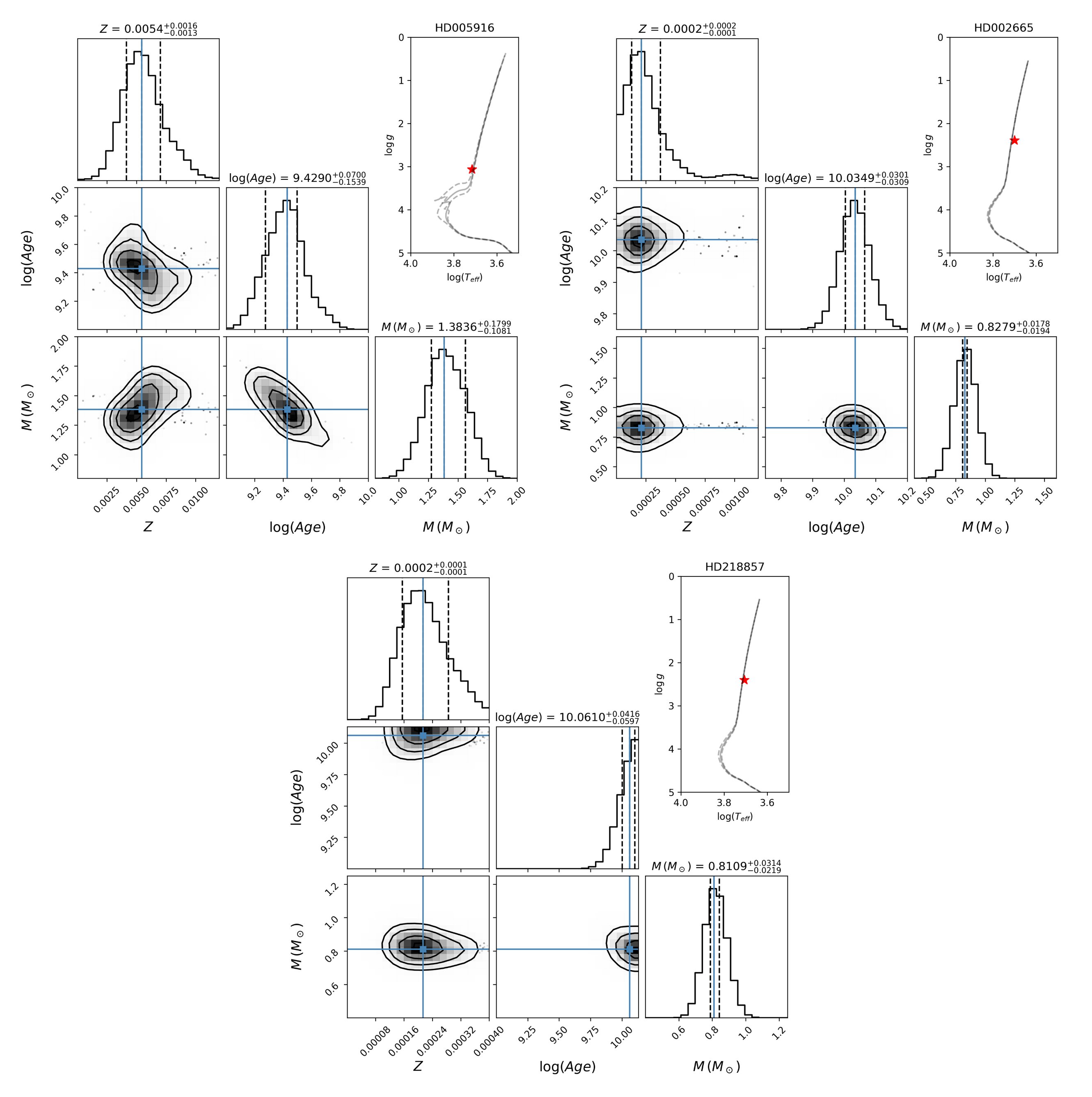}
    \caption{Corner plot displaying the posterior probability distributions for HD\,5916, HD\,2665, and HD\,218857, with confidence levels marked at 68\%, 90\%, and 95\%. The one-dimensional marginal distributions indicate the median values, along with the 16th and 84th percentiles. The right side of each panel shows the corresponding positions of the stars on the Kiel diagram.}
    \label{fig:age-corner}
\end{figure}

\begin{figure}
    \centering
    \includegraphics[width=0.98\linewidth]{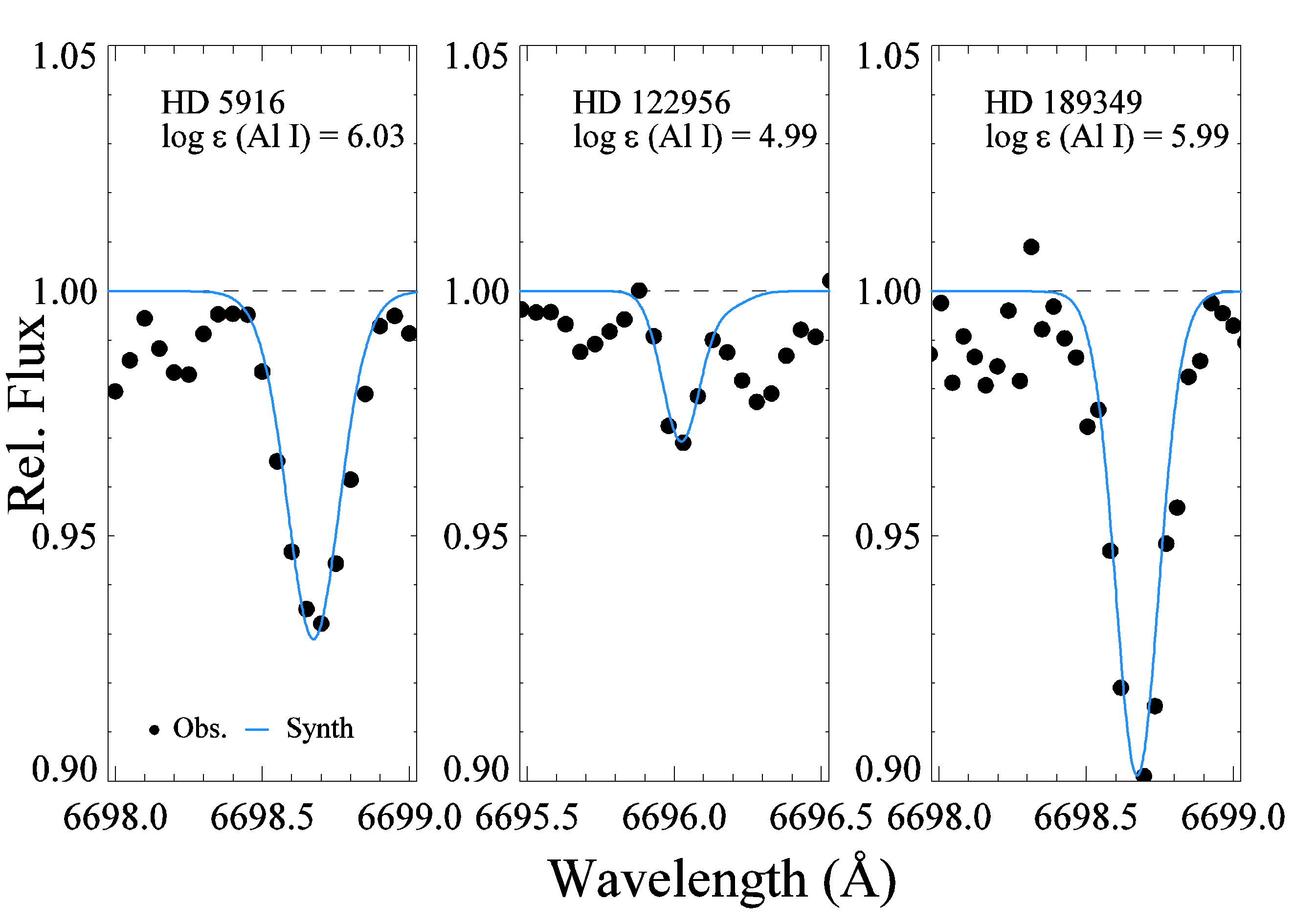}
    \caption{Observed (black) and synthetic (blue) spectra of HD\,5916, HD\,122956, and HD\,189349 near the Al \,{\sc i} $\lambda$6696-6699 \AA\, line. The derived aluminum abundances (log $\epsilon$(Al\,{\sc i})) are provided for each star.  The dashed horizontal lines denote the continuum level, and wavelength ranges are truncated for clarity.}
    \label{fig:alsynth}
\end{figure}

\end{appendix}
\end{document}